\PassOptionsToPackage{desactivate}{linenoaa}
\documentclass{aa}

\usepackage{graphicx}
\usepackage{txfonts}
\usepackage{lipsum}
\usepackage{subcaption}         
\usepackage{xcolor}             
\usepackage{lscape}             
\usepackage{placeins}           

\usepackage{amsmath}
\usepackage{makecell} 
\usepackage{arydshln}

\begin{document}

   \title{Resonant structures in exozodiacal clouds created by exo-Earths in the habitable zone of late-type stars}



 \author{Seung-Yoo Lee\inst{1,2}
        \and Masateru Ishiguro\inst{1,2}
        \and Hangbin Jo\inst{1,2}
        \and Sung-Chul Yoon\inst{1,2}}

   \institute{Department of Physics and Astronomy, Seoul National University, 1 Gwanak-ro, Gwanak-gu, Seoul 08826, Republic of Korea
   \and SNU Astronomy Research Center, Department of Physics and Astronomy, Seoul National University, 1 Gwanak-ro, Gwanak-gu, Seoul 08826, Republic of Korea\\
              \email{studymoon@snu.ac.kr, ishiguro@snu.ac.kr, hangbin9@naver.com, scyoon@snu.ac.kr}
             }



\abstract
   {Earth-like exoplanets can create resonant structures in exozodiacal dust through mean motion resonances (MMRs). These structures not only suggest the presence of such planets, but also act as potential noise sources in future mid-infrared (MIR) nulling interferometry observations.}
   {We aim to investigate how resonant structures in exozodiacal dust vary across stellar spectral types (F4--M4), and to evaluate how stellar wind drag affects their morphology and brightness in mature planetary systems.}
   {We conducted numerical simulations of dust dynamics, extending earlier studies by including spectral type variation in stellar wind drag in addition to Poynting-Robertson (PR) drag. Our models represented systems of a few Gyr hosting an Earth-like exoplanet in the habitable zone (HZ). We produced spatially resolved maps of optical depth and thermal emission for different stellar spectral types.}
   {Our simulations showed that resonant ring structures were formed for all stellar spectral types considered. In particular, we found that stellar wind drag played a critical role in shaping dust dynamics around old M-type stars, where it could dominate over PR drag by a factor of approximately 44. This reduced the contrast of resonant rings relative to the background disk, compared to cases without spectral type variation in stellar wind. Across different spectral types, the optical depth contrast of the resonant ring increased for lower-mass stars, assuming a fixed background level. Asymmetric thermal emission distributions were derived across all spectral types, which peaked for K-type stars under a dust size range of 0.1--300~$\mathrm{\mu m}$.}
   {Our findings highlight the importance of incorporating both resonant dynamics and stellar wind effects when modeling exozodiacal dust around stars of different spectral types.}


   \maketitle

\section{Introduction} \label{sec:intro}
Dust particles in debris disks can form resonant structures through interactions with planets. As the particles spiral inward under the influence of Poynting-Robertson (PR) drag and stellar wind drag, they may become trapped in mean motion resonances (MMRs), leading to enhanced densities along resonant orbits \citep{Kuchner2003,KH2003,Kucner2007,Pastor2009,Shannon2015}.
In the Solar System, such structures are exemplified by Earth’s resonant ring \citep{Dermott1994,LZ1999,KS2010,Sommer2020}. This structure has been observationally confirmed in the infrared by space telescopes such as IRAS and COBE/DIRBE \citep{Dermott1984,Reach1995}.
Similar processes can take place in extrasolar systems, producing asymmetric ring structures in exozodiacal dust (exozodi) near planetary orbits \citep{Kral2017,Pearce2024,currie2025}.

As the extrasolar analogs of the zodiacal dust, exozodis can be a critical foreground for the detection of potentially habitable exoplanets. Located near the habitable zone (HZ), warm exozodis ($T\sim 300 ~ \mathrm{K}$; \citealt{Pearce2024}) emit in the mid-infrared (MIR), producing thermal radiation that can hinder the detection of Earth-like planets—particularly for upcoming MIR nulling interferometry missions such as the Large Interferometer For Exoplanets (LIFE; \citealt{Quanz2022, Dannert2022, CG2023}).
While symmetric dust disks can be suppressed through phase chopping techniques and primarily contribute to shot noise \citep{Defrere2010,Dannert2022}, asymmetric features such as gaps and clumps in resonant structures \citep{SK2008, Stark2011} can produce signals that may be confused with planetary detections \citep{Lay2004, Defrere2010, Defrere2012, Quanz2022}.
This challenge is particularly acute for late-type stars, whose compact HZs (typically within a few tenths of an au) may necessitate nulling interferometry for the search for exo-Earths \citep{Absil2006,Defrere2010,Dannert2022}. Therefore, proper modeling of exozodi resonant structures is critical to minimize hindrance to planetary detections and optimize observational strategies.

Among many studies on exozodi structures \citep[e.g.,][]{Wyatt2003,Rodigas2014,KP2015,Kral2017}, \citet{SK2008} studied the resonant structures in systems with exo-Earths and super-Earths. Under a collisionless, steady-state disk model, they focused on the contrast of resonant ring structures and conducted simulations for cases with a single terrestrial planet around a Sun-like star---meaning that only G-type stars were considered in their research.
While \citet{SK2009} extended their earlier work by incorporating collisional processes---which can diminish resonant asymmetries---their simulations remained limited to Sun-like systems and did not account for spectral type variation in stellar wind. Subsequent collisional models of exozodis \citep{KS2010, Stark2011, Stark2015} retained these constraints. Although \citet{Stark2015} considered planets in the HZ for direct imaging with nulling interferometry such as the Large Binocular Telescope Interferometer (LBTI; \citealt{Kennedy2015}), their focus remained on edge-on, Sun-like systems observed in scattered light, again omitting stellar wind effects.
A more recent study by \citet{Currie2023} also emphasizes the role of resonant structures as potential noise sources in coronagraphic imaging for missions such as the Habitable Worlds Observatory \citep[HWO;][]{HWO}. Similar to \citet{Stark2015}, their work focuses on mitigating structured exozodi signals around Solar analogs, primarily in scattered light rather than thermal emission, and does not consider the spectral dependence of stellar wind.

The limitations on spectral type and the lack of prescription for spectral type variation in stellar wind remain in the works of \citet{Defrere2010, Defrere2012}, which examined the impact of exozodi resonant structures on the detection of exo-Earths with coronagraphs and interferometers.
Although the models of \citet{Kennedy2015} and \citet{RW2020} considered a range of spectral types (A to K) when assessing detectability with LBTI, they ignored stellar wind effects and did not consider resonant structures.
Additionally, while \citet{Stark2022} presented a tool for quickly generating debris disk models for various planetary systems, it does not rely on full numerical simulations and retains similar limitations, such as the absence of stellar wind and the lack of treatment for spectral type-dependent blowout sizes.

This study addresses the limitations of previous research by extending the \citet{SK2008} framework to explore resonant structures around a broader range of spectral types, from F4 to M4, and incorporating the effect of stellar wind drag with spectral type variation.
Considering the most favorable conditions in terms of habitability, we focus on systems with late-type stars at ages of a few Gyr similar to the Sun, and Earth-like planets placed in the conservative habitable zone (CHZ; \citealt{Kopparapu2014}).
While dust migration by stellar wind drag is known to be insignificant around G-type stars like our Sun at Gyr ages \citep{NK2023},  dust dynamics around later-type stars may experience significant influence from stellar winds. \citet{Plavchan2005} showed that stellar wind drag dominates for M-dwarfs at ages of tens to hundreds of Myr, and \citet{Reid2011} demonstrated its strong effect around a K-type star aged $\lesssim 1~\mathrm{Gyr}$. We explore whether this dominant role of stellar wind persists in more mature systems.
Previous studies on the significance of solar wind relative to the PR effect \citep[e.g.,][]{Klacka2014} further support the proper inclusion of stellar wind.
While similar research emphasizing the importance of stellar wind over radiation in debris disks around old late-type stars has been done for gas dynamics \citep{Kral2023}, simulations on dust resonant structures have yet to be done.

Overall, although stellar wind may play a non-negligible role in shaping dust dynamics, particularly around old, late-type stars, it remains largely absent from models of resonant structures. Therefore, we examine how different spectral types and corresponding stellar winds alter the morphology and brightness of resonant features, ultimately to assess their observational impact on future MIR interferometric missions. In particular, we aim to improve exozodi models used in performance tools like LIFEsim \citep{Dannert2022}, which currently assume homogeneous, symmetric dust distributions based on the disk models from \citet{Kennedy2015}. As explained previously, these models do not account for either resonant clumps or stellar wind effects.
Although \citet{Quanz2022} examined the impact of resonant structures on LIFE’s performance using the work of \citet{Defrere2010}, their analysis is based on the results from \citet{SK2008}, which are limited to a Solar twin system.
Our work fills this gap by providing updated dust structure predictions that incorporate both resonant dynamics and stellar wind physics, enabling more accurate exozodi noise modeling and optimized planet detection strategies.

Building on these motivations, we conduct two categories of simulations across spectral types: (1) a pilot study assessing the influence of stellar wind using a fixed dust setup, and (2) a main study examining resonant structures with multi-size dust relevant for exo-Earth observations.
While we do not perform detailed morphological analyses of resonant structures as in \citet{SK2008}, our simulations provide a foundation for integrating wind-influenced, resonant disk models into interferometric mission planning for late-type stars. In Section~\ref{sec:method}, we describe the methods of our study, followed by the results in Section~\ref{sec:results}. Detailed analyses and discussions are presented in Section~\ref{sec:discussions}, and we summarize our conclusions in Section~\ref{sec:conclusion}.

\section{Method} \label{sec:method}

\subsection{Dynamical modeling framework} \label{sec:method_framework}

\subsubsection{Basic equations of dust motion}  \label{sec:method_eom}
Setting up systems each consisting of a single star and planet, we conducted numerical simulations of dust particles using the N-body integrator MERCURY6 \citep{Chambers1999}. We used a modified version that includes the photon drag force from the central star, as described in \citet{Jo2024}. Gravitational and collisional interactions between particles were ignored. Since collisional effects can be critical for dusty debris disks, the justification and limitations of this assumption are given in Section~\ref{sec:dis_coll}. We also assumed that the dust clouds are in a steady state with constant replenishment of dust.
The Bulirsch--Stoer integrator \citep{BS1980} was used to solve the equations of motion in all simulations, with an initial integration timestep of about 1/25 of the planet's orbital period, assuming a circular Keplerian orbit with no inclination. Additional adjustments were made in the simulation code to ensure that the timestep remained below this initial value. This threshold was chosen based on the consideration that a timestep shorter than 1/20 of the planetary orbital period is required to avoid inhibiting the co-orbital resonant trapping of particles \citep{KJ2011, Sommer2020}.

We considered the effects of stellar and planetary gravity, radiation pressure, and stellar wind on the dust.  
The corresponding equation of motion can be expressed as \citep{Robertson1937, Burns1979, Mann2006, SK2008, Klacka2014}
\begin{align}  \label{eq:eq of motion}
\frac{\mathrm{d}^2 \mathbf{r}}{\mathrm{d}t^2} 
&= -\frac{G M_\ast}{r^2} \left( 1 - \beta_\text{PR} \left( 1 + \psi \frac{v_\text{SW}}{c} \right) \right) \hat{\mathbf{r}} \nonumber \\
& \quad - \frac{G M_\ast}{r^2} \frac{\beta_\text{PR} \left(1 + \psi\right)}{c} \left( \dot{r} \cdot \hat{\mathbf{r}} + \mathbf{v} \right)~,
\end{align}
where $\mathbf{r}$ and $\mathbf{v}$ are the astrocentric position and velocity of the particle, $M_\ast$ is the stellar mass, $G$ is the gravitational constant, $c$ is the speed of light, and $v_\text{SW}$ is the stellar wind speed, which we assume to be the typical solar wind value of $\sim 400~ \mathrm{km\ s^{-1}}$ \citep{Johnstone2015}.  
$\beta_\text{PR}$ is the ratio of the radiation pressure force $F_\text{rad}$ to stellar gravity $F_\text{grav}$. The ratio of the stellar wind force $F_\text{SW}$ to gravity, $\beta_\text{SW}$, is incorporated via the parameter $\psi$, which represents the ratio of stellar wind drag to PR drag.  

The parameters $\beta_\text{PR}$, $\beta_\text{SW}$, and $\psi$ can be further expressed using the following equations based on their definitions.
For a spherical dust particle with density $\rho$ and radius $s$, $\beta_\text{PR}$ and $\beta_\text{SW}$ are given by \citet{Mann2006}:
\begin{equation} \label{eq:beta_pr}
\beta_\text{PR} = \frac{3L_\ast \langle Q_\text{PR} \rangle}{16\pi G c M_\ast \rho s}, 
\end{equation}
\begin{equation} \label{beta_sw}
\beta_\text{SW} = \frac{3\dot{M}_\ast v_\text{SW} \langle Q_\text{SW} \rangle}{16\pi G M_\ast \rho s},
\end{equation}
where $L_\ast$ and $\dot{M}_\ast$ are the stellar luminosity and mass-loss rate, and $\langle Q_\text{PR} \rangle$ and $\langle Q_\text{SW} \rangle$ are the radiation pressure and stellar wind efficiencies averaged over the stellar spectrum and wind species, respectively. 
Although these efficiencies depend on the grain material and wavelength, we fix both $\langle Q_\text{PR} \rangle$ and $\langle Q_\text{SW} \rangle$ to 1 for simplicity, following common practice in similar studies \citep[e.g.,][]{SK2008, Sommer2020}.  However, since this assumption breaks down for very small grains---particularly at submicrometer sizes---we estimate numerical values for each assumed stellar spectral type and grain sizes used in the study (Fig.~\ref{fig:Qpr_beta}) and discuss the validity and treatment of this assumption in Section~\ref{sec:method_main}. 

Since $L_\ast \propto M_\ast^{3.5}$ for main-sequence stars \citep{Eddington1924}, it follows that $\beta_\text{PR} \propto M_\ast^{2.5}/s$, indicating that $\beta_\text{PR}$ generally increases with higher stellar mass and smaller dust size.
For the $\psi$ parameter \citep{Plavchan2005, Minato2006, Mann2006},
\begin{align} \label{eq: psi}
\psi &=  \frac{F_\text{SW}}{F_\text{PR}} \frac{c}{v_\text{SW}} = \frac{\beta_\text{SW}}{\beta_\text{PR}} \frac{c}{v_\text{SW}} 
\simeq 0.3 \left( \frac{\dot{M}_\ast}{\dot{M}_\odot} \right) \left( \frac{L_\ast}{L_\odot} \right)^{-1} \left( \frac{\langle Q_\text{SW} \rangle}{\langle Q_\text{PR} \rangle} \right),
\end{align}
where $\langle Q_\text{SW} \rangle$ and $\langle Q_\text{PR} \rangle$ are assumed to be unity, so they cancel out in the expression. As noted above, the implications of this assumption for smaller particles are explained in Section~\ref{sec:method_main}. The Sun is known to have a value of $\psi \sim 0.3$ \citep{Gustafson1994,Minato2006}. 

For the actual simulations, we used the following simplified form of Eq.~\eqref{eq:eq of motion}, neglecting the term proportional to $\psi \frac{v_{\text{SW}}}{c}$ since $v_{SW}/c \ll 1$ (see Section~\ref{sec:s_bo}):
\begin{equation} \label{eq:eom_simp}
\frac{\mathrm{d}^2 \mathbf{r}}{\mathrm{d}t^2} = 
-\frac{G M_\ast}{r^2} \left(1 - \beta_\text{PR} \right) \hat{\mathbf{r}} 
-\frac{G M_\ast}{r^2} \frac{\beta_\text{PR} \left(1 + \psi\right)}{c} \left(\dot{r} \cdot \hat{\mathbf{r}} + \mathbf{v} \right).
\end{equation}

\subsubsection{Effective radiation and wind parameters} \label{sec:eff_beta} 
To account for the effects of both radiation pressure and stellar wind, we introduce direction-dependent effective beta parameters, following the methodology described in \citet{Klacka2014}:
\begin{align} \label{eq:beta_r}
\beta_\text{r} &= \frac{ \mathbf{F}_{\text{rad}} \cdot \hat{\mathbf{r}} + \mathbf{F}_{\text{SW}} \cdot \hat{\mathbf{r}} }{ \mathbf{F}_{\text{grav}} \cdot \hat{\mathbf{r}} } = \beta_{\text{PR}} + \beta_{\text{SW}}= \beta_{\text{PR}} \left( 1 + \psi \frac{v_{\text{SW}}}{c} \right),
\end{align}
\begin{align} \label{eq:beta_t}
\beta_\text{t} &= \frac{ \mathbf{F}_{\text{rad}} \cdot \hat{\mathbf{t}} + \mathbf{F}_{\text{SW}} \cdot \hat{\mathbf{t}} }{ \mathbf{F}_{\text{grav}} \cdot \hat{\mathbf{r}} } = \beta_{\text{PR}} + \frac{c}{v_{\text{SW}}} \beta_{\text{SW}}= \beta_{\text{PR}} \left( 1 + \psi \right),
\end{align}
where we define the radial effective beta, $\beta_\text{r}$, and tangential effective beta, $\beta_\text{t}$.
These effective parameters provide a convenient framework for understanding how radiation and wind forces influence dust dynamics.

\subsubsection{Effective blowout size estimation} \label{sec:s_bo}
Grains smaller than a certain size can be entirely removed from the system due to radiation and stellar wind pressure. This defines the blowout size, which is the maximum grain size expelled from the system, whereas larger grains spiral inward due to PR and stellar wind drag \citep{Robertson1937, Burns1979}. 
The radial effective beta, $\beta_\text{r}$, primarily determines the blowout size of dust grains. Since $v_{\text{SW}}/c \approx 10^{-3} \ll 1$, the approximation $\beta_\text{r}\approx \beta_\text{PR}$ holds unless the value of $\psi$ is on the order of a thousand or higher (Eq.~\ref{eq:beta_r}). This condition is generally satisfied except for highly active young M-dwarfs (e.g., AU Mic; \citealt{Plavchan2005}), which is further discussed in Appendix~\ref{sec:dis_sbo}.  Based on the adopted stellar spectral types and ages of our simulation setup (Section~\ref{sec:sim_design}), and the resulting estimates of $\psi$ (Section~\ref{sec:psi}), the approximation is valid for our study. It is therefore applied in the derivation of the system’s effective blowout size ($s_\text{BO,eff}$).

Dust particles with low eccentricity ($e \ll 1$) are blown out of the system when $\beta_\text{r} \ge 0.5$ \citep{MD1999, Krivov2006, Kim2018}.  
Expressing $\beta_\text{r}$ as a function of particle size, the effective blowout size $s_\text{BO,eff}$ satisfies $\beta_\text{r}(s_\text{BO,eff}) \approx \beta_\text{PR}(s_\text{BO,PR}) = 0.5$, where $s_\text{BO,PR}$ is the radiative blowout size.  
We derive $s_\text{BO,eff}$ numerically by determining the largest particle size that meets this condition, using $\langle Q_\text{PR} \rangle$ values from Mie calculations \citep{KirchW2013, RW2020, miepy}.  
This approach was applied to determine the effective blowout size for each stellar spectral type in our simulations (Section~\ref{sec:method_main}) and to identify broader trends across late-type stars (Appendix~\ref{sec:dis_sbo}).

\subsubsection{Migration timescale and simulation duration} \label{sec:method_time}
The tangential effective beta, $\beta_\text{t}$, governs the inward drift of dust particles under the combined influence of PR and stellar wind drag. The corresponding migration timescale for circular orbits is expected to be reduced by a factor of $(1+\psi)$ compared to the case with PR drag alone \citep{MY1982, Kuchner2003, Plavchan2005, Minato2006}:
\begin{equation} \label{eq:t_eff}
t_\text{mig} = \frac{400}{\beta_\text{PR}(1+\psi)} \left(\frac{M_\ast}{M_\odot}\right)^{-1} \left(\frac{a_\text{d}}{1\,\mathrm{au}}\right)^2 \text{ years},
\end{equation}
where $a_\text{d}$ is the circumstellar distance at which the dust is released.

The total integration time was set to be about 10 times the effective migration time, accounting for the prolonged lifetime of dust grains due to resonant trapping \citep{JZ1989, SK2008}.

\subsubsection{Output and surface density visualization} \label{sec:method_surf_dens}
\begin{figure*}
    \centering
    \begin{subfigure}[t]{0.48\textwidth}
        \centering
        \includegraphics[height=6cm]{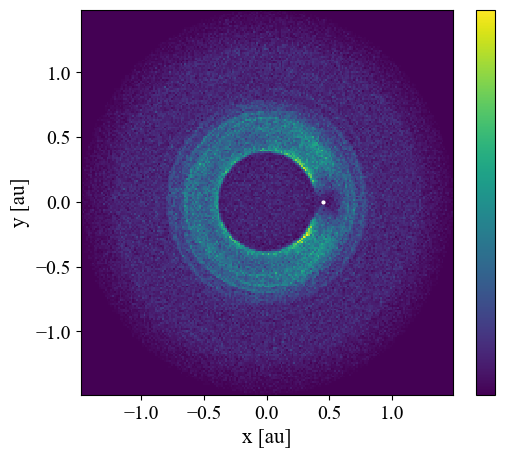}
    \end{subfigure}%
    \hfill
    \begin{subfigure}[t]{0.48\textwidth}
        \centering
        \includegraphics[height=6cm]{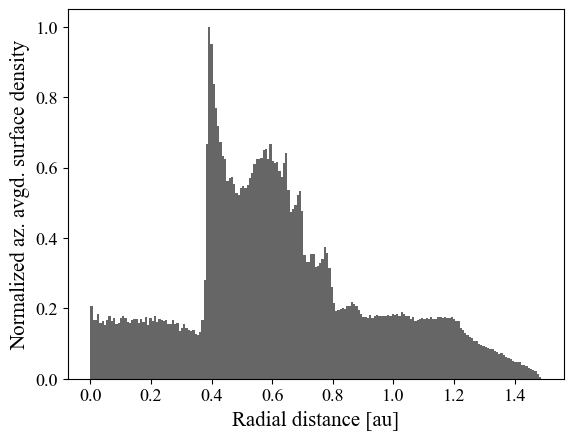}
    \end{subfigure}
\caption{(\textit{Left}) Example of face-on surface density distribution of dust in a planet's co-rotating frame, with a K4-type star located at the origin. The white dot indicates an Earth-like planet placed at the inner boundary of the CHZ. A total of 100 dust particles with a size of $50~\mathrm{\mu m}$ are used. The resonant ring structure is visible around the planet's position. The color scale indicates surface density in arbitrary units, with lighter colors corresponding to higher densities. (\textit{Right}) Azimuthally averaged surface density of dust in arbitrary normalized units, based on the distribution in the left panel.
The background disk appears as a nearly constant plateau, while the resonant ring structure shows a density enhancement.}
    \label{fig:surf_dens_dist}
\end{figure*}

For each simulation setup, the positions of the dust particles were recorded at regular intervals, typically about 20{,}000 times shorter than the total integration time.
The recorded positions at each epoch were presented in a 2D histogram in the co-rotating frame with the planet, using a grid resolution of about 1/200 of the total spatial scale. This is a commonly used method for obtaining surface density distributions of steady-state dust disks with a single planet \citep{SK2008, Defrere2012, Sommer2020}. An example of a surface density distribution and the corresponding azimuthally averaged surface density histogram is shown in Fig.~\ref{fig:surf_dens_dist}.  
Detailed analyses of geometric disk structures \citep[e.g.,][]{SK2008} are not discussed in this paper.

\subsection{Simulation design: Pilot and Main studies} \label{sec:sim_design}

\subsubsection{Pilot Study: Setup for evaluating the importance of stellar wind} \label{sec:method_pilot}

Based on the $\psi$ parameter, we examined two types of models assuming a radial, non-varying stellar wind within each system.
In Model~I, we adopted a solar-like stellar wind for all cases, where $\psi \sim 0.3$. In Model~II, distinct $\psi$ values were applied for cases with different spectral types.
For simplicity, we refer to the cases in Model~I as those without spectral type variation in stellar wind, as the $\psi$ value is fixed for all stars.
As mentioned in Section~\ref{sec:intro}, the spectral types for this study are chosen across F to M-type, motivated by their potential to host more habitable environments \citep{TW2022} and their relevance as candidate targets for the LIFE mission \citep{Menti2024}.

For both Model~I and Model~II, we considered 36 cases each: four types of late-type, main-sequence stars with ages similar to the Sun (F4, G4, K4, and M4), three planetary masses comparable to Earth ($M_\text{p} = 0.5, 1.0$, and $2.0 ~M_\oplus$), and three planetary semimajor axes ($a_\text{p}$) within the habitable zone (inner, middle, and outer boundaries of the CHZ). 
A stellar age of 4~Gyr was adopted for all simulations, based on the median age of stars known to host small planets \citep{Swastik2023}, as well as the estimated timescale for the emergence of complex life informed by Earth's evolutionary history \citep{Cuntz2014,Safonova2016}. This age enters our simulations through the stellar wind prescription used to estimate $\psi$ (see Section \ref{sec:res_pilot}), since stellar wind strength depends on stellar rotation and hence age.  For F4-type stars, whose main-sequence lifetime is shorter ($\sim 3~\mathrm{Gyr}$; \citealt{Choi2016}), the difference between the $\psi$ value for 4~Gyr ($\sim0.7$) and for nominal main-sequence ages of 2--3~Gyr ($\sim1$) is minor when using the same main-sequence stellar parameters (see Appendix~\ref{sec:dis_F_evol} for details)---particularly compared to the much higher values of M-type stars (see Section~\ref{sec:psi}). Therefore, although the input of 4~Gyr age is used for F4-type $\psi$ values for consistency in the simulations, the results are interpreted as representative of mid- to late-main-sequence F4 stars at 2--3~Gyr.

The exact values of $a_\text{p}$ vary depending on spectral type; for clarity, we denote them as $\text{HZ}_\text{in}$, $\text{HZ}_\text{mid}$, and $\text{HZ}_\text{out}$. We used $\text{HZ}_\text{mid}$ rather than the Earth-equivalent insolation distance (EEID), as the latter is nearly identical to $\text{HZ}_\text{in}$. Specific parameter values are given in Table~\ref{tab:parameters}, where stellar parameters are taken from the online table of Mamajek, E. E.\footnote{Mamajek, E. E., “A Modern Mean Dwarf Stellar Color and Effective Temperature Sequence,” available at \url{https://www.pas.rochester.edu/~emamajek/EEM_dwarf_UBVIJHK_colors_Teff.txt}} \citep{Pecaut2012, Pecaut2013}, and the HZ boundaries were derived using the method of \citet{Kopparapu2014}.
These stellar parameters represent average main-sequence values, appropriate for low-mass stars whose main-sequence lifetimes exceed the assumed age. Such stars evolve negligibly over Gyr timescales due to slow hydrogen fusion and convective interiors \citep{CB1997,Baraffe2015}. Thus, stellar age does not affect parameters other than via $\psi$ in our setup. As noted previously, since the results for F4-type stars can be interpreted as representative of typical or older main-sequence results, the use of these nominal parameters is acceptable (see Appendix~\ref{sec:dis_F_evol}). 

\begin{table}[h!]
\caption{Values of stellar parameters and $a_\text{p}$}
\label{tab:parameters}
\centering
\begin{tabular}{cccccc}
\hline\hline
\multicolumn{2}{c}{SpT} & F4 & G4 & K4 & M4 \\ 
\hline
\multicolumn{2}{c}{$M_\ast\ (M_{\odot})$} & 1.38 & 0.985& 0.730& 0.230\\ 
\multicolumn{2}{c}{$L_\ast\ (L_{\odot})$} & 4.17 & 0.910& 0.200& 0.00720\\ 
\multicolumn{2}{c}{$R_{\ast} \ (R_{\odot})$} & 1.53& 0.991& 0.713& 0.274\\
 \multicolumn{2}{c}{$T_\ast\ (\mathrm{K})$}& 6670 & 5680 & 4600 &3210 \\ 
& $\text{HZ}_\text{in}$ & 1.84& 0.912 & 0.451 & 0.0880\\ 
$a_\text{p}$\ (au) & $\text{HZ}_\text{mid}$ & 2.52& 1.26& 0.641& 0.131\\ 
& $\text{HZ}_\text{out}$ & 3.19& 1.61& 0.830& 0.174 \\
\hline
\end{tabular}
\end{table}

For each simulation, we used 100 dust particles of fixed size $s = 50~\mathrm{\mu m}$.
The number of particles was selected to balance the computational efficiency of MERCURY6 with the purpose of exploring general trends across different parameters prior to detailed modeling in the main study. The particle size was chosen as a convenient, computationally efficient baseline to capture general resonant behaviors of bound grains within the typical 1--100~$\mathrm{\mu m}$ range in the Solar System \citep{Grun1985, FD2002, KH2003}. Given the limited knowledge of exoplanetary grain size distributions and the pilot study’s focus on isolating the effect of $\psi$, we adopt this fixed size for all simulations. 

The initial semimajor axis of dust was set to be about 3~$a_\text{p}$, outside the strongest 2:1 and 3:2 MMRs according to \citet{SK2008}. We used initial eccentricities and inclinations of dust released from parent bodies of  $e_\text{p} \sim 0.1$ and inclination $i_\text{p} \sim 5^\circ$, considering that the creation of resonant ring structures is dominated by dynamically cold dust, i.e., $e_\text{d} \leq 0.2$ and $i_\text{d} \leq 20^\circ$ \citep{SK2008}. 
We set these fixed initial values of semimajor axis, eccentricity, and inclination for each simulation to simplify the setup and enhance the structures formed by MMRs. Accordingly, our simulations represent the most efficient case in terms of visibility of the resonant structures. The initial values of the longitude of ascending node $\Omega$, argument of pericenter $\omega$, and mean anomaly $M$ were independently drawn from uniform distributions between 0 and $2\pi$ for each particle. This ensures that each orbital element is initially random and evenly distributed, and any residual unevenness is quickly erased as the angles become isotropically distributed on relatively short timescales.

\subsubsection{Main Study: Setup for multi-size dust simulations } \label{sec:method_main} 
To investigate more realistic disk structures, we focused on Model~II, incorporating a range of dust sizes ($s = 0.1, 0.3, 1, 3, 10, 30, 100$, and $300~\mathrm{\mu m}$) and increasing the number of particles to 5{,}000 in each size bin. The chosen range covers typical Solar System values (1--100~$\mathrm{\mu m}$; Section~\ref{sec:method_pilot}) while extending toward both smaller and larger sizes to account for possible variations in other systems. 

\begin{figure*}
    \centering
    \includegraphics[width=0.8\linewidth]{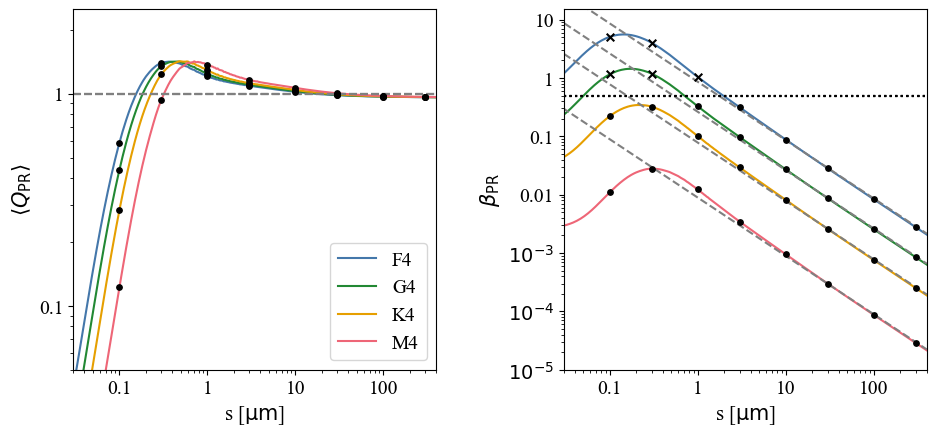}
    \caption{$\langle Q_\text{PR} \rangle$ values for grain sizes $s \sim 0.1$--$300~\mathrm{\mu m}$ (\textit{left}) and the corresponding  $\beta_\text{PR}$ values (\textit{right}; analogous to Figure~7 of \citealt{RW2020}). Solid lines show numerical values for different stellar spectral types used in this study, while the dashed lines in the right panel indicate the analytical $\beta_\text{PR}$ values using $\langle Q_\text{PR} \rangle$ = 1 (Eq.~\ref{eq:beta_pr}). Black dots mark the discrete grain sizes adopted in the main study. The black dotted line in the right panel denotes $\beta_\text{PR} = 0.5 $, and sizes above this threshold, corresponding to grains blown out of the system for circular orbits, are marked with '$\times$'. The size $\sim 50 ~ \mathrm{\mu m}$ used in the pilot study is safely higher than the blowout size and meets the assumption of $\langle Q_\text{PR} \rangle  \sim 1$.}
    \label{fig:Qpr_beta}
\end{figure*}

In Fig.~\ref{fig:Qpr_beta}, we show the numerically calculated $\langle Q_\text{PR} \rangle$ and $\beta_\text{PR}$ values across the size range based on the optical constants of astronomical silicates \citep{Draine2003}, using Mie theory \citep{miepy}. Since $\beta_\text{r} \approx \beta_\text{PR}$ (Section~\ref{sec:s_bo}), the effective blowout sizes are $s_\text{BO,eff} \sim 1.94$ and $0.71~\mathrm{\mu m}$ for F4 and G4 stars, respectively, while no blowout size exists for K4 and M4 stars. Accordingly, we excluded the 0.1--1~$\mathrm{\mu m}$ sizes for F4 and 0.1--0.3~$\mathrm{\mu m}$ sizes for G4 from the simulations. Further discussion and comparison with previous estimates are provided in Appendix~\ref{sec:dis_sbo}.

The right panel of Fig.~\ref{fig:Qpr_beta}  shows that the analytical $\beta_\text{PR}$ value assuming  $\langle Q_\text{PR} \rangle$ = 1 significantly deviates from the numerical value for $0.1~\mathrm{\mu m}$ particles, while the discrepancy is minor for larger sizes. We therefore adopt $\langle Q_\text{PR}\rangle=1$ for $s \ge 0.3~\mathrm{\mu m}$ and approximate the $0.1 ~ \mathrm{\mu m}$ cases using results from larger grains with comparable $\beta_\text{PR}$ values ($0.3~ \mathrm{\mu m}$ for K4 and $1~ \mathrm{\mu m}$ for M4). This approach is justified because the equation of motion depends directly on $\beta_\text{PR}$ rather than on $\langle Q_\text{PR}\rangle$ (Eq.~\ref{eq:eom_simp}).
For $\psi$, we retain the assumption $\langle Q_\text{SW} \rangle / \langle Q_\text{PR} \rangle \simeq 1$ across all grain sizes, as the behavior of $\langle Q_\text{SW} \rangle$ is analogous to that of $\langle Q_\text{PR} \rangle$, though not entirely identical \citep{Minato2004, Mann2006}.

The number of particles follows the setup of \citet{SK2008}, who demonstrated that 5{,}000 particles are sufficient for robust evaluation of resonant ring structures by reducing variations in MMR populations across repeated simulation runs.
The planetary configuration was fixed ($M_\text{p} = 1~ M_\oplus$, $a_\text{p} = \text{HZ}_\text{mid}$), with all other parameters identical to the pilot study. This results in a total of 32 cases (27 after excluding the blown-out sizes) across all stellar spectral types. 
These higher-resolution simulations provide more reliable contrast and density profiles \citep{SK2008} and are designed to represent plausible exozodi environments around Earth-analog systems. 
A summary of the parameter definitions and simulation setups for both the pilot and main studies is provided in Appendix~\ref{sec:app_setup}.

\section{Results} \label{sec:results}

In Section~\ref{sec:res_pilot}, we present our preliminary findings based on the pilot study approach described in Section~\ref{sec:method_pilot}, using a simplified simulation setup. These results highlight the importance of stellar wind in shaping exozodi resonant structures for various spectral types. Building on these findings, Section~\ref{sec:res_main} presents the main results based on the multi-size simulations outlined in Section~\ref{sec:method_main}, employing more realistic assumptions.  
We analyze the contrast of resonant rings in Section~\ref{sec:res_main_C}, and the optical depth and flux distributions of the clouds in Section~\ref{sec:res_tau_F}.

\subsection{Preliminary results based on the pilot study} \label{sec:res_pilot}

This pilot study presents the estimated ratio of stellar wind drag to PR drag ($\psi$) for various spectral types and examines its influence on dust dynamics and resonant structures. The $\psi$ prescription highlights the importance of stellar wind drag, while the simulations quantify its effect on resonant ring contrasts, which cannot be directly inferred from $\psi$ alone (see Section~\ref{sec:dis_spt_res}).  Based on the estimated $\psi$ values and comparison of two models---with and without stellar spectral type variation in stellar wind drag---we illustrate the importance of properly incorporating stellar wind in exozodi modeling, particularly around low-mass stars where $\psi$ attains high values. 

\subsubsection{Estimating stellar wind drag efficiency strength across spectral types} \label{sec:psi}

\begin{figure}
    \centering
    \includegraphics[width=1\linewidth]{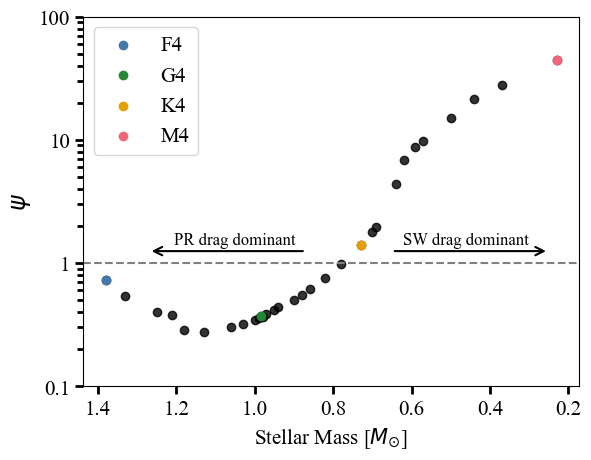}
    \caption{$\psi$ values given by Eqs.~\eqref{eq: psi} and~\eqref{eq:massloss} for main-sequence stars of spectral types F4 to M4 at an age of $\sim$ 4~Gyr (2--3~Gyr for F-type stars). The four spectral types used in this study are marked in blue, green, yellow, and red, respectively. The dashed line shows $\psi = 1$, where the stellar wind drag equals the PR drag.}
    \label{fig:psi}
\end{figure}

Following Eq.~\eqref{eq: psi}, we computed the $\psi$ values for different stellar spectral types. 
We used the relation from \citet{Johnstone2015,Johnstone2021} as input for stellar wind mass-loss rates, applied to main-sequence stars of the spectral types considered in this study:
\begin{equation} \label{eq:massloss}
\frac{\dot{M}_\ast}{\dot{M}_\odot} = \left( \frac{R_\ast}{R_\odot} \right)^2 \left( \frac{P_\ast}{P_\odot} \right)^{-1.33} \left( \frac{M_\ast}{M_\odot} \right)^{-3.36},
\end{equation}
where $R_\ast$ and $P_\ast$ are the stellar radius and rotation period, respectively. The value of $P_\ast$ was determined using gyrochronology, which depends on stellar effective temperature $T_\ast$ and age.
To cover the spectral range (F4--M4) and stellar age ($\sim$ 4~Gyr) used in this work, we combined results from different gyrochronology studies based on their applicable parameter ranges: for stars with  $T_\ast \geq 6280 ~ \mathrm{K}$ ($\sim$ F7 and earlier), $6280 ~ \mathrm{K} > T_\ast \geq 4200 ~ \mathrm{K}$ (F7--K6), and $4200 ~ \mathrm{K} > T_\ast \geq 3200 ~ \mathrm{K}$ (K6--M4), we adopted the gyrochronology models of \citet{Lu2024}, \citet{MH2008}, and \citet{Dungee2022}, respectively. 
Although the gyrochrones from \citet{Lu2024} span all spectral types and a broad stellar age range (0.67--14 Gyr), we chose the other two models for lower-mass stars because they generally yield more conservative estimates of $\psi$---especially for M4-type stars. Additionally, the gyrochrones from \citet{Dungee2022} were constructed specifically for 4~Gyr-old stars, matching the stellar age we assumed. The accuracy of this prescription is further discussed in Section~\ref{sec:dis_psi_acc_age}.

Fig.~\ref{fig:psi} presents the resulting $\psi$ values for all main-sequence spectral types in the range of F4 to M4 having an age of 4~Gyr (2--3 Gyr for F-type stars). 
The results indicate that at these ages of a few Gyr, systems with stellar masses below $\sim 0.8~ M_\odot$ (approximately K3 type) are more strongly influenced by stellar wind drag than by PR drag. The $\psi$ values for the F4, G4, K4, and M4 spectral types used in this research are approximately 0.73, 0.37, 1.40, and 44, respectively. Notably, the lowest-mass star, M4, exhibits a particularly high value, over an order of magnitude larger than those of earlier types. Using the gyrochrones from \cite{Lu2024} for all spectral types leads to an even higher value of $\sim 127$ for M4, whereas the values for the other spectral types remain largely unchanged. 
While we adopt the more conservative value of $\psi \sim 44$ for M4-type stars in our main analysis, these results demonstrate that M-type systems at even 4 Gyr ages can be dominated by stellar wind drag regardless of the choice of gyrochronology model.

\subsubsection{Impact of stellar wind drag: Model~I vs. Model~II} \label{sec:res_models_vs}

\begin{figure*}
    \centering
    \includegraphics[width=0.48\linewidth]{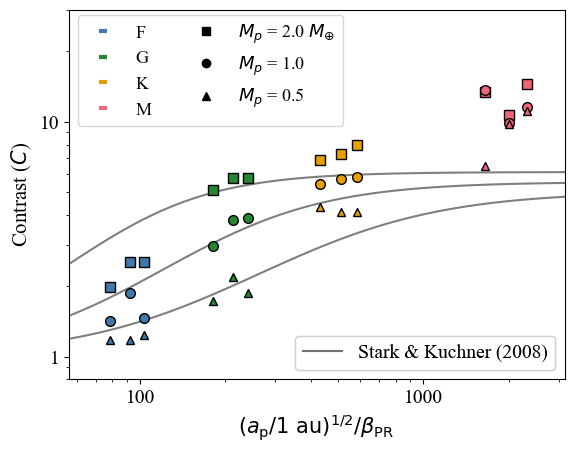}
    \includegraphics[width=0.48\linewidth]{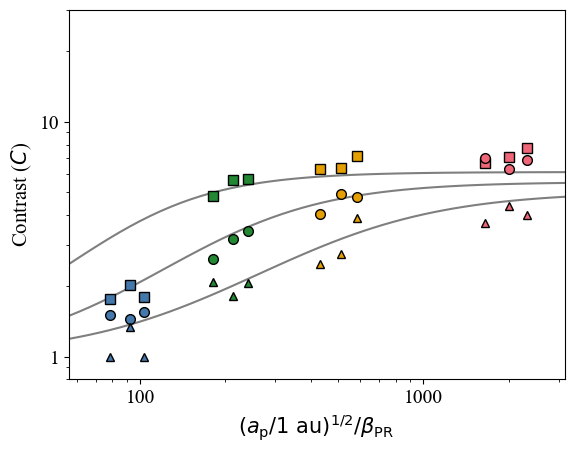}

    \caption{Contrast values from the pilot study, plotted against the $a_\text{p}^{1/2}/\beta_\text{PR}$ term for Model~I (\textit{left}) and Model~II (\textit{right}). Data corresponding to F4, G4, K4, and M4 are shown in different colors: blue, green, yellow, and red, respectively. Planetary masses of 0.5, 1.0, and 2.0~$M_\oplus$ are represented by different shapes, with solid lines indicating the fits from \cite{SK2008} for each $M_\text{p}$, increasing from bottom to top. For each color and symbol, the three points from left to right correspond to planets located at the inner, middle, and outer edges of the habitable zone. We note that these results are shown for $s = 50 ~ \mathrm{\mu m}$; the trends do not apply to grains smaller than $\sim  10 ~ \mathrm{\mu m}$ around F4, $\sim  3 ~ \mathrm{\mu m}$ around G4, and $\sim  1 ~ \mathrm{\mu m}$ around K4 and M4-type stars, which are blown out or fail to form resonant rings (see Section~\ref{sec:res_main_C}).}
    \label{fig:contrast_pilot}
\end{figure*}

In this subsection, we compare the results of Model~I and Model~II to evaluate the impact of stellar wind drag on resonant structure formation and model consistency. 
We examine the prominence of these ring structures, quantified by their contrast relative to the background distribution. 
Fig.~\ref{fig:contrast_pilot} shows the contrast values of resonant structures as a function of $a_\text{p}^{1/2}/\beta_\text{PR}$ , with results presented for different spectral types, planetary masses, and semimajor axes, assuming a fixed dust size.  

The contrast values are plotted against  $a_\text{p}^{1/2}/\beta_\text{PR}$ because, according to \cite{SK2008}, contrast depends only on this term and $M_\text{p}$. This expression is derived by the PR drag timescale over the libration timescale for resonance:
\begin{equation} \label{eq:x}
\frac{t_\text{PR}}{t_\text{lib}} \propto \frac{a_\text{d}^2/\beta_\text{PR}}{a_\text{p}^{3/2}} \propto a_\text{p}^{1/2}\beta_\text{PR}^{-1},
\end{equation}
where $a_\text{d} \sim a_\text{p}$ since the resonant trapping occurs near the orbit of the planet. 
Extensions of this formulation that incorporate spectral type and stellar wind effects are introduced later in Section~\ref{sec:dis_spt_res}, with the same results plotted against this modified term in Fig.~\ref{fig:contrast_pilot_mod}. 
For clarity and consistency with \citet{SK2008}, we use the original term $a_\text{p}^{1/2}/\beta_\text{PR}$ when comparing Model~I and~II. This allows us to isolate the effect of stellar wind drag and assess the differences between the two models under a shared baseline. 

For the quantitative calculation of contrast, we adopted $C_\text{AA,max}$ among the several definitions proposed by \cite{SK2008}, which is the maximum of the azimuthally averaged surface density of the resonant ring relative to the background disk density. To obtain this quantity, we constructed an azimuthally averaged surface density histogram from the 2D dust distribution in the planet’s co-rotating frame (see Section~\ref{sec:method_surf_dens} and Fig.~\ref{fig:surf_dens_dist}). The maximum of this profile was then divided by the average of the nearly constant background density to derive $C_\text{AA,max}$.
This choice reflects the fact that azimuthal asymmetries can be significant sources of noise for interferometric observations. For simplicity, we refer to this quantity ($C_\text{AA,max}$) as $C$ throughout the paper. 

When comparing different spectral types, Fig.~\ref{fig:contrast_pilot} reveals that M4-type stars exhibit the highest resonant ring contrasts overall, but their contrast values decrease significantly---from Model~I to Model~II---by roughly a factor of two. Other spectral types also show slight reductions, though these are less pronounced than in the case of M4-type.
More broadly, the results indicate a general trend of increasing resonant ring contrast with decreasing stellar mass across spectral types, regardless of model. However, using higher $\psi$ values based on an alternative gyrochronology model (see Section~\ref{sec:psi}) could further reduce the contrast for M4-type stars in Model~II, shifting the peak contrast toward K-type stars. While we base our analyses on the relatively conservative $\psi$ value throughout this study, the implications of adopting the higher value for M4-type are discussed in Appendix~\ref{sec:dis_gyro}. 
The findings here highlight two key points: (1) stellar wind drag significantly reduces the prominence of resonant structures around M-type stars, as indicated by the lower contrast of the red points in the right panel of Fig. 3 compared to the left; and (2) resonant ring contrast tends to increase toward lower-mass stars.
This confirms our expectation that stellar wind drag across various spectral types should be properly accounted for in exozodi modeling, even at old ages, an importance further discussed in Sections~\ref{sec:dis_psi} and~\ref{sec:dis_spt_res}.

Across both models, Fig.~\ref{fig:contrast_pilot} also demonstrates a general trend of increasing contrast with larger $M_\text{p}$ and $a_\text{p}$ within a given spectral type, consistent with the findings of \cite{SK2008}.  
Some deviations from this trend are likely due to statistical noise caused by the limited number of particles, as noted by \cite{SK2008}, who showed that simulations with only 100 particles can yield unreliable contrast estimates.  
To account for this, we conducted internal tests with 5{,}000 particles, which showed consistent trends with the 100-particle simulations, supporting the validity of our model comparison.
The agreement between the G-type star results and the contrast fit function from \cite{SK2008} further supports the consistency of our approach. Deviations from the fit for other spectral types are as expected, as \cite{SK2008} focused exclusively on Solar twins. 
Improved fits across spectral types, based on the modified parameter that includes spectral type and stellar wind effects, are also shown in Fig.~\ref{fig:contrast_pilot_mod}. The detailed fitting functions and their parameter values are introduced in Section~\ref{sec:dis_spt_res}, with additional analysis and discussion of their applications in Appendix~\ref{sec:dis_semianal}.

\subsection{Main results from multi-size simulations}\label{sec:res_main}

The main study expands upon the pilot results by systematically investigating resonant ring contrasts, as well as the spatial distributions of optical depth and thermal flux, across multiple dust sizes and stellar spectral types. Focusing on cases with an Earth-twin planet and properly accounting for spectral type variation in stellar wind effects, we quantify how dust size, stellar mass, and stellar wind influence the configuration of resonant structures. 

\subsubsection{Resonant ring contrasts} \label{sec:res_main_C}
\begin{figure}[htbp]
  \centering
  \includegraphics[width=\linewidth]{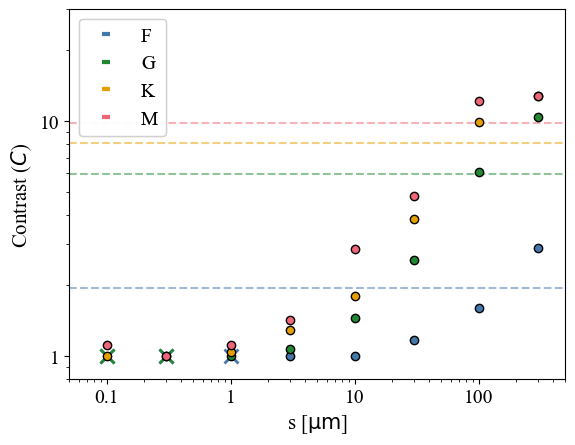}
  \caption{Contrast values from the main study plotted against grain sizes. Cases where $s < s_\text{BO,eff}$ are shown as $\times$ marks, which overlap with other data points up to $1~\mathrm{\mu m}$. Dashed lines in the upper panel indicate the size-combined contrast values of optical depth ($\langle C_\tau \rangle$; see Section~\ref{sec:res_tau_F}) for each spectral type. Colors and symbols follow those used in Fig.~\ref{fig:contrast_pilot}, but only cases with $M_\text{p} = 1~ M_\oplus$ and $a_\text{p} = \text{HZ}_\text{mid}$ are considered here, resulting in 8 points for each spectral type. The contrast for $300~\mathrm{\mu m}$ dust around the K-type star is only slightly lower than that for M-type stars, nearly overlapping and thus hidden from view.}
  \label{fig:contrast_main_s}
\end{figure}

Fig.~\ref{fig:contrast_main_s} shows the contrast values for each particle size from the multi-size simulations. The $a_\text{p}^{1/2}/\beta_\text{PR}$ term is not used for the plots here to facilitate comparison of contrast across different stellar spectral types for the same dust size. For results plotted with respect to the $a_\text{p}^{1/2}/\beta_\text{PR}$ term and the modified expression from Eq.~\eqref{eq:mod_x}, see Fig.~\ref{fig:contrast_main_mod}. 

At a fixed dust size, the contrast increases with decreasing stellar mass, consistent with the key results of the pilot study. 
For a given stellar type, the contrast also exhibits a general increase with grain size. Larger grains form more pronounced ring structures because they survive longer due to slower inward migration and are more likely to be captured in MMRs \citep{SK2008}.
Smaller grains tend to bypass resonant trapping, and the resulting contrast approaches unity, indicating little to no ring structure. Resonant rings fail to form for grains of $s \lesssim 1~\mathrm{\mu m}$ around K and M-type stars, $s \lesssim  3 ~ \mathrm{\mu m}$ around G-type stars, and $s \lesssim  10 ~ \mathrm{\mu m}$ around F-type stars. We note that the decrease in resonant contrast for Model II observed in the pilot study when accounting for spectral-type variations in stellar wind (Fig.~\ref{fig:contrast_pilot}), does not apply to grains in these size ranges. Detailed minimum grain sizes for resonant structure formation, derived from data fits, are given in Appendix~\ref{sec:dis_semianal}.

\subsubsection{Optical depth and flux distributions}
\label{sec:res_tau_F}
\begin{figure*} 
    \centering
    \includegraphics[width=\linewidth]{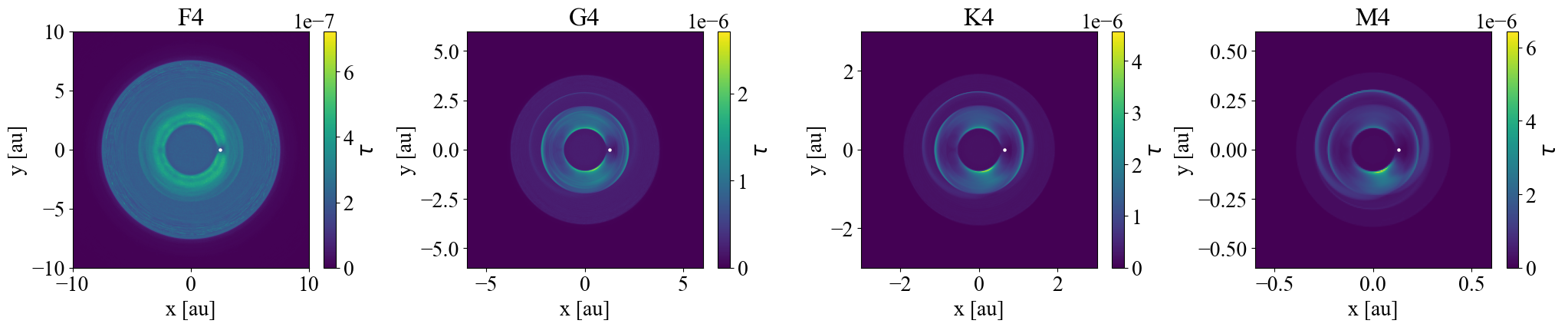}
    \includegraphics[width=\linewidth]{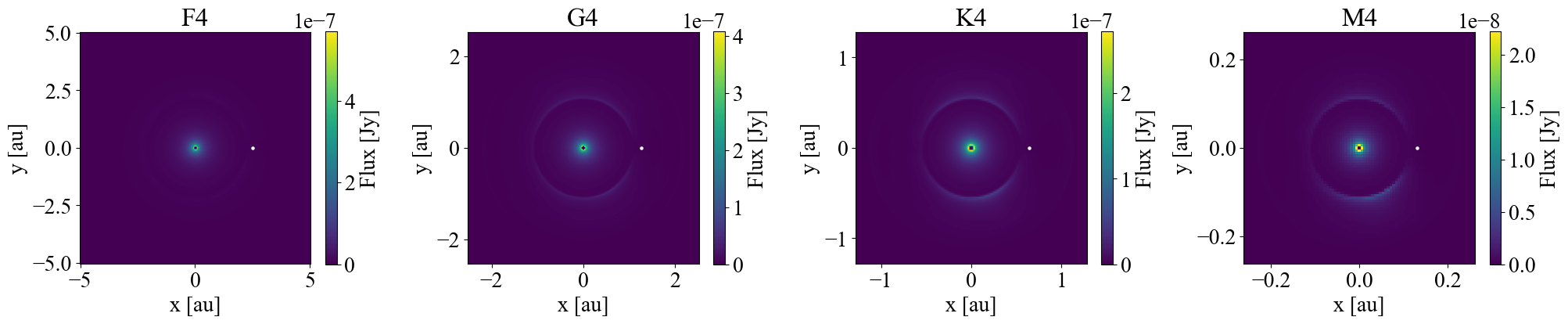}
    \caption{
    Distributions of optical depth (\textit{upper row}) and flux ($10~\mathrm{\mu m}$; \textit{lower row}) are presented for spectral types F4, G4, K4, and M4, from left to right, assuming a constant $\tau_\mathrm{BG} \sim 3$~zodis.
    The star is located at the center, and the planet is marked with a white dot to the right of the central star. The spatial scale differs between upper and lower panels: the optical depth maps show a larger region to capture the full disk structure,  while the flux maps zoom in on a smaller region ($\sim 2~a_\text{p}$) around the planet to emphasize resonant ring structures. The central region up to the sublimation distance is excluded from each flux map.
    The colorbar scales vary across spectral types to optimize visualization. For reference, the optical depth at $r = 2 ~ a_\text{p}$ is $\tau_\mathrm{BG} \approx 2.1 \times 10^{-7}$ for all spectral types. The spatial scale also varies with spectral type, with lower-mass stars having more compact configurations. These results correspond to Model~II only, as they are from the main study. We also note that the results are from grain sizes of 0.1--300~$\mathrm{\mu m}$, and may differ if larger grains are included (see Section~\ref{sec:dis_coll}).}
    \label{fig:maps}
\end{figure*}

The vertical geometrical optical depth, $\tau$, referred to simply as optical depth for clarity, was calculated from the surface density distributions of each grain size. It accounts for the combined contributions of all dust sizes, considering their total cross-sections.
Fig.~\ref{fig:maps} shows the distributions of exozodi optical depth and the corresponding thermal emission for F4, G4, K4, and M4 spectral types. 
Resonant structures were formed in all cases. In particular, larger dust grains contribute more significantly to these structures, as their dynamics are more strongly governed by gravity, as explained in Section~\ref{sec:res_main_C}. As a result, the optical depth distribution is primarily shaped by the largest grains, though it is diluted by the abundance of smaller particles produced by collisional cascade (Eq.~\ref{eq:crushing_law}) and retained due to the small blowout sizes of late-type stars (Fig. \ref{fig:s_bo}). This is consistent with the findings of \citet{SK2008}.

The effect of this dilution can be seen by comparing the size-combined contrast with the contrast from single-sized exozodis composed of only the largest grains.
The optical depth contrast, $\langle C_\tau \rangle$, represents the contrast of the size-combined resonant ring, and is calculated directly from the azimuthally averaged histograms of the optical depth distributions. It follows the same methodology as $C_\text{AA,max}$ (Section~\ref{sec:res_models_vs}), which measures the maximum azimuthally averaged contrast for single-grain-size distributions, but $\langle C_\tau \rangle$ incorporates the contributions of all grain sizes, naturally accounting for the dilution effect from smaller grains. 
The original semi-analytical definition from \citet{SK2008} is provided in Appendix~\ref{sec:dis_semianal}. The resulting $\langle C_\tau \rangle$ values are 1.95, 5.98, 8.07, and 9.85 for F4, G4, K4, and M4, respectively. Each of these values is lower than the estimate based solely on the largest grains, as marked in Fig.~\ref{fig:contrast_main_s}. Overall, $\langle C_\tau \rangle$ increases with decreasing stellar mass, consistent with trends observed in previous single-size cases. 
This confirms that resonant rings become more prominent around lower-mass stars, even when extended to more realistic multi-size dust distributions. 


The detailed calculation of optical depth was performed under the assumption that dust production follows a collisional cascade \citep{Dohnanyi1969},
\begin{equation} \label{eq:crushing_law}
\frac{dN}{ds} \propto s^{-\alpha}, \quad \alpha = 3.5.
\end{equation}
with particles not undergoing further collisions after release from their parent bodies \citep{SK2008}. Results from individual simulations are combined into composite models weighted by this size distribution. This approach is valid up to the critical grain size where the collisional timescale becomes comparable to the migration timescale \citep{KS2010} for a given dust level. Bearing this limitation in mind, the applicability and caveats of our simulation results are further discussed in Section~\ref{sec:dis_coll}.

The optical depth is then computed as:
\begin{equation}
\tau = \int \pi s^2 n(s) \, ds \approx \sum \pi s^2 N_0 \frac{N(s)}{\Delta A} \left( \frac{s}{s_0} \right)^{-3.5} \Delta s,
\end{equation}
where $n(s)$ is the number density of particles with size $s$, $N(s)$ is the number of dust particles in a given area $\Delta A$, $s_0$ is the reference size (1~$\mathrm{\mu m}$), $N_0$ is the normalization number of particles at the reference size, and $\Delta s$ is the bin width in particle size, corresponding to a constant logarithmic interval in s. 
$N(s)$ is calculated considering the different output timesteps of the simulations, and $\Delta A$ corresponds to the projected area of each simulation grid cell. The value of $N_0$ is determined by the dust production rate of the parent body, which can vary randomly between systems. We scaled $N_0$ according to a specified background optical depth in zodis, where 1~zodi corresponds to the optical depth of the zodiacal dust of the Solar System at 1~au, $\sim 7.12 \times 10^{-8}$ \citep{Kelsall1998}.  
For the nominal background level, we adopt $\tau_\text{BG}\sim3$~zodis for all stars, which corresponds to the median best-fit value inferred from the Hunt for Observable Signatures of Terrestrial Systems (HOSTS) survey population modeling, independent of spectral type and age \citep{Ertel2020}. Although M-type stars were not included in the survey, we assume the same value due to the lack of confirmed observational constraints (see Sections~\ref{sec:dis_psi} and~\ref{sec:dis_spt_res} for details). For reference, the 95\% confidence upper limit of 27~zodis reported by HOSTS is also considered in the comparison of collision and migration timescales (Section~\ref{sec:dis_coll}) to assess the validity of our results for dustier systems. 

Because of their higher ring contrast, lower-mass stars exhibit more optically thick resonant structures for the same background level in the optical depth maps (upper row of Fig.~\ref{fig:maps}). 
The flux distributions (lower row), however, do not necessarily follow the same trend due to differences in spatial scale, which will be discussed further in Section~\ref{sec:dis_flux}.
The flux maps were derived from the optical depth distributions by calculating the brightness ($I_\lambda$) and the corresponding flux ($F_\lambda$) as
\begin{equation} \label{eq:flux}
   \Delta F_\lambda\left(x,y\right) = I_\lambda\left(x,y\right) \, \Delta\Omega = \tau\left(x,y\right) \, B_\lambda\left(T\left(x,y\right)\right) \, \frac{\Delta A}{d^2},
\end{equation}
where $(x,y)$ is the position in astrocentric coordinates, with the planet located at $(a_\text{p}, 0)$. $\Delta \Omega$ is the solid angle subtended by the grid area $\Delta A$ at distance $d$ from the observer. $B_\lambda$ is the Planck function, and $T$ is the equilibrium temperature at that location, assuming that the dust behaves as a perfect blackbody. 
We assumed a fixed distance of $d = 10$~pc for all spectral types, and adopted the basic observation parameters for the LIFE instrument \citep{Dannert2022,CG2023}: we considered a mirror diameter of $D = 2~\mathrm{m}$, the wavelength coverage of 4--18.5~$\mathrm{\mu m}$, baseline minimum and maximum lengths of $B_\text{min} = 10~\mathrm{m}$ and $B_\text{max} = 100~\mathrm{m}$, a constant spectral resolution of $R = 20$, an inner working angle (IWA) of $\sim 0.5 ~\lambda/B$, and a field of view (FoV) of $\sim \lambda/D$.
At a distance of 10~pc, the angular distances of the planet are 0.25, 0.13, 0.064, and 0.013$^{\prime\prime}$ for systems with F4, G4, K4, and M4 stars, respectively. Considering a nominal MIR wavelength of $10~\mathrm{\mu m}$, the IWA has a value of $\sim0.01^{\prime\prime}$, which is marginally smaller than the innermost planet position around the M-type star (0.013$^{\prime\prime}$), when using $B_\text{max} = 100~\mathrm{m}$. Therefore, we set the baseline length to 100~m to reflect the most favorable observing configuration. Flux maps for $10~\mathrm{\mu m}$ were calculated accordingly using Eq.~\eqref{eq:flux}.

To show the intrinsic emission morphology of each disk, the flux maps presented in Fig.~\ref{fig:maps} display emission down to the dust sublimation distance, without applying the IWA cutoff. This allows the bright inner regions originating from hot dust located close to the star to be fully represented. The ring structures are faint relative to the intense central emission, particularly for higher-mass stars. The same flux maps with masked IWA regions, which reproduce the observational constraints expected for LIFE, are provided separately in Appendix~\ref{sec:app_IWA} for reference. 
Detailed analyses of the flux maps are presented in Section~\ref{sec:dis_flux}, where we focus on the resonant structures and exclude these central regions.
Overall, the optical depth and flux maps presented here characterize exozodi distributions around late-type stars, incorporating stellar wind drag and the resonant structures induced by exo-Earths.

\section{Discussion} \label{sec:discussions}

This section discusses the broader implications of our results for the modeling of exozodi distributions, focusing on the influence of stellar wind drag, spectral type, and stellar age on resonant structures and planet detectability. We first evaluate the accuracy and the age dependence of the adopted $\psi$ prescription (Section~\ref{sec:dis_psi_acc_age}) and then examine how stellar wind drag influences dust dynamics and ring contrasts, with particular emphasis on M-type stars (Section~\ref{sec:dis_psi}). We then explain how the contrast and optical depth vary with spectral type (Section~\ref{sec:dis_spt_res}), and explore their observational impact by analyzing flux and asymmetric features that could be confused with planetary signals (Section~\ref{sec:dis_flux}). The validity of the collisionless assumption is lastly considered, along with other model limitations (Section~\ref{sec:dis_coll}). Additional analyses and caveats are briefly introduced here, with detailed discussions provided in Appendix~\ref{sec:app_anal}.

\subsection{Reliability and age dependence of the $\psi$ prescription} \label{sec:dis_psi_acc_age}

The $\psi$ values in this study are estimated via a stepwise prescription: stellar rotation periods ($P_\ast$) are derived from gyrochronology models based on $T_\ast$ and age, mass-loss rates ($\dot{M}_\ast$) from Eq.~\eqref{eq:massloss}, and $\psi$ from Eq.~\eqref{eq: psi}.  
Uncertainties mainly stem from $P_\ast$ and $\dot{M}_\ast$, while errors in basic stellar parameters and the functional form of Eq.~\eqref{eq: psi} are neglected.

We adopt three gyrochronology models---\citet{Lu2024}, \citet{MH2008}, and \citet{Dungee2022}---applied sequentially from F4 to M4 stars (Section~\ref{sec:psi}). The \citet{Lu2024} relation provides consistent calibration across the full spectral range up to $\sim$14~Gyr, with typical $P_\ast$ uncertainties of $\sim10\%$. For G--K stars, \citet{MH2008} has minimum $P_\ast$ uncertainties of $\sim0.12$~days. \citet{Dungee2022} yields results consistent with \citet{Lu2024} with similar uncertainties but offers a more conservative calibration of $\psi$ for M-type stars, exclusive for 4~Gyr-old ages.  
Mass-loss rates follow \citet{Johnstone2015,Johnstone2021}, which are constrained for 0.1--1.2~$M_\odot$ main-sequence stars but less certain for fully convective M-dwarfs. The exponents in the scaling with $P_\ast$ and $M_\ast$ have typical uncertainties of order a few tens of percent.

\begin{figure*}
    \centering
    \includegraphics[width=1\linewidth]{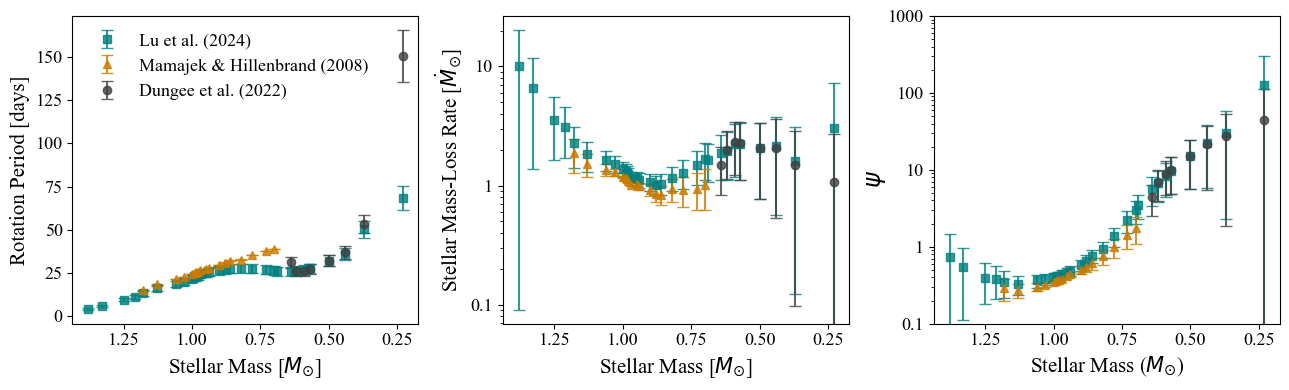}
    \caption{Stellar rotation period ($P_\ast$, left), mass-loss rate ($\dot{M}_\ast$, middle), and $\psi$ (right) for main-sequence stars of spectral types F4--M4, at age 4~Gyr (2--3~Gyr for F-type). Different symbols/colors show results from the three gyrochronology models, with approximate uncertainties.}
    \label{fig:pres_err}
\end{figure*}

The propagated uncertainties in $\psi$ (Fig.~\ref{fig:pres_err}) are largest for F4 and M4 stars, mainly reflecting limited observational constraints and magnetic dynamo variability in fully convective stars. Nevertheless, the overall trend of increasing $\psi$ toward late K--M types is consistent across all models. Across the varying $\psi$ values from different gyrochronology models, we have adopted the conservative estimates for each stellar type (Fig.~\ref{fig:psi}). 
The particularly large $\psi$ value for M4-type stars primarily arises from their low stellar luminosities: at 4~Gyr, Fig.~\ref{fig:pres_err} shows that an M4 star has a long rotation period of $\sim$151~days \citep{Dungee2022} and a modest stellar mass-loss rate of $\sim$1.06~$\dot{M}_{\odot}$, comparable to G--K stars. However, its extremely low luminosity ($\sim$0.0072~$L_{\odot}$; Table~\ref{tab:parameters}) greatly enhances the importance of stellar wind drag relative to radiation, producing the higher $\psi$ via Eq.~\eqref{eq: psi}. 

\begin{figure}
    \centering
    \includegraphics[width=1\linewidth]{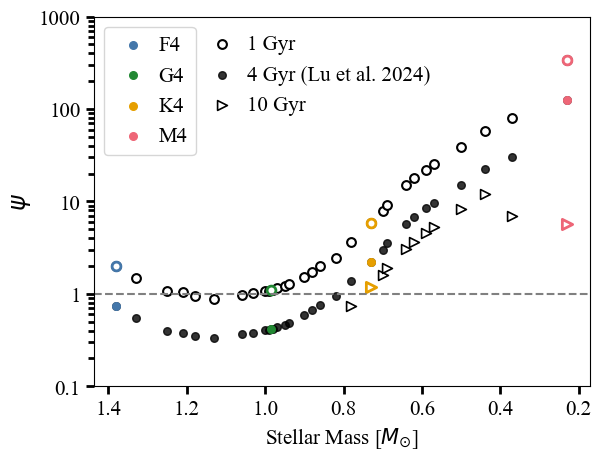}
    \caption{$\psi$ values given by Eqs.~\eqref{eq: psi} and~\eqref{eq:massloss} for main-sequence stars of spectral types F4--M4 at an age of 1~Gyr (empty circles), 4~Gyr (2--3~Gyr for F-type stars; filled circles), and 10 Gyr (empty triangles) using gyrochrones from \citet{Lu2024}. For 10 Gyr, only values for stars with masses below $\sim$ 0.8~$M_\odot$ are shown, as more massive stars have evolved off the main sequence by that age \citep{Choi2016}. The 4~Gyr results are almost identical to those in Fig.~\ref{fig:psi}, except for the larger value for M4 due to the change in gyrochronology model.}
    \label{fig:psi_age}
\end{figure}

To assess age dependence, we recomputed $\psi$ at 1, 4, and 10~Gyr using the \citet{Lu2024} gyrochrones (Fig.~\ref{fig:psi_age}). Younger systems show systematically higher $\psi$ across all spectral types, with most exceeding unity and M4 stars reaching $\sim$343 at 1~Gyr compared to $\sim$127 at 4~Gyr. While we do not extend our analysis to even younger ages due to gyrochronology limits, the increasing dominance of stellar wind drag with decreasing age is consistent with previous findings \citep[e.g.,][]{NK2023}. The K2-type star $\epsilon$~Eridani, with an age of $\lesssim$1~Gyr, has a value of $\psi\sim28$ derived from observed mass-loss rates \citep{Reid2011}, higher than the value ($\sim$2) for 1 Gyr in Fig.~\ref{fig:psi_age}. Likewise, the M0-type star AU~Mic at $\sim$15~Myr is estimated to have $\psi$ values of a few thousand (\citealt{Plavchan2005}; see Appendix~\ref{sec:dis_sbo}), further illustrating the expected steep age dependence. At older ages ($\sim$10~Gyr), $\psi$ decreases but remains above unity for K--M stars that stay on the main sequence, implying that stellar wind drag can remain relevant even in such old, low-mass systems.

Overall, this prescription provides a practical means to estimate stellar wind drag strength for main-sequence stars between F4 and M4 types and ages of $\sim$0.7--14~Gyr. Outside this range, $\dot{M}_\ast$ should be derived from direct observational methods. Despite the underlying uncertainties, the combined stellar-type and age trends of $\psi$ are physically supported and justify the $\psi$ values adopted in our simulations.

\subsection{Influence of stellar wind drag on debris disk structures} \label{sec:dis_psi}
The $\psi$ values used in this study (Fig.~\ref{fig:psi}) indicate that stellar wind drag can substantially influence dust dynamics around K--M-type stars, even at Gyr ages comparable to that of the Solar System. This complements earlier studies \citep[e.g.,][]{SK2008,Defrere2010}, which primarily considered PR drag as the main dust migration mechanism and used a constant solar wind value of $\psi \sim 0.3$. 
In particular, the high value of $\psi \sim 44$ for the M4-type star demonstrates that stellar wind drag can remain dynamically important over a wide range of evolutionary stages.  
This finding builds on the age-dependent trends discussed in Section~\ref{sec:dis_psi_acc_age}, showing that while the stellar wind dominance inferred for young early K-type stars (e.g., $\epsilon$~Eridani; \citealt{Reid2011}) weakens with age, late K to M-type stars can retain strong drag effects throughout their main-sequence lifetimes.  

The high $\psi$ values underscore the importance of incorporating stellar wind drag into models of dust distributions around old, low-mass stars---a significance further supported by our pilot study comparing ring contrast between the two models (see Section~\ref{sec:res_models_vs}). 
Since $\beta_t$ increases with $\psi$, a higher value of $\psi$ results in a shorter migration timescale, as can be inferred from Section \ref{sec:eff_beta} \citep{Mann2006, Plavchan2005}. This faster migration reduces the probability of resonant trapping, thereby lowering the resulting ring contrast.
Consequently, an increase in $\psi$ value leads to weaker resonant ring structures (see Eq.~\ref{eq:C}), as illustrated most clearly by the contrast change for the M-type star between Model~I and Model~II (Fig.~\ref{fig:contrast_pilot}). 
The contrasts for higher-mass stars exhibit smaller variations since the differences in their respective $\psi$ values between the two models are less pronounced.

This reduction in ring contrast may have important observational implications.
Using constant solar wind drag values regardless of spectral type variation can lead to an overestimation of the resonant ring optical depth by nearly a factor of two in the M4-type case (see Section~\ref{sec:res_models_vs}). This highlights the need to revisit observational strategies for M-dwarfs with updated models that properly account for stellar wind. Including spectral-dependent stellar wind drag can alter expectations for the detectability of exozodiacal structures around M-type stars, potentially broadening the range of viable targets for general exo-Earth searches. Accurate modeling of stellar wind is thus essential for realistic exozodi predictions around M-type stars.

Similarly, in smooth disks without resonant structures \citep[e.g.,][]{Kennedy2015}, accounting for stellar wind could lower noise from dust when replenishment is limited, rather than maintaining a fixed background optical depth. In such cases, stellar wind lowers optical depths by speeding up dust migration ($n_\text{BG} \propto t_\text{mig} \propto \left(1+\psi\right)^{-1}$).
This implication aligns with the analytical work of  \cite{Plavchan2005}, who attributed the lack of infrared excess detections around M-dwarfs aged ten to hundreds of Myr to the dominance of stellar wind drag. 
To explain this, they constructed a debris disk model in which collisional dust production is balanced by removal via stellar wind and PR drag. Unlike our model, which assumes continuous dust replenishment, their model includes gradual depletion of planetesimal reservoirs over time. Under such conditions, strong stellar wind drag rapidly removes dust, consistent with our study where higher $\psi$ values correspond to shorter effective migration timescales. 
\citet{NK2023} reached similar conclusions for young (<1~Myr) Sun-like systems undergoing terrestrial planet formation, under the assumption of strong stellar wind drag.

Further support for stellar wind-driven dust removal may come from the lack of warm excess in the AKARI 18~$\mathrm{\mu m}$ survey \citep{Fujiwara2013}, although current surveys targeting colder dust around M-dwarfs \citep{CC2023,Lestrade2025} suggest that such non-detections are a product of observational bias due to limited sensitivity. 
The studies by \citet{ML2014} and \citet{Binks2016} point to both possibilities---the dearth of dust around M-dwarfs resulting from either stellar wind drag removal or observational limitations. 
While further data with improved sensitivity are needed for confirmation, the estimations of $\psi$ suggest that in systems with limited dust replenishment, old M-type stars are likely to lack dust due to rapid inward migration. 
Our results thus extend the work of \citet{Plavchan2005} and \citet{NK2023} by explicitly presenting estimates of stellar wind drag strengths across spectral types from F to M, focusing on older systems.
However, predictions may differ when assuming continuous dust production and the presence of resonant structures, as considered in our model (see Section~\ref{sec:dis_spt_res}).

\subsection{Stellar-type dependence of resonant structures} \label{sec:dis_spt_res}
Despite the influence of stellar wind, our results show that the ring contrast increases for lower-mass stars, with M-type stars exhibiting the highest values (Fig.~\ref{fig:contrast_pilot}). 
This trend arises because the reduction in $M_\ast$ has a stronger influence than the increase in $\psi$ in determining the contrast. 
The contrast can be described by extending the semi-analytical functional form of Eq.~4 from \cite{SK2008}, following the approach of \cite{Wyatt2003} to fit an empirical function with power-law dependencies on the relevant physical parameters:
\begin{equation} \label{eq:C}
C = 1 + p_1 \left[ 1 + \left( \frac{p_2}{a_\text{p}^{1/2} \left[ \beta_\text{PR} \left(1+\psi\right) \right]^{-1} M_\ast^{-1/2}} \right)^{p_3} \right]^{-1},
\end{equation}
where $p_i = p_{i,1} \left( M_\text{p}/M_\ast \right)^{p_{i,2}}$, with $M_\text{p}$ and $M_\ast$ in units of $M_\oplus$ and $M_\odot$, respectively. 
This form shows that the contrast depends on $M_\text{p}/M_\ast$ and the modified term $a_\text{p}^{1/2} \left[ \beta_\text{PR} \left(1+\psi\right) \right]^{-1} M_\ast^{-1/2} $, expanding from the original parameters from \citet{SK2008} (see Section~\ref{sec:res_models_vs}).  The latter term is derived similarly as Eq.~\eqref{eq:x} \citep{Wyatt2003, Kuchner2003, Shannon2015}:
\begin{equation} \label{eq:mod_x}
\frac{t_\text{mig}}{t_\text{lib}} \propto \frac{a_\text{d}^2/\left( \beta_\text{t} M_\ast \right)}{a_\text{p}^{3/2} M_\ast^{-1/2}} \propto a_\text{p}^{1/2} \left[ \beta_\text{PR} \left(1+\psi\right) \right]^{-1} M_\ast^{-1/2},
\end{equation}
where the additional drag due to stellar wind is included in the effective migration timescale (see Eq.~\ref{eq:t_eff}).
The best-fit parameters $p_i$ are
\begin{equation}
p_1 \approx 9.08 \left( \frac{M_\text{p}}{M_\ast} \right)^{0.13},
p_2 \approx 388 \left( \frac{M_\text{p}}{M_\ast} \right)^{-0.88},
p_3 \approx 1.29 \left( \frac{M_\text{p}}{M_\ast} \right)^{-0.22},
\label{eq:contrast_p}
\end{equation}
where the coefficients were obtained through non-linear least-squares fitting to all simulated contrast values across different stellar types, planetary masses and dust sizes from both the pilot and main study. 
Most parameters are constrained to 10--20\%, though the exponents $p_{1,2}$ and $p_{3,2}$ have larger uncertainties ($\sim50\%$). These values offer a first-order estimation of the contrast rather than an exact analytical solution, with further discussion provided in Appendix~\ref{sec:dis_semianal}.

For fixed values of $M_\text{p}$ and $s$, the denominator term $a_\text{p}^{0.5} \beta_\text{PR}^{-1}\left(1+\psi\right)^{-1}M_\ast ^{-0.5}$ of Eq.~\eqref{eq:C} scales as $ M_\ast^{-2.125} \left(1+\psi\right)^{-1}$ , using $a_\text{p} \propto \sqrt{L_\ast} $, $\beta_\text{PR} \propto L_\ast / M_\ast$, and $L_\ast \propto M_\ast^{3.5} $ (see Section~ \ref{sec:method_eom} and eq.~\ref{eq:beta_pr}). Although $\psi$ increases up to $\sim$ 44  for M4-type stars, this increase is insufficient to offset the stronger effect of decreasing $M_\ast$.
The parameters in the contrast function (Eq.~\ref{eq:contrast_p}) vary in a way that further enhances the contrast as $M_\ast$ decreases, with $p_1$ increasing and $p_2$, $p_3$ decreasing. As a result, the resonant ring contrast overall increases with decreasing stellar mass. 
This relation can also be approximately estimated from the original term $a_\text{p}^{1/2}/\beta_\text{PR}$, although a more precise derivation requires extra $M_\ast$ terms and the inclusion of $\psi$ as done here. 
When ignoring $\psi$, this trend can also be understood intuitively from the definition of $\beta_\text{PR}$, where lower stellar mass leads to a smaller $\beta_\text{PR}$ (Eq.~\ref{eq:beta_pr}). This implies that dust particles in systems with lower-mass stars remain more gravitationally bound and are more likely to be trapped in MMRs, assuming a fixed dust size. 
Therefore, for the same level of background exozodi, lower-mass stars are expected to host more optically thick resonant structures---though not to the extent predicted by models without spectral type variation in $\psi$ (e.g., Model~I). As mentioned in Section~\ref{sec:res_models_vs}, a possible inversion of this trend between K and M-type due to even higher $\psi$ values for M4-type is discussed in Appendix~\ref{sec:dis_gyro}.

Since the maximum optical depths appear as part of the asymmetric structure around the inner edge of the resonant ring as seen in the optical depth distributions, the azimuthally averaged maximum contrast $C$ serves as a useful proxy for roughly estimating the peak exozodi signal most threatening to planet detection. Using the size-combined contrast values, we can approximate the strongest contamination level as $\tau_\text{BG} \cdot \langle C_\tau \rangle$ (see Section~\ref{sec:res_tau_F} for definition of $\langle C_\tau \rangle$), offering a crude estimate of the exozodi noise.
For instance, assuming a fixed background level of 3~zodis, the resulting resonant ring optical depths may reach up to $3 \cdot 1.95 \approx 6$~zodis for F-type stars, $3 \cdot 5.98 \approx 18$~zodis for G-type, $3 \cdot 8.07 \approx 24$~zodis for K-type, and $3 \cdot 9.85 \approx 30$~zodis for M-type stars (Section \ref{sec:res_tau_F}).  

Consequently, the optical depth of resonant structures would be higher for systems with stellar masses lower than the Sun, and vice versa for systems with higher stellar masses, compared to the results expected from the resonant models of  \cite{SK2008}. 
This strengthens the importance for considering different spectral types for resonant structure modeling, rather than relying solely on Solar-twin assumptions as done in previous studies \citep[e.g.,][]{Defrere2010,Stark2015,Currie2023}.
The increase of optical depth up to a factor of $\sim$ 2--10 relative to the background level (see Section~\ref{sec:res_tau_F}) also suggests the need to include resonant structures in exozodi models, when aiming to utilize it for estimating exoplanet observations. 

While our models predict systematically higher optical depths for resonant structures around lower-mass stars, current observational surveys of warm dust do not yet reveal a clear dependence on stellar type. For instance, the HOSTS survey with LBTI \citep{Ertel2020} did not find a statistically significant increase in exozodi levels toward low-mass stars. This apparent discrepancy likely stems from sample limitations, which included only A to K-type stars and mostly lacked known planets in their HZs, with $\tau$ Ceti being an exception \citep{Tuomi2013}. 
Recent work by \citet{Huang2025} using the Wide-field Infrared Survey Explorer (WISE; \citealt{Wright2010}), found that extreme warm-dust excesses (>21\%) occur in fewer than 1\% of M-type stars, consistent with previous results for FGK stars \citep{KW2013}. They emphasize that current facilities such as WISE and LBTI lack the sensitivity to detect fainter exozodis---such as those considered in this work---leaving most LIFE-resolvable exo-Earth targets unconstrained. Similarly, these survey results do not directly constrain planet-associated dust structures, limiting their relevance to our resonant model predictions. 

Rather than contradicting our findings, these observational gaps address the need for dedicated studies of systems with known planets in the HZ. 
Our models suggest that, in such systems, resonant structures induced by planets could result in more prevalent warm exozodis around lower-mass (K and M-type) stars, potentially at levels below the sensitivity of current surveys. Thus, while existing data do not constrain the low-density disks relevant to our study, our results may help inform expectations for future observations targeting faint exozodiacal dust in the vicinity of planetary orbits.

\subsection{Exozodi flux distributions across spectral types} \label{sec:dis_flux}

While contrast-based estimates offer a useful diagnostic of resonant ring significance, the simulated optical depth maps presented in this work (Fig.~\ref{fig:maps}) provide spatially resolved input for observational simulations, such as LIFEsim. These maps enable a more comprehensive evaluation of how exozodis affect the detectability of planets.
\begin{figure*} 
    \centering
    \includegraphics[width=\linewidth]{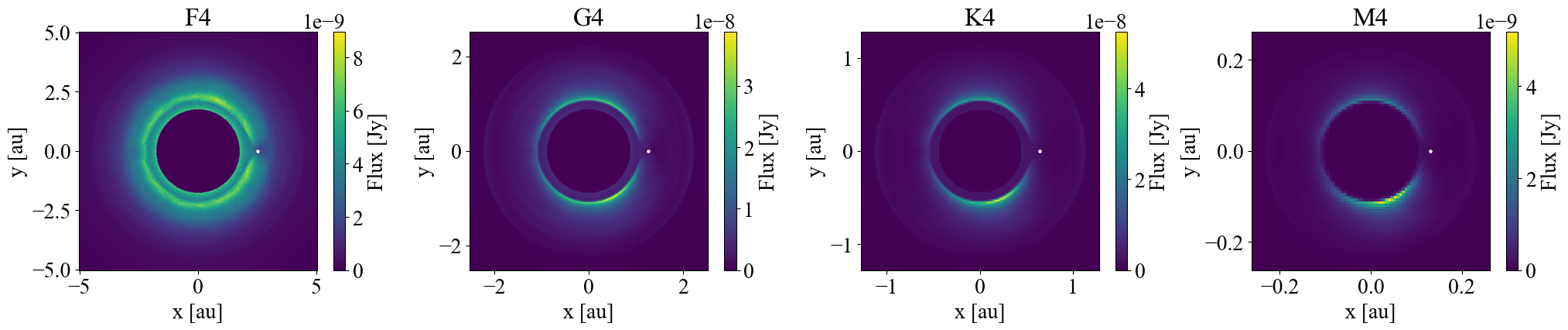}
    \includegraphics[width=1\linewidth]{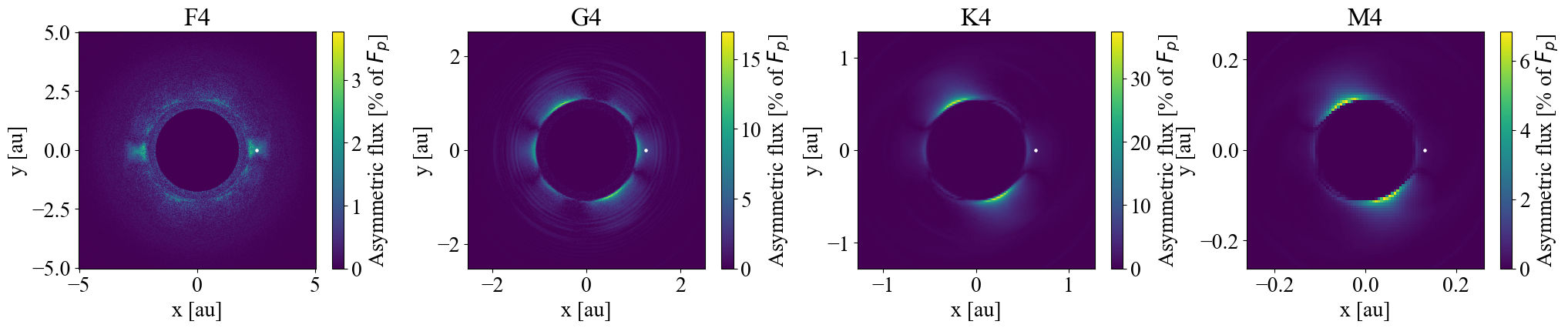}
    \caption{Flux with central regions masked (10~$\mathrm{\mu m}$; \textit{upper row}), and corresponding asymmetric flux relative to the planet flux $F_\text{p}$ (\textit{lower row}), following the format of Fig.~\ref{fig:maps}. This masking emphasizes resonant structures, and all maps are shown within a spatial scale of $\sim 2 ~ a_\text{p}$. Asymmetric flux panels follow the approach of Fig.~11 in \cite{Defrere2010}.}
    \label{fig:asym_flux}
\end{figure*}

To explore asymmetric structures, which can be critical sources of confusion with planetary signals, we compute asymmetric flux distributions by subtracting the flux of the centrally symmetric counterpart from each pixel of the original flux map and taking the absolute value of the residuals, following the method illustrated in Fig.~11 of \cite{Defrere2010}:
\begin{equation}
F_\mathrm{asym}(x,y) = \left| F(x,y) - F(-x,-y) \right|,
\end{equation}
where $F(x,y)$ is the flux at each pixel and $F(-x,-y)$ is the flux at the pixel symmetric with respect to the star. The results are given in Fig.~\ref{fig:asym_flux} (lower row), along with corresponding flux distributions (upper row) adapted from Fig.~\ref{fig:maps}.
Although some central residuals initially appeared in the asymmetric maps due to imperfect numerical cancellation, they were confirmed to be nearly symmetric and will likely contribute to shot noise under interferometric observations, rather than representing signals that could mimic planetary detections. For all types of stars, we masked out the central regions where flux dominates over that of the resonant ring (see Fig.~\ref{fig:maps}), to highlight the relevant asymmetric structures. 
The asymmetric fluxes are presented relative to the emission from the planet, where we assumed planetary thermal equilibrium with albedo A $\approx$ 0.3 similar to Earth \citep{Charbonneau2005}.

\begin{figure}
    \centering
    \includegraphics[width=1\linewidth]{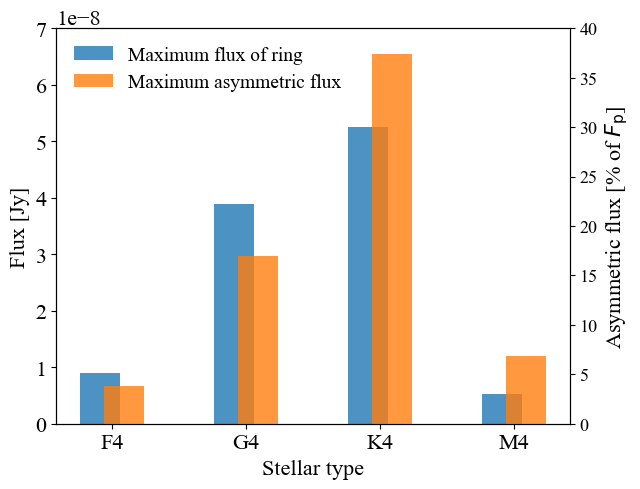}
    \caption{Maximum value of the resonant ring flux at $\sim10~\mathrm{\mu m}$ (blue, left axis) and the asymmetric flux (orange, right axis) for each spectral type, assuming a constant background level of $\sim3$~zodis and grain sizes of 0.1--300~$\mathrm{\mu m}$. Higher dust levels or the inclusion of larger grains may alter the fluxes (see Section~\ref{sec:dis_coll}). Note that the two values are not in scale.}
    \label{fig:flux_max}
\end{figure}

Focusing on the resonant ring region, the asymmetric flux maps reveal structures that could be confused with the planetary signal, with peak values reaching $\sim3.28\%, 14.1\%, 30.8\%, \text{and } 1.40\%$ of the planetary flux for the F4, G4, K4, and M4 stars respectively.
These values are shown in Fig.~\ref{fig:flux_max} along with the maximum flux values of the ring itself, ignoring the central region mentioned previously. Since the asymmetry structures originate from the resonant ring, the two values follow a similar trend with spectral type: K-type stars yield the strongest asymmetric signal relative to the planetary thermal emission, followed by G, M, and F-type stars.

This trend arises from the competing effects of ring contrast and system size, according to Eq.~\eqref{eq:flux}. As discussed in Section~\ref{sec:dis_spt_res}, lower stellar mass increases the contrast and thus the optical depth of the ring for a fixed background level. While the equilibrium temperature is similar regardless of spectral type since the ring structure is formed around the HZ, the emitting area of the system decreases with lower stellar mass since the system is more compact. Therefore, the increase in optical depth and decrease in grid area leads to a peak in the resonant ring flux (and thus asymmetric flux) for K-type stars, rather than a monotonic increase toward lower stellar masses.
This behavior reflects the fundamental interplay between stellar gravity and radiation in shaping the resonant structures. Dust grains around lower-mass stars are more gravitationally bound (as discussed in Section~\ref{sec:dis_spt_res}), which enhances resonant trapping and boosts the resulting ring contrast and optical depth. At the same time, the spatial extent of the resonant structure shrinks since lower-mass stars emit less radiation and have closer-in HZs. K-type stars represent a balance point---gravity dominates enough to enable high resonant optical depth, while radiation remains sufficient to avoid the tight spatial compactness that limits total flux in M-type systems. As a result, they naturally produce the strongest asymmetric flux in our models, assuming a constant background exozodi level. 

Although \citet{Quanz2022} concluded that exozodi resonant structures are unlikely to pose a significant obstacle to planet detection, their analysis relied on the studies of \citet{Defrere2010}, which are limited to Solar-twin systems. Our results show that the maximum asymmetric flux compared to the planetary flux for a K-type star can be nearly twice that of a G-type (Fig.~\ref{fig:flux_max}), emphasizing the need for further examination of stars beyond Solar analogs.
These results suggest that resonant structures should be considered across late-type stars (F4 to M4) when modeling exozodis for interferometric observations, with special attention given to K-type systems. While this conclusion is based on a relatively low background level of 3~zodis, implications for higher background exozodi levels are discussed in Section~\ref{sec:dis_coll}.

The flux and asymmetric flux maps for the G4-type star closely resemble those shown in Fig.~11 of \cite{Defrere2010}, as both are based on the work of \cite{SK2008}.  However, our simulations continue to track dust inward to the sublimation zone, whereas \cite{SK2008} truncate the disk at half the planet's position. While \cite{Defrere2010} note that this absent inner disk component is likely insignificant due to its expected central symmetry---an assumption our simulations confirm---our results indicate that including this region can still improve shot noise estimates, particularly given the high flux levels found in the area. 

Although strict conclusions regarding the modulated signal and cross-correlation for interferometry will require further analysis, these results provide valuable direction for refining future studies, particularly for the LIFE mission.

\subsection{Collisional effects and model limitations} \label{sec:dis_coll}
While this work represents a scenario where the impact of resonant structures are maximized, introducing more realistic factors such as dust grain collisions, dust with more dynamically active initial conditions, or the presence of additional planets, would likely reduce the contrast and ease the observational constraints \citep{SK2008, SK2009, KS2010, Defrere2010}. 

To evaluate the validity of our results neglecting particle collisions, we compare the timescales for effective migration by PR and stellar wind drag against the collisional timescale.
The effective migration timescale for a grain at a distance $a$ from the star is given by
\begin{equation} \label{eq:t_eff_cmp}
t_\text{mig} \simeq \frac{693}{\left(1+\psi\right)}  \left(\frac{\rho}{\mathrm{g \ cm^{-3}}}\right) \left(\frac{s}{\mathrm{\mu m}}\right)\left(\frac{L_\ast}{L_\odot}\right)^{-1} \left(\frac{a}{1\,\mathrm{au}}\right)^2 \text{ years}, 
\end{equation}
using Eqs.~\eqref{eq:beta_pr} and \eqref{eq:t_eff}. 
Following the methods of \citet{NK2023} and adapting their Eq. (6), the collisional timescale is
\begin{equation} \label{eq:t_coll}
\begin{aligned}
t_\text{coll} &= \frac{P}{4\pi} \left(\frac{\tau}{4}\right)^{-1} \left(\frac{s}{s_\text{BO,eff}}\right)^{1/2} \\
&\simeq 4.5 \times 10^6  
\left(\frac{s}{s_\text{BO,eff}}\right)^{1/2}
\left(\frac{M_\ast}{M_\odot}\right)^{-1/2}
\left(\frac{a}{1\,\mathrm{au}}\right)^{3/2}
\left(\frac{\tau}{\mathrm{zodi}}\right)^{-1} \, \mathrm{years},
\end{aligned}
\end{equation}
assuming collisions between grains of size $s$ and larger particles, in an exozodi with face-on optical depth $\tau$. $P$ represents the orbital period of the particle. If $s_\text{BO,eff}$ does not exist, the smallest size in the system is used instead.

\begin{figure*}
    \centering
    \includegraphics[width=0.8\linewidth]{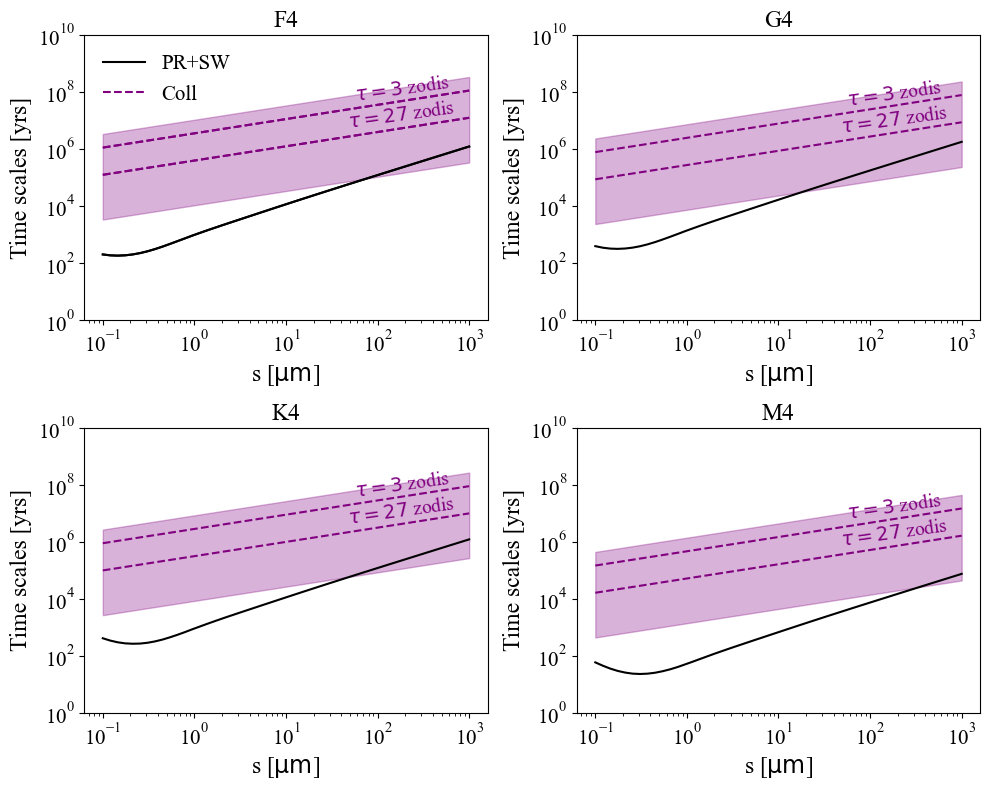}
    \caption{Comparison of timescales for effective migration driven by PR and stellar wind drag (solid black lines), with the collisional timescale corresponding to the optical depths $\tau_\text{BG}=3$ and 27~zodis referenced from the HOSTS survey (purple dashed lines; see Sections~\ref{sec:res_tau_F} and~\ref{sec:dis_spt_res}). The comparison is made near the position of the planet $\sim a_\text{p}$ ($ \text{HZ}_\text{mid}$). Collisional timescales for a range of face-on optical depths, $\tau$ = 1--1000~zodis, are also shown as a purple band (top to bottom). Results are shown for each spectral type, as a function of grain sizes of 0.1--1000~$\mathrm{\mu m}$. Adapted from Fig.~3 of \citet{NK2023}.}
    \label{fig:times_no_planet}
\end{figure*}

Following Fig.~3 of \citet{NK2023},  $t_\text{mig}$ and $t_\text{coll}$ are shown in Fig.~\ref{fig:times_no_planet} for a smooth disk without planetary perturbations, evaluated at $\sim a_\text{p}$ (HZ$_\text{mid}$) for grain sizes of 0.1--1000 $\mathrm{\mu m}$. Across all grain sizes and stellar types (FGKM), $t_\text{coll}> t_\text{mig}$ even for disks up to a few hundred zodis, confirming that the collisionless assumption is valid for the nominal 3~zodis and the 27~zodis upper limit reported by HOSTS. This likely produces a steady-state size distribution $\propto s^{-2.5}$ in the absence of planetary perturbations, as smaller grains are preferentially removed by drag forces \citep{SK2008}. Stellar wind-driven migration dominates over collisions particularly for M-type stars, extending the findings of \citet{Plavchan2005} to Gyr-age systems. 
While dust around GJ 581 (2--8~Gyr, M3) is known to be more collisionally influenced \citep{Lestrade2012}, this applies to its cold disk at $\sim25$~au ($\sim1000$~zodis), not the lower levels of closer-in warm exozodis considered here.
\begin{figure*}
    \centering
    \includegraphics[width=0.8\linewidth]{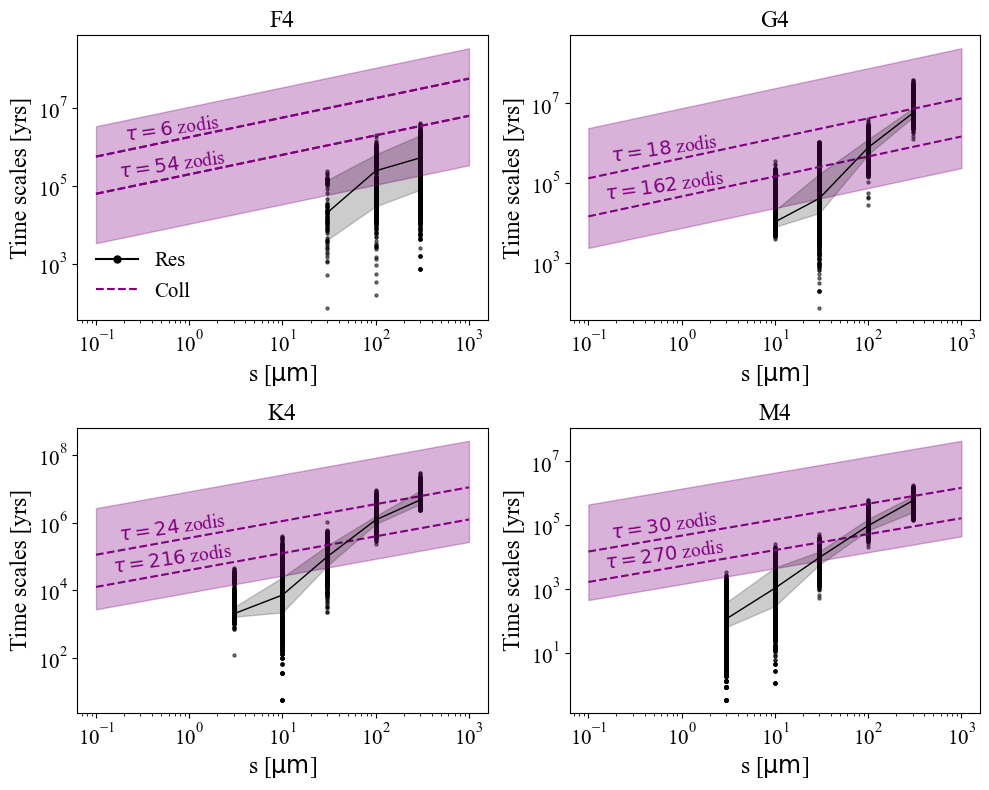}
    \caption{Comparison of timescales for resonant trapping from simulation data (black dots) with the median values (solid black lines) and 1$\sigma$ uncertainties (gray bands), against the collisional timescale corresponding to the maximum resonant ring optical depth of $\tau_\text{BG} \cdot \langle C_\tau \rangle$ (purple dashed lines). For each $\langle C_\tau \rangle$ value of $\sim$ 2, 6, 8, and 10 for F, G, K, and M-types stars, $\tau_\text{BG}$ of 3 (upper dashed-line) and 27~zodis (lower dashed-line) are examined (see Section~\ref{sec:res_tau_F}). Other plotting conventions are identical to those in Fig.~\ref{fig:times_no_planet}.}
    \label{fig:times_res}
\end{figure*}

For the densest regions, we compare the resonant trapping timescale $t_\text{res}$ (total survival time minus $t_\text{mig}$) with $t_\text{coll}$ at the azimuthal maximum of ring $\tau_\text{BG}\cdot\langle C_\tau\rangle$ (Fig.~\ref{fig:times_res}). At $\tau_\text{BG}=3$~zodis, $t_\text{res} \lesssim t_\text{coll}$ across all grains producing resonant structures, confirming the collisionless assumption. For $300~\mathrm{\mu m}$ grains (GKM), $t_\text{res}$ approaches $t_\text{coll}$, defining the critical size up to which the largest grains dominate the ring \citep{KS2010}.
F-type stars could support cm-sized or larger grains, potentially producing resonant rings with contrasts comparable to M-type stars (estimated from Fig.~\ref{fig:contrast_main_s} and Eq.~\ref{eq:C}). This may result in F-type stars exhibiting the highest ring fluxes due to both higher optical depths and larger spatial scales (Section~\ref{sec:dis_flux}). While this scenario is unlikely if exozodis share the typical size range of 1--100~$\mathrm{\mu}$m of zodiacal dust, we caution against directly applying our results to dust populations extending beyond $300~\mathrm{\mu m}$.

At higher background levels ($\tau_\text{BG}=27$~zodis), collisions become relevant for grains $\gtrsim100~ \mathrm{\mu m}$ (G-type) or $\gtrsim50~ \mathrm{\mu m}$ (K and M-type). Pilot study results for $s\sim50~\mathrm{\mu m}$ remain valid, while main study results for 100--300~$\mathrm{\mu m}$ may not. Under these conditions, $\langle C_\tau\rangle$ is expected to be lower than in Fig.~\ref{fig:contrast_main_s}: G stars would follow $\sim100~\mathrm{\mu m}$ contrasts ($\sim6$), K and M stars $\sim50~ \mathrm{\mu m}$ ($\sim$5--6). For F stars, the collisionless assumption still holds, allowing grains up to roughly $1000~ \mathrm{\mu m}$ and producing comparable contrasts. Overall, at higher dust levels and broader size distributions, we tentatively expect similar resonant ring contrasts across FGKM stars, with ring flux declining toward lower-mass stars as spatial scales shrink.

While our simulations assume face-on viewing geometry, the results from \cite{Defrere2010} have shown that exozodiacal noise increases with system inclination. Thus, more conservative estimates would be necessary regardless of spectral type when dealing with highly inclined targets \citep{Quanz2022}.
In light of these limitations, the models presented in this work could be further improved by incorporating particle collisions and extending the analysis to inclined systems.
Future work may also focus on more rigorously assessing the influence of resonant structures on the detection of exo-Earths around different spectral types using nulling interferometry, by deriving modulated signals and applying cross-correlation techniques based on the optical depth and flux distributions obtained here. 
Modeling the scattered-light flux from our optical depth maps would also benefit coronagraphic exoplanet imaging missions such as HWO.

\subsection{Additional analyses and caveats}
To aid the interpretation and application of our results, we highlight several underlying assumptions and extended analyses, detailed in the appendices. These include discussion on the effective blowout size across spectral types (Appendix~\ref{sec:dis_sbo}), semi-analytical contrast functions for efficiently estimating resonant ring strengths across spectral types (Appendix~\ref{sec:dis_semianal}), and the impact of alternative gyrochronology models on $\psi$ and resonant structures (Appendix~\ref{sec:dis_gyro}).

We also discuss further caveats of the study, regarding assumptions related to stellar evolution, particularly for F-type stars (Appendix~\ref{sec:dis_F_evol}).
Along with the limitations introduced in Section~\ref{sec:dis_coll}, such factors should be considered when applying our results to observational predictions or interpreting resonant structures in exozodi modeling. Acknowledging these considerations will help guide more comprehensive follow-up studies and ensure robust application to future interferometric missions. 

\section{Summary and conclusion} \label{sec:conclusion}

We simulated dust in F, G, K, M-type systems, modeling resonant rings around exo-Earths in the HZ and extending the work of \cite{SK2008}. Stellar wind plays a significant role in shaping dust dynamics, dominating over PR drag in M-type stars ($\psi$ $\sim$ 44) even at an old age of a few Gyr. 
This high value of $\psi$ leads to a decrease in resonant ring contrast to roughly half of that in models without spectral type variation in stellar wind, thereby reducing the estimated ring optical depth for M-type stars at a fixed background exozodi level. 

Resonant ring contrast and optical depth generally increases with decreasing stellar mass because these systems are more gravitationally dominated. While the exact K--M trend depends on rotational evolution and wind environment of M-type stars, the significance of stellar wind drag and the overall flux trends remain unchanged.
Our optical depth and flux maps at $10~\mathrm{\mu m}$ which lies within the LIFE band range (4--18.5~$\mathrm{\mu m}$), provide spatially resolved exozodi distributions that improve upon smooth-disk assumptions \citep{Kennedy2015} and expand previous Solar-analog studies \citep{SK2008, Defrere2010}.
Asymmetric flux distributions remain a key concern for nulling interferometry, particularly for K-type stars under a dust size range of 0.1--300~$\mathrm{\mu m}$, where peak flux relative to planetary emission is highest.

By accounting for spectral type variations in stellar wind drag, this work refines exozodi models for old, late-type stars and enables more realistic simulations for MIR interferometers such as LIFE, providing a foundation for improved target prioritization and detection strategies. 

\begin{acknowledgements}
This research was supported by a grant from the Korean National Research Foundation (NRF) (MEST), funded by the Korean government (No. 2023R1A2C1006180). We acknowledge the use of the computational facility \texttt{gmunu} and thank Prof. Hyung Mok Lee, Prof. Woong-Tae Kim, and Da-jung Jang for generously sharing access, as well as Dr. Hee-il Kim for technical support. 
We also thank Y. Lu for kindly sharing a Jupyter notebook that supplemented our implementation of the gyrochronology model presented in \citet{Lu2024}.
We are especially grateful to the referee for their insightful and constructive comments, which substantially improved the quality of this paper.
\end{acknowledgements}

%

\bibliographystyle{aa} 
\bibliography{aa_main} 

@ARTICLE{Dermott1984,
       author = {{Dermott}, S.~F. and {Nicholson}, P.~D. and {Burns}, J.~A. and {Houck}, J.~R.},
        title = "{Origin of the Solar System dust bands discovered by IRAS}",
      journal = {\nat},
         year = 1984,
        month = dec,
       volume = {312},
       number = {5994},
        pages = {505-509},
          doi = {10.1038/312505a0},
       adsurl = {https://ui.adsabs.harvard.edu/abs/1984Natur.312..505D}
}

@ARTICLE{Reach1995,
       author = {{Reach}, W.~T. and {Franz}, B.~A. and {Weiland}, J.~L. and {Hauser}, M.~G. and {Kelsall}, T.~N. and {Wright}, E.~L. and {Rawleyt}, G. and {Stemwedel}, S.~W. and {Spiesman}, W.~J.},
        title = "{Observational confirmation of a circumsolar dust ring by the COBE satellite}",
      journal = {\nat},
         year = 1995,
        month = apr,
       volume = {374},
       number = {6522},
        pages = {521-523},
          doi = {10.1038/374521a0},
       adsurl = {https://ui.adsabs.harvard.edu/abs/1995Natur.374..521R}
}

@ARTICLE{SK2008,
       author = {{Stark}, Christopher C. and {Kuchner}, Marc J.},
        title = "{The Detectability of Exo-Earths and Super-Earths Via Resonant Signatures in Exozodiacal Clouds}",
      journal = {\apj},
         year = 2008,
        month = oct,
       volume = {686},
       number = {1},
        pages = {637-648},
          doi = {10.1086/591442},
archivePrefix = {arXiv},
       eprint = {0810.2702},
 primaryClass = {astro-ph},
       adsurl = {https://ui.adsabs.harvard.edu/abs/2008ApJ...686..637S}
}

@ARTICLE{Burns1979,
       author = {{Burns}, J.~A. and {Lamy}, P.~L. and {Soter}, S.},
        title = "{Radiation forces on small particles in the solar system}",
      journal = {\icarus},
         year = 1979,
        month = oct,
       volume = {40},
       number = {1},
        pages = {1-48},
          doi = {10.1016/0019-1035(79)90050-2},
       adsurl = {https://ui.adsabs.harvard.edu/abs/1979Icar...40....1B}
}

@BOOK{MD1999,
       author = {{Murray}, Carl D. and {Dermott}, Stanley F.},
        title = "{Solar System Dynamics}",
         year = 1999,
        publisher = {Cambridge University Press},
          doi = {10.1017/CBO9781139174817},
       adsurl = {https://ui.adsabs.harvard.edu/abs/1999ssd..book.....M}
}

@ARTICLE{Chambers1999,
       author = {{Chambers}, J.~E.},
        title = "{A hybrid symplectic integrator that permits close encounters between massive bodies}",
      journal = {\mnras},
         year = 1999,
        month = apr,
       volume = {304},
       number = {4},
        pages = {793-799},
          doi = {10.1046/j.1365-8711.1999.02379.x},
       adsurl = {https://ui.adsabs.harvard.edu/abs/1999MNRAS.304..793C}
}

@ARTICLE{Jo2024,
       author = {{Jo}, Hangbin and {Ishiguro}, Masateru},
        title = "{Dynamical study of Geminid formation assuming a rotational instability scenario}",
      journal = {\aap},
         year = 2024,
        month = mar,
       volume = {683},
          eid = {A68},
        pages = {A68},
          doi = {10.1051/0004-6361/202347898},
archivePrefix = {arXiv},
       eprint = {2401.03682},
 primaryClass = {astro-ph.EP},
       adsurl = {https://ui.adsabs.harvard.edu/abs/2024A&A...683A..68J}
}

@ARTICLE{KJ2011,
       author = {{Kortenkamp}, Stephen J. and {Joseph}, Emily C.~S.},
        title = "{Transformation of Trojans into quasi-satellites during planetary migration and their subsequent close-encounters with the host planet}",
      journal = {\icarus},
         year = 2011,
        month = oct,
       volume = {215},
       number = {2},
        pages = {669-681},
          doi = {10.1016/j.icarus.2011.07.019},
archivePrefix = {arXiv},
       eprint = {1102.2211},
 primaryClass = {astro-ph.EP},
       adsurl = {https://ui.adsabs.harvard.edu/abs/2011Icar..215..669K}
}

@ARTICLE{Sommer2020,
       author = {{Sommer}, M. and {Yano}, H. and {Srama}, R.},
        title = "{Effects of neighbouring planets on the formation of resonant dust rings in the inner Solar System}",
      journal = {\aap},
         year = 2020,
        month = mar,
       volume = {635},
          eid = {A10},
        pages = {A10},
          doi = {10.1051/0004-6361/201936676},
archivePrefix = {arXiv},
       eprint = {2001.07611},
 primaryClass = {astro-ph.EP},
       adsurl = {https://ui.adsabs.harvard.edu/abs/2020A&A...635A..10S}
}

@ARTICLE{Gustafson1994,
       author = {{Gustafson}, B.~A.~S.},
        title = "{Physics of Zodiacal Dust}",
      journal = {Annual Review of Earth and Planetary Sciences},
         year = 1994,
        month = jan,
       volume = {22},
        pages = {553-595},
          doi = {10.1146/annurev.ea.22.050194.003005},
       adsurl = {https://ui.adsabs.harvard.edu/abs/1994AREPS..22..553G}
}

@ARTICLE{Robertson1937,
       author = {{Robertson}, H.~P.},
        title = "{Dynamical effects of radiation in the solar system}",
      journal = {\mnras},
         year = 1937,
        month = apr,
       volume = {97},
        pages = {423},
          doi = {10.1093/mnras/97.6.423},
       adsurl = {https://ui.adsabs.harvard.edu/abs/1937MNRAS..97..423R}
}

@ARTICLE{Minato2006,
       author = {{Minato}, T. and {K{\"o}hler}, M. and {Kimura}, H. and {Mann}, I. and {Yamamoto}, T.},
        title = "{Momentum transfer to fluffy dust aggregates from stellar winds}",
      journal = {\aap},
         year = 2006,
        month = jun,
       volume = {452},
       number = {2},
        pages = {701-707},
          doi = {10.1051/0004-6361:20054774},
       adsurl = {https://ui.adsabs.harvard.edu/abs/2006A&A...452..701M}
}

@ARTICLE{Mann2006,
       author = {{Mann}, Ingrid and {K{\"o}hler}, Melanie and {Kimura}, Hiroshi and {Cechowski}, Andrzej and {Minato}, Tetsunori},
        title = "{Dust in the solar system and in extra-solar planetary systems}",
      journal = {\aapr},
         year = 2006,
        month = jun,
       volume = {13},
       number = {3},
        pages = {159-228},
          doi = {10.1007/s00159-006-0028-0},
       adsurl = {https://ui.adsabs.harvard.edu/abs/2006A&ARv..13..159M}
}

@ARTICLE{Klacka2014,
       author = {{Kla{\v{c}}ka}, J.},
        title = "{Solar wind dominance over the Poynting-Robertson effect in secular orbital evolution of dust particles}",
      journal = {\mnras},
         year = 2014,
        month = sep,
       volume = {443},
       number = {1},
        pages = {213-229},
          doi = {10.1093/mnras/stu1133},
archivePrefix = {arXiv},
       eprint = {1401.0581},
 primaryClass = {astro-ph.EP},
       adsurl = {https://ui.adsabs.harvard.edu/abs/2014MNRAS.443..213K}
}

@book{BS1980,
  title={Introduction to Numerical Analysis},
  author={Stoer, J. and Bartels, R. and Gautschi, W. and Bulirsch, R. and Witzgall, C.},
  year={1980},
  publisher={Springer New York}
}

@ARTICLE{Pecaut2013,
       author = {{Pecaut}, Mark J. and {Mamajek}, Eric E.},
        title = "{Intrinsic Colors, Temperatures, and Bolometric Corrections of Pre-main-sequence Stars}",
      journal = {\apjs},
         year = 2013,
        month = sep,
       volume = {208},
       number = {1},
          eid = {9},
        pages = {9},
          doi = {10.1088/0067-0049/208/1/9},
archivePrefix = {arXiv},
       eprint = {1307.2657},
 primaryClass = {astro-ph.SR},
       adsurl = {https://ui.adsabs.harvard.edu/abs/2013ApJS..208....9P}
}

@ARTICLE{Pecaut2012,
       author = {{Pecaut}, Mark J. and {Mamajek}, Eric E. and {Bubar}, Eric J.},
        title = "{A Revised Age for Upper Scorpius and the Star Formation History among the F-type Members of the Scorpius-Centaurus OB Association}",
      journal = {\apj},
         year = 2012,
        month = feb,
       volume = {746},
       number = {2},
          eid = {154},
        pages = {154},
          doi = {10.1088/0004-637X/746/2/154},
archivePrefix = {arXiv},
       eprint = {1112.1695},
 primaryClass = {astro-ph.SR},
       adsurl = {https://ui.adsabs.harvard.edu/abs/2012ApJ...746..154P}
}

@ARTICLE{Kopparapu2014,
       author = {{Kopparapu}, Ravi Kumar and {Ramirez}, Ramses M. and {SchottelKotte}, James and {Kasting}, James F. and {Domagal-Goldman}, Shawn and {Eymet}, Vincent},
        title = "{Habitable Zones around Main-sequence Stars: Dependence on Planetary Mass}",
      journal = {\apjl},
         year = 2014,
        month = jun,
       volume = {787},
       number = {2},
          eid = {L29},
        pages = {L29},
          doi = {10.1088/2041-8205/787/2/L29},
archivePrefix = {arXiv},
       eprint = {1404.5292},
 primaryClass = {astro-ph.EP},
       adsurl = {https://ui.adsabs.harvard.edu/abs/2014ApJ...787L..29K}
}

@ARTICLE{Kuchner2003,
       author = {{Kuchner}, Marc J.},
        title = "{Planetary Perturbers in Debris Disks}",
      journal = {Earth Moon and Planets},
         year = 2003,
        month = jun,
       volume = {92},
       number = {1},
        pages = {435-445},
          doi = {10.1023/B:MOON.0000031957.71828.d3},
       adsurl = {https://ui.adsabs.harvard.edu/abs/2003EM&P...92..435K}
}

@ARTICLE{JZ1989,
       author = {{Jackson}, A.~A. and {Zook}, H.~A.},
        title = "{A Solar System dust ring with the Earth as its shepherd}",
      journal = {\nat},
         year = 1989,
        month = feb,
       volume = {337},
       number = {6208},
        pages = {629-631},
          doi = {10.1038/337629a0},
       adsurl = {https://ui.adsabs.harvard.edu/abs/1989Natur.337..629J}
}

@ARTICLE{MY1982,
       author = {{Mukai}, T. and {Yamamoto}, T.},
        title = "{Solar wind pressure on interplanetary dust}",
      journal = {\aap},
         year = 1982,
        month = mar,
       volume = {107},
       number = {1},
        pages = {97-100},
       adsurl = {https://ui.adsabs.harvard.edu/abs/1982A&A...107...97M}
}

@ARTICLE{KirchW2013,
       author = {{Kirchschlager}, F. and {Wolf}, S.},
        title = "{Porous dust grains in debris disks}",
      journal = {\aap},
         year = 2013,
        month = apr,
       volume = {552},
          eid = {A54},
        pages = {A54},
          doi = {10.1051/0004-6361/201220486},
archivePrefix = {arXiv},
       eprint = {1302.5275},
 primaryClass = {astro-ph.SR},
       adsurl = {https://ui.adsabs.harvard.edu/abs/2013A&A...552A..54K}
}

@ARTICLE{Krivov2006,
       author = {{Krivov}, A.~V. and {L{\"o}hne}, T. and {Srem{\v{c}}evi{\'c}}, M.},
        title = "{Dust distributions in debris disks: effects of gravity, radiation pressure and collisions}",
      journal = {\aap},
         year = 2006,
        month = aug,
       volume = {455},
       number = {2},
        pages = {509-519},
          doi = {10.1051/0004-6361:20064907},
       adsurl = {https://ui.adsabs.harvard.edu/abs/2006A&A...455..509K}
}

@ARTICLE{Kim2018,
       author = {{Kim}, M. and {Wolf}, S. and {L{\"o}hne}, T. and {Kirchschlager}, F. and {Krivov}, A.~V.},
        title = "{Impact of planetesimal eccentricities and material strength on the appearance of eccentric debris disks}",
      journal = {\aap},
         year = 2018,
        month = oct,
       volume = {618},
          eid = {A38},
        pages = {A38},
          doi = {10.1051/0004-6361/201833061},
archivePrefix = {arXiv},
       eprint = {1806.02391},
 primaryClass = {astro-ph.EP},
       adsurl = {https://ui.adsabs.harvard.edu/abs/2018A&A...618A..38K}
}

@ARTICLE{Defrere2010,
       author = {{Defr{\`e}re}, D. and {Absil}, O. and {den Hartog}, R. and {Hanot}, C. and {Stark}, C.},
        title = "{Nulling interferometry: impact of exozodiacal clouds on the performance of future life-finding space missions}",
      journal = {\aap},
         year = 2010,
        month = jan,
       volume = {509},
          eid = {A9},
        pages = {A9},
          doi = {10.1051/0004-6361/200912973},
archivePrefix = {arXiv},
       eprint = {0910.3486},
 primaryClass = {astro-ph.IM},
       adsurl = {https://ui.adsabs.harvard.edu/abs/2010A&A...509A...9D}
}

@INPROCEEDINGS{Defrere2012,
       author = {{Defr{\`e}re}, D. and {Stark}, C. and {Cahoy}, K. and {Beerer}, I.},
        title = "{Direct imaging of exoEarths embedded in clumpy debris disks}",
    booktitle = {Space Telescopes and Instrumentation 2012: Optical, Infrared, and Millimeter Wave},
         year = 2012,
       editor = {{Clampin}, Mark C. and {Fazio}, Giovanni G. and {MacEwen}, Howard A. and {Oschmann}, Jacobus M., Jr.},
       series = {Society of Photo-Optical Instrumentation Engineers (SPIE) Conference Series},
       volume = {8442},
        month = sep,
          eid = {84420M},
        pages = {84420M},
          doi = {10.1117/12.926324},
       adsurl = {https://ui.adsabs.harvard.edu/abs/2012SPIE.8442E..0MD}
}

@ARTICLE{Dermott1994,
       author = {{Dermott}, Stanley F. and {Jayaraman}, Sumita and {Xu}, Y.~L. and {Gustafson}, B. {\r{A}}. S. and {Liou}, J.~C.},
        title = "{A circumsolar ring of asteroidal dust in resonant lock with the Earth}",
      journal = {\nat},
         year = 1994,
        month = jun,
       volume = {369},
       number = {6483},
        pages = {719-723},
          doi = {10.1038/369719a0},
       adsurl = {https://ui.adsabs.harvard.edu/abs/1994Natur.369..719D}
}

@ARTICLE{Kucner2007,
       author = {{Kuchner}, M. and {Stark}, C. and {Absil}, O. and {Augereau}, J. -C. and {Thebault}, P.},
        title = "{Dynamics of Exozodiacal Clouds}",
      journal = {arXiv e-prints},
         year = 2007,
        month = jul,
          eid = {arXiv:0707.1280},
        pages = {arXiv:0707.1280},
          doi = {10.48550/arXiv.0707.1280},
archivePrefix = {arXiv},
       eprint = {0707.1280},
 primaryClass = {astro-ph},
       adsurl = {https://ui.adsabs.harvard.edu/abs/2007arXiv0707.1280K}
}

@ARTICLE{Kral2017,
       author = {{Kral}, Quentin and {Krivov}, Alexander V. and {Defr{\`e}re}, Denis and {van Lieshout}, Rik and {Bonsor}, Amy and {Augereau}, Jean-Charles and {Th{\'e}bault}, Philippe and {Ertel}, Steve and {Lebreton}, J{\'e}r{\'e}my and {Absil}, Olivier},
        title = "{Exozodiacal clouds: hot and warm dust around main sequence stars}",
      journal = {The Astronomical Review},
         year = 2017,
        month = apr,
       volume = {13},
       number = {2},
        pages = {69-111},
          doi = {10.1080/21672857.2017.1353202},
archivePrefix = {arXiv},
       eprint = {1703.02540},
 primaryClass = {astro-ph.EP},
       adsurl = {https://ui.adsabs.harvard.edu/abs/2017AstRv..13...69K}
}

@ARTICLE{TW2022,
       author = {{Tuchow}, Noah W. and {Wright}, Jason T.},
        title = "{Potential Habitability as a Stellar Property: Effects of Model Uncertainties and Measurement Precision}",
      journal = {\apj},
         year = 2022,
        month = may,
       volume = {930},
       number = {1},
          eid = {78},
        pages = {78},
          doi = {10.3847/1538-4357/ac65ea},
archivePrefix = {arXiv},
       eprint = {2112.02745},
 primaryClass = {astro-ph.EP},
       adsurl = {https://ui.adsabs.harvard.edu/abs/2022ApJ...930...78T}
}

@ARTICLE{Stark2011,
       author = {{Stark}, Christopher C.},
        title = "{The Transit Light Curve of an Exozodiacal Dust Cloud}",
      journal = {\aj},
         year = 2011,
        month = oct,
       volume = {142},
       number = {4},
          eid = {123},
        pages = {123},
          doi = {10.1088/0004-6256/142/4/123},
archivePrefix = {arXiv},
       eprint = {1108.1396},
 primaryClass = {astro-ph.SR},
       adsurl = {https://ui.adsabs.harvard.edu/abs/2011AJ....142..123S}
}

@ARTICLE{Ertel2020,
       author = {{Ertel}, S. and {Defr{\`e}re}, D. and {Hinz}, P. and {Mennesson}, B. and {Kennedy}, G.~M. and {Danchi}, W.~C. and {Gelino}, C. and {Hill}, J.~M. and {Hoffmann}, W.~F. and {Mazoyer}, J. and {Rieke}, G. and {Shannon}, A. and {Stapelfeldt}, K. and {Spalding}, E. and {Stone}, J.~M. and {Vaz}, A. and {Weinberger}, A.~J. and {Willems}, P. and {Absil}, O. and {Arbo}, P. and {Bailey}, V.~P. and {Beichman}, C. and {Bryden}, G. and {Downey}, E.~C. and {Durney}, O. and {Esposito}, S. and {Gaspar}, A. and {Grenz}, P. and {Haniff}, C.~A. and {Leisenring}, J.~M. and {Marion}, L. and {McMahon}, T.~J. and {Millan-Gabet}, R. and {Montoya}, M. and {Morzinski}, K.~M. and {Perera}, S. and {Pinna}, E. and {Pott}, J. -U. and {Power}, J. and {Puglisi}, A. and {Roberge}, A. and {Serabyn}, E. and {Skemer}, A.~J. and {Su}, K.~Y.~L. and {Vaitheeswaran}, V. and {Wyatt}, M.~C.},
        title = "{The HOSTS Survey for Exozodiacal Dust: Observational Results from the Complete Survey}",
      journal = {\aj},
         year = 2020,
        month = apr,
       volume = {159},
       number = {4},
          eid = {177},
        pages = {177},
          doi = {10.3847/1538-3881/ab7817},
archivePrefix = {arXiv},
       eprint = {2003.03499},
 primaryClass = {astro-ph.SR},
       adsurl = {https://ui.adsabs.harvard.edu/abs/2020AJ....159..177E}
}

@ARTICLE{Currie2023,
       author = {{Currie}, Miles H. and {Stark}, Christopher C. and {Kammerer}, Jens and {Juanola-Parramon}, Roser and {Meadows}, Victoria S.},
        title = "{Mitigating Worst-case Exozodiacal Dust Structure in High-contrast Images of Earth-like Exoplanets}",
      journal = {\aj},
         year = 2023,
        month = nov,
       volume = {166},
       number = {5},
          eid = {197},
        pages = {197},
          doi = {10.3847/1538-3881/acfda7},
archivePrefix = {arXiv},
       eprint = {2309.14234},
 primaryClass = {astro-ph.EP},
       adsurl = {https://ui.adsabs.harvard.edu/abs/2023AJ....166..197C}
}

@misc{currie2025,
      title={Exozodiacal dust as a limitation to exoplanet imaging and spectroscopy}, 
      author={Miles H. Currie and John Debes and Yasuhiro Hasegawa and Isabel Rebollido and Virginie Faramaz and Steve Ertel and William Danchi and Bertrand Mennesson and Mark Wyatt and NASA SAG23 Members},
      year={2025},
      eprint={2503.19932},
      archivePrefix={arXiv},
      primaryClass={astro-ph.IM},
      url={https://arxiv.org/abs/2503.19932}, 
}

@ARTICLE{Plavchan2005,
       author = {{Plavchan}, Peter and {Jura}, M. and {Lipscy}, S.~J.},
        title = "{Where Are the M Dwarf Disks Older Than 10 Million Years?}",
      journal = {\apj},
         year = 2005,
        month = oct,
       volume = {631},
       number = {2},
        pages = {1161-1169},
          doi = {10.1086/432568},
archivePrefix = {arXiv},
       eprint = {astro-ph/0506132},
 primaryClass = {astro-ph},
       adsurl = {https://ui.adsabs.harvard.edu/abs/2005ApJ...631.1161P}
}

@ARTICLE{Rodigas2014,
       author = {{Rodigas}, Timothy J. and {Malhotra}, Renu and {Hinz}, Philip M.},
        title = "{Predictions for Shepherding Planets in Scattered Light Images of Debris Disks}",
      journal = {\apj},
         year = 2014,
        month = jan,
       volume = {780},
       number = {1},
          eid = {65},
        pages = {65},
          doi = {10.1088/0004-637X/780/1/65},
archivePrefix = {arXiv},
       eprint = {1311.1207},
 primaryClass = {astro-ph.EP},
       adsurl = {https://ui.adsabs.harvard.edu/abs/2014ApJ...780...65R}
}

@ARTICLE{FD2002,
       author = {{Fixsen}, D.~J. and {Dwek}, Eli},
        title = "{The Zodiacal Emission Spectrum as Determined by COBE and Its Implications}",
      journal = {\apj},
         year = 2002,
        month = oct,
       volume = {578},
       number = {2},
        pages = {1009-1014},
          doi = {10.1086/342658},
       adsurl = {https://ui.adsabs.harvard.edu/abs/2002ApJ...578.1009F}
}

@ARTICLE{Wyatt2003,
       author = {{Wyatt}, M.~C.},
        title = "{Resonant Trapping of Planetesimals by Planet Migration: Debris Disk Clumps and Vega's Similarity to the Solar System}",
      journal = {\apj},
         year = 2003,
        month = dec,
       volume = {598},
       number = {2},
        pages = {1321-1340},
          doi = {10.1086/379064},
archivePrefix = {arXiv},
       eprint = {astro-ph/0308253},
 primaryClass = {astro-ph},
       adsurl = {https://ui.adsabs.harvard.edu/abs/2003ApJ...598.1321W}
}

@ARTICLE{Shannon2015,
       author = {{Shannon}, Andrew and {Mustill}, Alexander J. and {Wyatt}, Mark},
        title = "{Capture and evolution of dust in planetary mean-motion resonances: a fast, semi-analytic method for generating resonantly trapped disc images}",
      journal = {\mnras},
         year = 2015,
        month = mar,
       volume = {448},
       number = {1},
        pages = {684-702},
          doi = {10.1093/mnras/stv045},
archivePrefix = {arXiv},
       eprint = {1501.01631},
 primaryClass = {astro-ph.EP},
       adsurl = {https://ui.adsabs.harvard.edu/abs/2015MNRAS.448..684S}
}

@ARTICLE{Dohnanyi1969,
       author = {{Dohnanyi}, J.~S.},
        title = "{Collisional Model of Asteroids and Their Debris}",
      journal = {\jgr},
         year = 1969,
        month = may,
       volume = {74},
        pages = {2531-2554},
          doi = {10.1029/JB074i010p02531},
       adsurl = {https://ui.adsabs.harvard.edu/abs/1969JGR....74.2531D}
}

@ARTICLE{Quanz2022,
       author = {{Quanz}, S.~P. and {Ottiger}, M. and {Fontanet}, E. and {Kammerer}, J. and {Menti}, F. and {Dannert}, F. and {Gheorghe}, A. and {Absil}, O. and {Airapetian}, V.~S. and {Alei}, E. and {Allart}, R. and {Angerhausen}, D. and {Blumenthal}, S. and {Buchhave}, L.~A. and {Cabrera}, J. and {Carri{\'o}n-Gonz{\'a}lez}, {\'O}. and {Chauvin}, G. and {Danchi}, W.~C. and {Dandumont}, C. and {Defr{\'e}re}, D. and {Dorn}, C. and {Ehrenreich}, D. and {Ertel}, S. and {Fridlund}, M. and {Garc{\'\i}a Mu{\~n}oz}, A. and {Gasc{\'o}n}, C. and {Girard}, J.~H. and {Glauser}, A. and {Grenfell}, J.~L. and {Guidi}, G. and {Hagelberg}, J. and {Helled}, R. and {Ireland}, M.~J. and {Janson}, M. and {Kopparapu}, R.~K. and {Korth}, J. and {Kozakis}, T. and {Kraus}, S. and {L{\'e}ger}, A. and {Leedj{\"a}rv}, L. and {Lichtenberg}, T. and {Lillo-Box}, J. and {Linz}, H. and {Liseau}, R. and {Loicq}, J. and {Mahendra}, V. and {Malbet}, F. and {Mathew}, J. and {Mennesson}, B. and {Meyer}, M.~R. and {Mishra}, L. and {Molaverdikhani}, K. and {Noack}, L. and {Oza}, A.~V. and {Pall{\'e}}, E. and {Parviainen}, H. and {Quirrenbach}, A. and {Rauer}, H. and {Ribas}, I. and {Rice}, M. and {Romagnolo}, A. and {Rugheimer}, S. and {Schwieterman}, E.~W. and {Serabyn}, E. and {Sharma}, S. and {Stassun}, K.~G. and {Szul{\'a}gyi}, J. and {Wang}, H.~S. and {Wunderlich}, F. and {Wyatt}, M.~C. and {LIFE Collaboration}},
        title = "{Large Interferometer For Exoplanets (LIFE). I. Improved exoplanet detection yield estimates for a large mid-infrared space-interferometer mission}",
      journal = {\aap},
         year = 2022,
        month = aug,
       volume = {664},
          eid = {A21},
        pages = {A21},
          doi = {10.1051/0004-6361/202140366},
archivePrefix = {arXiv},
       eprint = {2101.07500},
 primaryClass = {astro-ph.EP},
       adsurl = {https://ui.adsabs.harvard.edu/abs/2022A&A...664A..21Q}
}

@ARTICLE{Dannert2022,
       author = {{Dannert}, Felix A. and {Ottiger}, Maurice and {Quanz}, Sascha P. and {Laugier}, Romain and {Fontanet}, Emile and {Gheorghe}, Adrian and {Absil}, Olivier and {Dandumont}, Colin and {Defr{\`e}re}, Denis and {Gasc{\'o}n}, Carlos and {Glauser}, Adrian M. and {Kammerer}, Jens and {Lichtenberg}, Tim and {Linz}, Hendrik and {Loicq}, Jer{\^o}me and {LIFE Collaboration}},
        title = "{Large Interferometer For Exoplanets (LIFE). II. Signal simulation, signal extraction, and fundamental exoplanet parameters from single-epoch observations}",
      journal = {\aap},
         year = 2022,
        month = aug,
       volume = {664},
          eid = {A22},
        pages = {A22},
          doi = {10.1051/0004-6361/202141958},
archivePrefix = {arXiv},
       eprint = {2203.00471},
 primaryClass = {astro-ph.EP},
       adsurl = {https://ui.adsabs.harvard.edu/abs/2022A&A...664A..22D}
}

@ARTICLE{Kennedy2015,
       author = {{Kennedy}, Grant M. and {Wyatt}, Mark C. and {Bailey}, Vanessa and {Bryden}, Geoffrey and {Danchi}, William C. and {Defr{\`e}re}, Denis and {Haniff}, Chris and {Hinz}, Philip M. and {Lebreton}, J{\'e}r{\'e}my and {Mennesson}, Bertrand and {Millan-Gabet}, Rafael and {Morales}, Farisa and {Pani{\'c}}, Olja and {Rieke}, George H. and {Roberge}, Aki and {Serabyn}, Eugene and {Shannon}, Andrew and {Skemer}, Andrew J. and {Stapelfeldt}, Karl R. and {Su}, Katherine Y.~L. and {Weinberger}, Alycia J.},
        title = "{Exo-zodi Modeling for the Large Binocular Telescope Interferometer}",
      journal = {\apjs},
         year = 2015,
        month = feb,
       volume = {216},
       number = {2},
          eid = {23},
        pages = {23},
          doi = {10.1088/0067-0049/216/2/23},
archivePrefix = {arXiv},
       eprint = {1412.0675},
 primaryClass = {astro-ph.EP},
       adsurl = {https://ui.adsabs.harvard.edu/abs/2015ApJS..216...23K}
}

@ARTICLE{Kelsall1998,
       author = {{Kelsall}, T. and {Weiland}, J.~L. and {Franz}, B.~A. and {Reach}, W.~T. and {Arendt}, R.~G. and {Dwek}, E. and {Freudenreich}, H.~T. and {Hauser}, M.~G. and {Moseley}, S.~H. and {Odegard}, N.~P. and {Silverberg}, R.~F. and {Wright}, E.~L.},
        title = "{The COBE Diffuse Infrared Background Experiment Search for the Cosmic Infrared Background. II. Model of the Interplanetary Dust Cloud}",
      journal = {\apj},
         year = 1998,
        month = nov,
       volume = {508},
       number = {1},
        pages = {44-73},
          doi = {10.1086/306380},
archivePrefix = {arXiv},
       eprint = {astro-ph/9806250},
 primaryClass = {astro-ph},
       adsurl = {https://ui.adsabs.harvard.edu/abs/1998ApJ...508...44K}
}

@ARTICLE{Johnstone2015,
       author = {{Johnstone}, C.~P. and {G{\"u}del}, M. and {Brott}, I. and {L{\"u}ftinger}, T.},
        title = "{Stellar winds on the main-sequence. II. The evolution of rotation and winds}",
      journal = {\aap},
         year = 2015,
        month = may,
       volume = {577},
          eid = {A28},
        pages = {A28},
          doi = {10.1051/0004-6361/201425301},
archivePrefix = {arXiv},
       eprint = {1503.07494},
 primaryClass = {astro-ph.SR},
       adsurl = {https://ui.adsabs.harvard.edu/abs/2015A&A...577A..28J}
}

@ARTICLE{Lu2024,
       author = {{Lu}, Yuxi(Lucy) and {Angus}, Ruth and {Foreman-Mackey}, Daniel and {Hattori}, Soichiro},
        title = "{In This Day and Age: An Empirical Gyrochronology Relation for Partially and Fully Convective Single Field Stars}",
      journal = {\aj},
         year = 2024,
        month = apr,
       volume = {167},
       number = {4},
          eid = {159},
        pages = {159},
          doi = {10.3847/1538-3881/ad28b9},
archivePrefix = {arXiv},
       eprint = {2310.14990},
 primaryClass = {astro-ph.SR},
       adsurl = {https://ui.adsabs.harvard.edu/abs/2024AJ....167..159L}
}

@ARTICLE{Dungee2022,
       author = {{Dungee}, Ryan and {van Saders}, Jennifer and {Gaidos}, Eric and {Chun}, Mark and {Garc{\'\i}a}, Rafael A. and {Magnier}, Eugene A. and {Mathur}, Savita and {Santos}, {\^A}ngela R.~G.},
        title = "{A 4 Gyr M-dwarf Gyrochrone from CFHT/MegaPrime Monitoring of the Open Cluster M67}",
      journal = {\apj},
         year = 2022,
        month = oct,
       volume = {938},
       number = {2},
          eid = {118},
        pages = {118},
          doi = {10.3847/1538-4357/ac90be},
archivePrefix = {arXiv},
       eprint = {2211.01377},
 primaryClass = {astro-ph.SR},
       adsurl = {https://ui.adsabs.harvard.edu/abs/2022ApJ...938..118D}
}

@ARTICLE{MH2008,
       author = {{Mamajek}, Eric E. and {Hillenbrand}, Lynne A.},
        title = "{Improved Age Estimation for Solar-Type Dwarfs Using Activity-Rotation Diagnostics}",
      journal = {\apj},
         year = 2008,
        month = nov,
       volume = {687},
       number = {2},
        pages = {1264-1293},
          doi = {10.1086/591785},
archivePrefix = {arXiv},
       eprint = {0807.1686},
 primaryClass = {astro-ph},
       adsurl = {https://ui.adsabs.harvard.edu/abs/2008ApJ...687.1264M}
}

@ARTICLE{Lay2004,
       author = {{Lay}, Oliver P.},
        title = "{Systematic Errors in Nulling Interferometers}",
      journal = {\ao},
         year = 2004,
        month = nov,
       volume = {43},
       number = {33},
        pages = {6100-6123},
          doi = {10.1364/AO.43.006100},
       adsurl = {https://ui.adsabs.harvard.edu/abs/2004ApOpt..43.6100L}
}

@ARTICLE{NK2023,
       author = {{Najita}, Joan R. and {Kenyon}, Scott J.},
        title = "{Takeout and Delivery: Erasing the Dusty Signature of Late-stage Terrestrial Planet Formation}",
      journal = {\apj},
         year = 2023,
        month = feb,
       volume = {944},
       number = {2},
          eid = {125},
        pages = {125},
          doi = {10.3847/1538-4357/acac8f},
archivePrefix = {arXiv},
       eprint = {2301.05719},
 primaryClass = {astro-ph.EP},
       adsurl = {https://ui.adsabs.harvard.edu/abs/2023ApJ...944..125N}
}

@ARTICLE{Cuntz2014,
       author = {{Cuntz}, M.},
        title = "{S-type and P-type Habitability in Stellar Binary Systems: A Comprehensive Approach. I. Method and Applications}",
      journal = {\apj},
         year = 2014,
        month = jan,
       volume = {780},
       number = {1},
          eid = {14},
        pages = {14},
          doi = {10.1088/0004-637X/780/1/14},
archivePrefix = {arXiv},
       eprint = {1303.6645},
 primaryClass = {astro-ph.EP},
       adsurl = {https://ui.adsabs.harvard.edu/abs/2014ApJ...780...14C}
}

@ARTICLE{Safonova2016,
       author = {{Safonova}, M. and {Murthy}, J. and {Shchekinov}, Yu. A.},
        title = "{Age aspects of habitability}",
      journal = {International Journal of Astrobiology},
         year = 2016,
        month = apr,
       volume = {15},
       number = {2},
        pages = {93-105},
          doi = {10.1017/S1473550415000208},
archivePrefix = {arXiv},
       eprint = {1404.0641},
 primaryClass = {astro-ph.EP},
       adsurl = {https://ui.adsabs.harvard.edu/abs/2016IJAsB..15...93S}
}

@ARTICLE{Swastik2023,
       author = {{Swastik}, C. and {Banyal}, Ravinder K. and {Narang}, Mayank and {Unni}, Athira and {Banerjee}, Bihan and {Manoj}, P. and {Sivarani}, T.},
        title = "{Age Distribution of Exoplanet Host Stars: Chemical and Kinematic Age Proxies from GAIA DR3}",
      journal = {\aj},
         year = 2023,
        month = sep,
       volume = {166},
       number = {3},
          eid = {91},
        pages = {91},
          doi = {10.3847/1538-3881/ace782},
archivePrefix = {arXiv},
       eprint = {2307.11442},
 primaryClass = {astro-ph.SR},
       adsurl = {https://ui.adsabs.harvard.edu/abs/2023AJ....166...91S}
}

@ARTICLE{Kral2023,
       author = {{Kral}, Q. and {Pringle}, J.~E. and {Matr{\`a}}, L. and {Th{\'e}bault}, P.},
        title = "{Potential effects of stellar winds on gas dynamics in debris disks leading to observable belt winds}",
      journal = {\aap},
         year = 2023,
        month = jan,
       volume = {669},
          eid = {A116},
        pages = {A116},
          doi = {10.1051/0004-6361/202243729},
archivePrefix = {arXiv},
       eprint = {2211.04191},
 primaryClass = {astro-ph.EP},
       adsurl = {https://ui.adsabs.harvard.edu/abs/2023A&A...669A.116K}
}

@ARTICLE{Lestrade2012,
       author = {{Lestrade}, J. -F. and {Matthews}, B.~C. and {Sibthorpe}, B. and {Kennedy}, G.~M. and {Wyatt}, M.~C. and {Bryden}, G. and {Greaves}, J.~S. and {Thilliez}, E. and {Moro-Mart{\'\i}n}, A. and {Booth}, M. and {Dent}, W.~R.~F. and {Duch{\^e}ne}, G. and {Harvey}, P.~M. and {Horner}, J. and {Kalas}, P. and {Kavelaars}, J.~J. and {Phillips}, N.~M. and {Rodriguez}, D.~R. and {Su}, K.~Y.~L. and {Wilner}, D.~J.},
        title = "{A DEBRIS disk around the planet hosting M-star GJ 581 spatially resolved with Herschel}",
      journal = {\aap},
         year = 2012,
        month = dec,
       volume = {548},
          eid = {A86},
        pages = {A86},
          doi = {10.1051/0004-6361/201220325},
archivePrefix = {arXiv},
       eprint = {1211.4898},
 primaryClass = {astro-ph.EP},
       adsurl = {https://ui.adsabs.harvard.edu/abs/2012A&A...548A..86L}
}

@ARTICLE{Stark2015,
       author = {{Stark}, Christopher C. and {Kuchner}, Marc J. and {Lincowski}, Andrew},
        title = "{The Pseudo-Zodi Problem for Edge-On Planetary Systems}",
      journal = {\apj},
         year = 2015,
        month = mar,
       volume = {801},
       number = {2},
          eid = {128},
        pages = {128},
          doi = {10.1088/0004-637X/801/2/128},
archivePrefix = {arXiv},
       eprint = {1502.02040},
 primaryClass = {astro-ph.SR},
       adsurl = {https://ui.adsabs.harvard.edu/abs/2015ApJ...801..128S}
}

@ARTICLE{RW2020,
       author = {{Rigley}, Jessica K. and {Wyatt}, Mark C.},
        title = "{Dust size and spatial distributions in debris discs: predictions for exozodiacal dust dragged in from an exo-Kuiper belt}",
      journal = {\mnras},
         year = 2020,
        month = sep,
       volume = {497},
       number = {1},
        pages = {1143-1165},
          doi = {10.1093/mnras/staa2029},
archivePrefix = {arXiv},
       eprint = {2007.01313},
 primaryClass = {astro-ph.EP},
       adsurl = {https://ui.adsabs.harvard.edu/abs/2020MNRAS.497.1143R}
}

@ARTICLE{Pearce2024,
       author = {{Pearce}, Tim D.},
        title = "{Debris disks around main-sequence stars}",
      journal = {arXiv e-prints},
         year = 2024,
        month = mar,
          eid = {arXiv:2403.11804},
        pages = {arXiv:2403.11804},
          doi = {10.48550/arXiv.2403.11804},
archivePrefix = {arXiv},
       eprint = {2403.11804},
 primaryClass = {astro-ph.EP},
       adsurl = {https://ui.adsabs.harvard.edu/abs/2024arXiv240311804P}
}

@ARTICLE{CG2023,
       author = {{Carri{\'o}n-Gonz{\'a}lez}, {\'O}scar and {Kammerer}, Jens and {Angerhausen}, Daniel and {Dannert}, Felix and {Garc{\'\i}a Mu{\~n}oz}, Antonio and {Quanz}, Sascha P. and {Absil}, Olivier and {Beichman}, Charles A. and {Girard}, Julien H. and {Mennesson}, Bertrand and {Meyer}, Michael R. and {Stapelfeldt}, Karl R. and {LIFE Collaboration}},
        title = "{Large Interferometer For Exoplanets (LIFE). X. Detectability of currently known exoplanets and synergies with future IR/O/UV reflected-starlight imaging missions}",
      journal = {\aap},
         year = 2023,
        month = oct,
       volume = {678},
          eid = {A96},
        pages = {A96},
          doi = {10.1051/0004-6361/202347027},
archivePrefix = {arXiv},
       eprint = {2308.09646},
 primaryClass = {astro-ph.EP},
       adsurl = {https://ui.adsabs.harvard.edu/abs/2023A&A...678A..96C}
}

@ARTICLE{KS2010,
       author = {{Kuchner}, Marc J. and {Stark}, Christopher C.},
        title = "{Collisional Grooming Models of the Kuiper Belt Dust Cloud}",
      journal = {\aj},
         year = 2010,
        month = oct,
       volume = {140},
       number = {4},
        pages = {1007-1019},
          doi = {10.1088/0004-6256/140/4/1007},
archivePrefix = {arXiv},
       eprint = {1008.0904},
 primaryClass = {astro-ph.EP},
       adsurl = {https://ui.adsabs.harvard.edu/abs/2010AJ....140.1007K}
}

@ARTICLE{KH2003,
       author = {{Kuchner}, Marc J. and {Holman}, Matthew J.},
        title = "{The Geometry of Resonant Signatures in Debris Disks with Planets}",
      journal = {\apj},
         year = 2003,
        month = may,
       volume = {588},
       number = {2},
        pages = {1110-1120},
          doi = {10.1086/374213},
archivePrefix = {arXiv},
       eprint = {astro-ph/0209261},
 primaryClass = {astro-ph},
       adsurl = {https://ui.adsabs.harvard.edu/abs/2003ApJ...588.1110K}
}

@article{Pastor2009,
   title={Motion of dust in mean motion resonances with planets},
   volume={103},
   ISSN={1572-9478},
   url={http://dx.doi.org/10.1007/s10569-009-9202-9},
   DOI={10.1007/s10569-009-9202-9},
   number={4},
   journal={Celestial Mechanics and Dynamical Astronomy},
   publisher={Springer Science and Business Media LLC},
   author={Pástor, Pavol and Klačka, Jozef and Kómar, Ladislav},
   year={2009},
   month=mar, pages={343–364} }

@ARTICLE{KP2015,
       author = {{Kennedy}, Grant M. and {Piette}, Anjali},
        title = "{Warm exo-Zodi from cool exo-Kuiper belts: the significance of P-R drag and the inference of intervening planets}",
      journal = {\mnras},
         year = 2015,
        month = may,
       volume = {449},
       number = {3},
        pages = {2304-2311},
          doi = {10.1093/mnras/stv453},
archivePrefix = {arXiv},
       eprint = {1503.01460},
 primaryClass = {astro-ph.EP},
       adsurl = {https://ui.adsabs.harvard.edu/abs/2015MNRAS.449.2304K}
}

@PHDTHESIS{Absil2006,
       author = {{Absil}, Olivier},
        title = "{Astrophysical Studies of Extrasolar Planetary Systems Using Infrared Interferometric Techniques}",
       school = {University of Liege, Belgium},
         year = 2006,
        month = jan,
       adsurl = {https://ui.adsabs.harvard.edu/abs/2006PhDT.......249A}
}

@ARTICLE{Charbonneau2005,
       author = {{Charbonneau}, David and {Allen}, Lori E. and {Megeath}, S. Thomas and {Torres}, Guillermo and {Alonso}, Roi and {Brown}, Timothy M. and {Gilliland}, Ronald L. and {Latham}, David W. and {Mandushev}, Georgi and {O'Donovan}, Francis T. and {Sozzetti}, Alessandro},
        title = "{Detection of Thermal Emission from an Extrasolar Planet}",
      journal = {\apj},
         year = 2005,
        month = jun,
       volume = {626},
       number = {1},
        pages = {523-529},
          doi = {10.1086/429991},
archivePrefix = {arXiv},
       eprint = {astro-ph/0503457},
 primaryClass = {astro-ph},
       adsurl = {https://ui.adsabs.harvard.edu/abs/2005ApJ...626..523C}
}

@ARTICLE{LZ1999,
       author = {{Liou}, Jer-Chyi and {Zook}, Herbert A.},
        title = "{Signatures of the Giant Planets Imprinted on the Edgeworth-Kuiper Belt Dust Disk}",
      journal = {\aj},
         year = 1999,
        month = jul,
       volume = {118},
       number = {1},
        pages = {580-590},
          doi = {10.1086/300938},
       adsurl = {https://ui.adsabs.harvard.edu/abs/1999AJ....118..580L}
}

@ARTICLE{Eddington1924,
       author = {{Eddington}, A.~S.},
        title = "{On the relation between the masses and luminosities of the stars}",
      journal = {\mnras},
         year = 1924,
        month = mar,
       volume = {84},
        pages = {308-332},
          doi = {10.1093/mnras/84.5.308},
       adsurl = {https://ui.adsabs.harvard.edu/abs/1924MNRAS..84..308E}
}

@ARTICLE{Menti2024,
       author = {{Menti}, Franziska and {Caballero}, Jos{\'e} A. and {Wyatt}, Mark C. and {Garc{\'\i}a Mu{\~n}oz}, Antonio and {Stassun}, Keivan G. and {Alei}, Eleonora and {Demleitner}, Markus and {Kennedy}, Grant and {Lichtenberg}, Tim and {Schmitt}, Uwe and {Schonhut-Stasik}, Jessica S. and {Wang}, Haiyang S. and {Quanz}, Sascha P. and {LIFE Collaboration}},
        title = "{Database of Candidate Targets for the LIFE Mission}",
      journal = {Research Notes of the American Astronomical Society},
         year = 2024,
        month = oct,
       volume = {8},
       number = {10},
          eid = {267},
        pages = {267},
          doi = {10.3847/2515-5172/ad887e},
archivePrefix = {arXiv},
       eprint = {2410.23892},
 primaryClass = {astro-ph.IM},
       adsurl = {https://ui.adsabs.harvard.edu/abs/2024RNAAS...8..267M}
}

@ARTICLE{Grun1985,
       author = {{Gr\"un}, E. and {Zook}, H.~A. and {Fechtig}, H. and {Giese}, R.~H.},
        title = "{Collisional balance of the meteoritic complex}",
      journal = {\icarus},
         year = 1985,
        month = may,
       volume = {62},
       number = {2},
        pages = {244-272},
          doi = {10.1016/0019-1035(85)90121-6},
       adsurl = {https://ui.adsabs.harvard.edu/abs/1985Icar...62..244G}
}

@ARTICLE{SK2009,
       author = {{Stark}, Christopher C. and {Kuchner}, Marc J.},
        title = "{A New Algorithm for Self-consistent Three-dimensional Modeling of Collisions in Dusty Debris Disks}",
      journal = {\apj},
         year = 2009,
        month = dec,
       volume = {707},
       number = {1},
        pages = {543-553},
          doi = {10.1088/0004-637X/707/1/543},
archivePrefix = {arXiv},
       eprint = {0909.2227},
 primaryClass = {astro-ph.SR},
       adsurl = {https://ui.adsabs.harvard.edu/abs/2009ApJ...707..543S}
}

@ARTICLE{Stark2022,
       author = {{Stark}, Christopher C.},
        title = "{ExoVista: A Suite of Planetary System Models for Exoplanet Studies}",
      journal = {\aj},
         year = 2022,
        month = mar,
       volume = {163},
       number = {3},
          eid = {105},
        pages = {105},
          doi = {10.3847/1538-3881/ac45f5},
archivePrefix = {arXiv},
       eprint = {2201.02652},
 primaryClass = {astro-ph.EP},
       adsurl = {https://ui.adsabs.harvard.edu/abs/2022AJ....163..105S}
}

@BOOK{HWO,
  author    = {{National Academies of Sciences, Engineering, and Medicine}},
  title     = "Pathways to Discovery in Astronomy and Astrophysics for the 2020s",
  isbn      = "978-0-309-46734-6",
  doi       = "10.17226/26141",
  url       = "https://nap.nationalacademies.org/catalog/26141/pathways-to-discovery-in-astronomy-and-astrophysics-for-the-2020s",
  year      = 2023,
  publisher = "The National Academies Press",
  address   = "Washington, DC"
}

@ARTICLE{MS2024,
       author = {{Mamajek}, Eric and {Stapelfeldt}, Karl},
        title = "{NASA Exoplanet Exploration Program (ExEP) Mission Star List for the Habitable Worlds Observatory (2023)}",
      journal = {arXiv e-prints},
         year = 2024,
        month = feb,
          eid = {arXiv:2402.12414},
        pages = {arXiv:2402.12414},
          doi = {10.48550/arXiv.2402.12414},
archivePrefix = {arXiv},
       eprint = {2402.12414},
 primaryClass = {astro-ph.IM},
       adsurl = {https://ui.adsabs.harvard.edu/abs/2024arXiv240212414M}
}

@ARTICLE{Fujiwara2013,
       author = {{Fujiwara}, H. and {Ishihara}, D. and {Onaka}, T. and {Takita}, S. and {Kataza}, H. and {Yamashita}, T. and {Fukagawa}, M. and {Ootsubo}, T. and {Hirao}, T. and {Enya}, K. and {Marshall}, J.~P. and {White}, G.~J. and {Nakagawa}, T. and {Murakami}, H.},
        title = "{AKARI/IRC 18 {\ensuremath{\mu}}m survey of warm debris disks}",
      journal = {\aap},
         year = 2013,
        month = feb,
       volume = {550},
          eid = {A45},
        pages = {A45},
          doi = {10.1051/0004-6361/201219841},
archivePrefix = {arXiv},
       eprint = {1211.6365},
 primaryClass = {astro-ph.EP},
       adsurl = {https://ui.adsabs.harvard.edu/abs/2013A&A...550A..45F}
}

@ARTICLE{Tuomi2013,
       author = {{Tuomi}, M. and {Jones}, H.~R.~A. and {Jenkins}, J.~S. and {Tinney}, C.~G. and {Butler}, R.~P. and {Vogt}, S.~S. and {Barnes}, J.~R. and {Wittenmyer}, R.~A. and {O'Toole}, S. and {Horner}, J. and {Bailey}, J. and {Carter}, B.~D. and {Wright}, D.~J. and {Salter}, G.~S. and {Pinfield}, D.},
        title = "{Signals embedded in the radial velocity noise. Periodic variations in the {\ensuremath{\tau}} Ceti velocities}",
      journal = {\aap},
         year = 2013,
        month = mar,
       volume = {551},
          eid = {A79},
        pages = {A79},
          doi = {10.1051/0004-6361/201220509},
archivePrefix = {arXiv},
       eprint = {1212.4277},
 primaryClass = {astro-ph.EP},
       adsurl = {https://ui.adsabs.harvard.edu/abs/2013A&A...551A..79T}
}

@ARTICLE{Morales2009,
       author = {{Morales}, Farisa Y. and {Werner}, M.~W. and {Bryden}, G. and {Plavchan}, P. and {Stapelfeldt}, K.~R. and {Rieke}, G.~H. and {Su}, K.~Y.~L. and {Beichman}, C.~A. and {Chen}, C.~H. and {Grogan}, K. and {Kenyon}, S.~J. and {Moro-Martin}, A. and {Wolf}, S.},
        title = "{Spitzer Mid-IR Spectra of Dust Debris Around A and Late B Type Stars: Asteroid Belt Analogs and Power-Law Dust Distributions}",
      journal = {\apj},
         year = 2009,
        month = jul,
       volume = {699},
       number = {2},
        pages = {1067-1086},
          doi = {10.1088/0004-637X/699/2/1067},
       adsurl = {https://ui.adsabs.harvard.edu/abs/2009ApJ...699.1067M}
}

@ARTICLE{Lestrade2025,
       author = {{Lestrade}, J. -F. and {Matthews}, B.~C. and {Kennedy}, G.~M. and {Sibthorpe}, B. and {Wyatt}, M.~C. and {Booth}, M. and {Greaves}, J.~S. and {Duch{\^e}ne}, G. and {Moro-Mart{\'\i}n}, A. and {Jobic}, C.},
        title = "{Debris disks around M dwarfs: The Herschel DEBRIS survey}",
      journal = {\aap},
         year = 2025,
        month = feb,
       volume = {694},
          eid = {A123},
        pages = {A123},
          doi = {10.1051/0004-6361/202451673},
archivePrefix = {arXiv},
       eprint = {2502.04441},
 primaryClass = {astro-ph.SR},
       adsurl = {https://ui.adsabs.harvard.edu/abs/2025A&A...694A.123L}
}

@INPROCEEDINGS{Binks2016,
       author = {{Binks}, Alex},
        title = "{The Debris Disk Fraction for M-dwarfs in Nearby Young Moving Groups}",
    booktitle = {Young Stars \& Planets Near the Sun},
         year = 2016,
       editor = {{Kastner}, J.~H. and {Stelzer}, B. and {Metchev}, S.~A.},
       series = {IAU Symposium},
       volume = {314},
        month = jan,
        pages = {159-162},
          doi = {10.1017/S1743921315006158},
archivePrefix = {arXiv},
       eprint = {1601.06835},
 primaryClass = {astro-ph.SR},
       adsurl = {https://ui.adsabs.harvard.edu/abs/2016IAUS..314..159B}
}

@ARTICLE{CC2023,
       author = {{Cronin-Coltsmann}, Patrick F. and {Kennedy}, Grant M. and {Kral}, Quentin and {Lestrade}, Jean-Fran{\c{c}}ois and {Marino}, Sebastian and {Matr{\`a}}, Luca and {Wyatt}, Mark C.},
        title = "{An ALMA Survey of M-dwarfs in the Beta Pictoris Moving Group with two new debris disc detections}",
      journal = {\mnras},
         year = 2023,
        month = dec,
       volume = {526},
       number = {4},
        pages = {5401-5417},
          doi = {10.1093/mnras/stad3083},
archivePrefix = {arXiv},
       eprint = {2310.15255},
 primaryClass = {astro-ph.SR},
       adsurl = {https://ui.adsabs.harvard.edu/abs/2023MNRAS.526.5401C}
}

@ARTICLE{ML2014,
       author = {{Morey}, {\'E}tienne and {Lestrade}, Jean-Fran{\c{c}}ois},
        title = "{On the steady state collisional evolution of debris disks around M dwarfs}",
      journal = {\aap},
         year = 2014,
        month = may,
       volume = {565},
          eid = {A58},
        pages = {A58},
          doi = {10.1051/0004-6361/201322567},
archivePrefix = {arXiv},
       eprint = {1404.1954},
 primaryClass = {astro-ph.EP},
       adsurl = {https://ui.adsabs.harvard.edu/abs/2014A&A...565A..58M}
}

@ARTICLE{CB1997,
       author = {{Chabrier}, Gilles and {Baraffe}, Isabelle},
        title = "{Structure and evolution of low-mass stars}",
      journal = {\aap},
         year = 1997,
        month = nov,
       volume = {327},
        pages = {1039-1053},
          doi = {10.48550/arXiv.astro-ph/9704118},
archivePrefix = {arXiv},
       eprint = {astro-ph/9704118},
 primaryClass = {astro-ph},
       adsurl = {https://ui.adsabs.harvard.edu/abs/1997A&A...327.1039C}
}

@ARTICLE{Baraffe2015,
       author = {{Baraffe}, Isabelle and {Homeier}, Derek and {Allard}, France and {Chabrier}, Gilles},
        title = "{New evolutionary models for pre-main sequence and main sequence low-mass stars down to the hydrogen-burning limit}",
      journal = {\aap},
         year = 2015,
        month = may,
       volume = {577},
          eid = {A42},
        pages = {A42},
          doi = {10.1051/0004-6361/201425481},
archivePrefix = {arXiv},
       eprint = {1503.04107},
 primaryClass = {astro-ph.SR},
       adsurl = {https://ui.adsabs.harvard.edu/abs/2015A&A...577A..42B}
}

@ARTICLE{Choi2016,
       author = {{Choi}, Jieun and {Dotter}, Aaron and {Conroy}, Charlie and {Cantiello}, Matteo and {Paxton}, Bill and {Johnson}, Benjamin D.},
        title = "{Mesa Isochrones and Stellar Tracks (MIST). I. Solar-scaled Models}",
      journal = {\apj},
         year = 2016,
        month = jun,
       volume = {823},
       number = {2},
          eid = {102},
        pages = {102},
          doi = {10.3847/0004-637X/823/2/102},
archivePrefix = {arXiv},
       eprint = {1604.08592},
 primaryClass = {astro-ph.SR},
       adsurl = {https://ui.adsabs.harvard.edu/abs/2016ApJ...823..102C}
}

@misc{miepy,
  author       = {Prahl, Scott},
  title        = {{miepython: Pure python calculation of Mie 
                   scattering}},
  month        = may,
  year         = 2024,
  publisher    = {Zenodo},
  version      = {2.5.4},
  doi          = {10.5281/zenodo.11135148},
  url          = {https://doi.org/10.5281/zenodo.11135148}
}

@ARTICLE{KW2024,
       author = {{Kim}, M. and {Wolf}, S.},
        title = "{Impact of discontinuous grain size distributions on the spectral energy distribution of debris disks}",
      journal = {\aap},
         year = 2024,
        month = mar,
       volume = {683},
          eid = {A148},
        pages = {A148},
          doi = {10.1051/0004-6361/202347168},
archivePrefix = {arXiv},
       eprint = {2312.15283},
 primaryClass = {astro-ph.EP},
       adsurl = {https://ui.adsabs.harvard.edu/abs/2024A&A...683A.148K}
}

@ARTICLE{Minato2004,
       author = {{Minato}, T. and {K{\"o}hler}, M. and {Kimura}, H. and {Mann}, I. and {Yamamoto}, T.},
        title = "{Momentum transfer to interplanetary dust from the solar wind}",
      journal = {\aap},
         year = 2004,
        month = sep,
       volume = {424},
        pages = {L13-L16},
          doi = {10.1051/0004-6361:200400037},
       adsurl = {https://ui.adsabs.harvard.edu/abs/2004A&A...424L..13M}
}

@ARTICLE{Matthews2015,
       author = {{Matthews}, Brenda C. and {Kennedy}, Grant and {Sibthorpe}, Bruce and {Holland}, Wayne and {Booth}, Mark and {Kalas}, Paul and {MacGregor}, Meredith and {Wilner}, David and {Vandenbussche}, Bart and {Olofsson}, G{\"o}ran and {Blommaert}, Joris and {Brandeker}, Alexis and {Dent}, W.~R.~F. and {de Vries}, Bernard L. and {Di Francesco}, James and {Fridlund}, Malcolm and {Graham}, James R. and {Greaves}, Jane and {Heras}, Ana M. and {Hogerheijde}, Michiel and {Ivison}, R.~J. and {Pantin}, Eric and {Pilbratt}, G{\"o}ran L.},
        title = "{The AU Mic Debris Disk: Far-infrared and Submillimeter Resolved Imaging}",
      journal = {\apj},
         year = 2015,
        month = oct,
       volume = {811},
       number = {2},
          eid = {100},
        pages = {100},
          doi = {10.1088/0004-637X/811/2/100},
archivePrefix = {arXiv},
       eprint = {1509.06415},
 primaryClass = {astro-ph.SR},
       adsurl = {https://ui.adsabs.harvard.edu/abs/2015ApJ...811..100M}
}

@ARTICLE{Metchev2005,
       author = {{Metchev}, Stanimir A. and {Eisner}, Joshua A. and {Hillenbrand}, Lynne A. and {Wolf}, Sebastian},
        title = "{Adaptive Optics Imaging of the AU Microscopii Circumstellar Disk: Evidence for Dynamical Evolution}",
      journal = {\apj},
         year = 2005,
        month = mar,
       volume = {622},
       number = {1},
        pages = {451-462},
          doi = {10.1086/427869},
archivePrefix = {arXiv},
       eprint = {astro-ph/0412143},
 primaryClass = {astro-ph},
       adsurl = {https://ui.adsabs.harvard.edu/abs/2005ApJ...622..451M}
}

@ARTICLE{Wright2010,
       author = {{Wright}, Edward L. and {Eisenhardt}, Peter R.~M. and {Mainzer}, Amy K. and {Ressler}, Michael E. and {Cutri}, Roc M. and {Jarrett}, Thomas and {Kirkpatrick}, J. Davy and {Padgett}, Deborah and {McMillan}, Robert S. and {Skrutskie}, Michael and {Stanford}, S.~A. and {Cohen}, Martin and {Walker}, Russell G. and {Mather}, John C. and {Leisawitz}, David and {Gautier}, III, Thomas N. and {McLean}, Ian and {Benford}, Dominic and {Lonsdale}, Carol J. and {Blain}, Andrew and {Mendez}, Bryan and {Irace}, William R. and {Duval}, Valerie and {Liu}, Fengchuan and {Royer}, Don and {Heinrichsen}, Ingolf and {Howard}, Joan and {Shannon}, Mark and {Kendall}, Martha and {Walsh}, Amy L. and {Larsen}, Mark and {Cardon}, Joel G. and {Schick}, Scott and {Schwalm}, Mark and {Abid}, Mohamed and {Fabinsky}, Beth and {Naes}, Larry and {Tsai}, Chao-Wei},
        title = "{The Wide-field Infrared Survey Explorer (WISE): Mission Description and Initial On-orbit Performance}",
      journal = {\aj},
         year = 2010,
        month = dec,
       volume = {140},
       number = {6},
        pages = {1868-1881},
          doi = {10.1088/0004-6256/140/6/1868},
archivePrefix = {arXiv},
       eprint = {1008.0031},
 primaryClass = {astro-ph.IM},
       adsurl = {https://ui.adsabs.harvard.edu/abs/2010AJ....140.1868W}
}

@ARTICLE{KW2013,
       author = {{Kennedy}, G.~M. and {Wyatt}, M.~C.},
        title = "{The bright end of the exo-Zodi luminosity function: disc evolution and implications for exo-Earth detectability}",
      journal = {\mnras},
         year = 2013,
        month = aug,
       volume = {433},
       number = {3},
        pages = {2334-2356},
          doi = {10.1093/mnras/stt900},
archivePrefix = {arXiv},
       eprint = {1305.6607},
 primaryClass = {astro-ph.EP},
       adsurl = {https://ui.adsabs.harvard.edu/abs/2013MNRAS.433.2334K}
}

@ARTICLE{Huang2025,
       author = {{Huang}, Dong and {Liu}, Qiong and {Wyatt}, Mark C. and {Kennedy}, Grant M.},
        title = "{WISE 12 {\ensuremath{\mu}}m search for exozodi candidates within 10 parsecs}",
      journal = {\aap},
         year = 2025,
        month = jun,
       volume = {698},
          eid = {A246},
        pages = {A246},
          doi = {10.1051/0004-6361/202554746},
archivePrefix = {arXiv},
       eprint = {2505.07602},
 primaryClass = {astro-ph.EP},
       adsurl = {https://ui.adsabs.harvard.edu/abs/2025A&A...698A.246H}
}

@ARTICLE{Johnstone2021,
       author = {{Johnstone}, C.~P. and {Bartel}, M. and {G{\"u}del}, M.},
        title = "{The active lives of stars: A complete description of the rotation and XUV evolution of F, G, K, and M dwarfs}",
      journal = {\aap},
         year = 2021,
        month = may,
       volume = {649},
          eid = {A96},
        pages = {A96},
          doi = {10.1051/0004-6361/202038407},
archivePrefix = {arXiv},
       eprint = {2009.07695},
 primaryClass = {astro-ph.SR},
       adsurl = {https://ui.adsabs.harvard.edu/abs/2021A&A...649A..96J}
}

@ARTICLE{Lohne2017,
       author = {{L{\"o}hne}, T. and {Krivov}, A.~V. and {Kirchschlager}, F. and {Sende}, J.~A. and {Wolf}, S.},
        title = "{Collisions and drag in debris discs with eccentric parent belts}",
      journal = {\aap},
         year = 2017,
        month = aug,
       volume = {605},
          eid = {A7},
        pages = {A7},
          doi = {10.1051/0004-6361/201630297},
archivePrefix = {arXiv},
       eprint = {1704.08085},
 primaryClass = {astro-ph.EP},
       adsurl = {https://ui.adsabs.harvard.edu/abs/2017A&A...605A...7L}
}

@ARTICLE{Draine2003,
       author = {{Draine}, B.~T.},
        title = "{Interstellar Dust Grains}",
      journal = {\araa},
         year = 2003,
        month = jan,
       volume = {41},
        pages = {241-289},
          doi = {10.1146/annurev.astro.41.011802.094840},
archivePrefix = {arXiv},
       eprint = {astro-ph/0304489},
 primaryClass = {astro-ph},
       adsurl = {https://ui.adsabs.harvard.edu/abs/2003ARA&A..41..241D}
}

@ARTICLE{Reid2011,
       author = {{Reidemeister}, M. and {Krivov}, A.~V. and {Stark}, C.~C. and {Augereau}, J.-C. and {L{\"o}hne}, T. and {M{\"u}ller}, S.},
        title = "{The cold origin of the warm dust around ɛ Eridani}",
      journal = {\aap},
         year = 2011,
        month = mar,
       volume = {527},
          eid = {A57},
        pages = {A57},
          doi = {10.1051/0004-6361/201015328},
archivePrefix = {arXiv},
       eprint = {1011.4882},
 primaryClass = {astro-ph.EP},
       adsurl = {https://ui.adsabs.harvard.edu/abs/2011A&A...527A..57R}
}

\begin{appendix}




%
\section{Parameters and simulation setup} \label{sec:app_setup}
Table~\ref{tab:params_def} summarizes the parameters and symbols used throughout this study, followed by Table~\ref{tab:setup_comparison}, which lists the simulation setups for both the pilot and main studies.
\begin{table}[htbp]
\caption{Summary of parameters and definitions}
\label{tab:params_def}
\centering
\begin{tabular}{ll}
\hline\hline
Symbol & \multicolumn{1}{p{0.8\linewidth}}{Definition} \\
\hline
$M_\ast$ & Stellar mass \\
$R_\ast$ & Stellar radius \\
$L_\ast$ & Stellar luminosity \\
$T_\ast$ & Stellar effective temperature \\
$\dot{M}_\ast$ & Stellar mass-loss rate \\
$v_\text{SW}$ & \parbox[t]{0.8\linewidth}{Stellar wind speed (fixed to $\sim 400~\mathrm{km\,s^{-1}}$ for all simulations)} \\
$P_\text{rot}$ & Stellar rotation period \\
$M_\text{p}$ & Planetary mass \\
$a_\text{p}$ & Planetary semimajor axis \\
$\rho$ & Dust particle density \\
$s$ & Dust particle radius \\
$\langle Q_\text{PR} \rangle$ & \parbox[t]{0.8\linewidth}{Radiation pressure efficiency averaged over the wind species}\\
$\langle Q_\text{SW} \rangle$ & \parbox[t]{0.8\linewidth}{Stellar wind efficiency averaged over stellar spectrum}\\
$\beta_\text{PR}$ & Ratio of radiation pressure to stellar gravity \\
$\beta_\text{SW}$ & Ratio of stellar wind force to stellar gravity \\
$\psi$ & Ratio of stellar wind drag to PR drag \\
$\beta_\text{r}$ & Radial effective beta \\
$\beta_\text{t}$ & Tangential effective beta \\
$t_\text{mig}$ & Effective migration time \\
$s_\text{BO,eff}$ & Effective blowout size \\
 $s_\text{BO,PR}$&Radiative blowout size\\
$s_\text{C,min}$ & \parbox[t]{0.8\linewidth}{Minimum particle size required to form resonant ring structures} \\
$n$ & Number density of dust \\
$\tau$ & Vertical geometrical optical depth \\
$C$ & \parbox[t]{0.8\linewidth}{Contrast between the maximum and background of the azimuthally averaged ring density} \\
$\langle C_\tau \rangle$ & \parbox[t]{0.8\linewidth}{Optical depth contrast of the size-integrated resonant ring} \\
\hline
\end{tabular}
\end{table}

\begin{table}[ht]
\centering
\caption{Summary of simulation setup}
\label{tab:setup_comparison}
\begin{tabular}{lll}
\hline\hline
Parameter & Pilot Study & Main Study\\
\hline
Number of particles & 100 & 5000 \\
$s$ & 50~$\mathrm{\mu m}$& 0.1--300~$\mathrm{\mu m}$\\
$M_\text{p}$ & 0.5, 1.0, 2.0~$M_\oplus$ & 1.0~$M_\oplus$ \\
$a_\text{p}$ & HZ$_\text{in}$, HZ$_\text{mid}$, HZ$_\text{out}$ & HZ$_\text{mid}$ \\
Stellar wind model & Model~I and II & Model~II only \\
\hline
Spectral types & \multicolumn{2}{c}{F4, G4, K4, M4} \\
$a_\text{d}$ & \multicolumn{2}{c}{$\sim$3~$a_\text{p}$} \\
$e_\text{p}$& \multicolumn{2}{c}{0.1} \\
$i_\text{p}$& \multicolumn{2}{c}{$5^\circ$} \\
\hline
\end{tabular}
\end{table}

\section{Extended analyses and additional caveats} \label{sec:app_anal}

\subsection{Effective blowout size across spectral types} \label{sec:dis_sbo}

\begin{figure*}
     \centering
     \includegraphics[width=0.47 \linewidth]{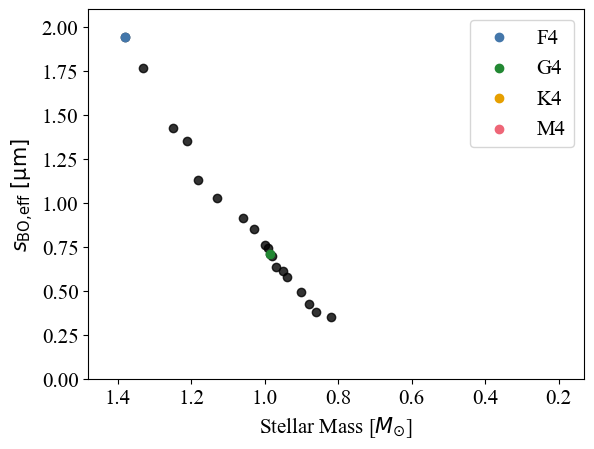}
     \includegraphics[width=0.47 \linewidth]{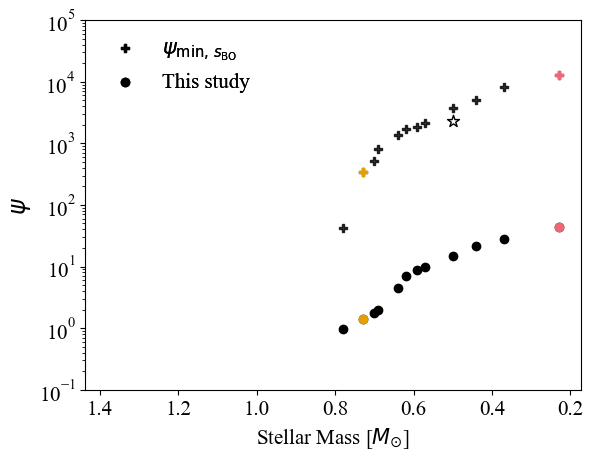}
     
     \caption{(\textit{Left}) Blowout size ($s_\text{BO,eff}$) derived from numerical calculations (see Section~\ref{sec:s_bo}) for main-sequence stars F4--M4 at an age of 4~Gyr (2--3~Gyr for F-type stars). The values for stars later than $\sim$~K2-type are not shown as they do not exist. (\textit{Right}) Minimum values of $\psi$ required for $s_\text{BO,eff}$ to exist ($\psi_{\text{min},\,s_\text{BO}}$) for stars without radiative blowout size, and their $\psi$ values estimated in this study (Fig.~\ref{fig:psi}). The star shape indicates the value corresponding to AU Mic from \citet{Plavchan2005}.}
     \label{fig:s_bo}
 \end{figure*}

The small blowout size for late-type stars provides implications for debris disk dust size distributions across spectral types. Given that smaller grains are produced abundantly by collisional cascades (Eq.~\ref{eq:crushing_law}), their retention or absence can influence the exozodi distributions. Including the specific values of $s_\text{BO, eff}$ presented in Section~\ref{sec:method_main}, estimates for all F4--M4 main-sequence stars at 4~Gyr (2--3~Gyr for F-type stars) are shown in the left panel of Fig.~\ref{fig:s_bo}. The $s_\text{BO,eff}$ values decrease almost monotonically toward lower-mass stars, where it becomes non-existent for spectral type later than $\sim$ K2. 
This trend of lower-mass stars retaining smaller grains of even submicrometer sizes is consistent with the results of \cite{KirchW2013}, where systems with stars $<$ 5250 K do not have any radiative blowout size when stellar wind is ignored. 

For systems with sufficiently strong stellar wind such that $\psi \frac{v_{\text{SW}}}{c}$ is non-negligible,  $\beta_\text{r} > \beta_\text{PR}$ can lead to the presence of effective blowout size even in cases where a radiative blowout size does not exist (see Section~\ref{sec:s_bo}). Solving for the criteria of $\beta_\text{r} = 0.5$, the minimum value of $\psi$ for a blowout size to exist can be described as follows (see Eq.~\ref{eq:beta_r}):
\begin{align} \label{eq:psi_sbo}
\psi_{\text{min},\,s_\text{BO}} &=  \left(\frac{ 0.5}{\beta_\text{PR, max}}-1 \right) \frac{c}{v_{\text{SW}}},
\end{align}
where $\beta_\text{PR, max}$ denotes the maximum value of $\beta_\text{PR}$ for a given star across different grain sizes (see Fig.~\ref{fig:Qpr_beta}). While the assumption of $v_\text{SW}\sim 400$ km/s is retained, inclusion of possible estimates of variation across stellar spectral types could lead to more accurate results.
The right panel of Fig.~\ref{fig:s_bo} shows the estimated $\psi_{\text{min},\,s_\text{BO}}$ for systems without radiative blowout sizes along with the $\psi$ values used in this study (Fig.~\ref{fig:psi}). The $\psi_{\text{min},\,s_\text{BO}}$ values are an order of $\sim$ 2--3 magnitudes higher, confirming the validity of the approximation $\beta_\text{r}\approx \beta_\text{PR}$ and thus  $s_\text{BO, eff}\approx s_\text{BO, PR}$ utilized in deriving the results. This also holds for a higher estimate $\psi \sim 127$ for M4-type stars derived using an alternative gyrochronology model (Section~\ref{sec:dis_gyro}). 

Many studies considering the stellar-wind blowout sizes  \citep[e.g.,][]{Lohne2017,Kim2018, KW2024} use the minimum grain size of $\sim 0.25 ~ \mathrm{\mu m}$ for systems without radiative blowout sizes. This value is derived from observations \citep[e.g.,][]{Metchev2005,Matthews2015} and the estimation by \citet{Plavchan2005} on a young M0 dwarf AU Mic of $\sim$~15 Myr age with $M_\ast = 0.5 ~ M_\odot$, $L_\ast = 1.3 ~ L_\odot$, assuming $v_\text{SW} \sim$ 400 km/s and $\dot{M}_\ast \sim 1000 ~ \dot{M}_\odot$. This corresponds to a value of $\psi \sim 2308$ (Eq.~\ref{eq: psi}) which is comparable to the threshold  $\psi_{\text{min},\,s_\text{BO}}$ as shown in Fig.~\ref{fig:s_bo}. The discrepancy arises because the presented $\psi_{\text{min},\,s_\text{BO}}$ values are based on typical main-sequence stellar masses and luminosities, whereas AU Mic is a pre-main-sequence star. Nonetheless, the comparison demonstrates that the high $\psi$ value for AU Mic is sufficient to produce an effective blowout size near $0.25 ~ \mathrm{\mu m}$ (see Section~\ref{sec:s_bo} and Fig.~\ref{fig:Qpr_beta}). 
Therefore, the commonly adopted minimum grain size of $\sim 0.25 ~ \mathrm{\mu m}$ for the dust distributions around K and M-type systems is valid for active young stars, unlike old systems of a few Gyr assumed in our simulations. This does not contradict the significance of stellar wind suggested in this study (Sections~\ref{sec:psi} and \ref{sec:res_pilot}), as the effect of $\psi$ on migration timescale and resonant ring contrast remains non-negligible, particularly for M-type stars (see Eq.~\ref{eq:t_eff} and Section~\ref{sec:dis_spt_res}). 

Despite discrepancies in estimated blowout sizes for M-type stars due to differing age assumptions, our results show that lower-mass stars retain smaller grains in their systems, consistent with the prediction of \citet{Plavchan2005} that M-dwarf debris disks have an excess of small grains and, consequently, a bluer disk color than earlier-type stars.
This is further supported by \citet{Fujiwara2013}, whose observations imply that FGK-type stars tend to host warm, inner debris dust that may include $\mathrm{\mu}$m-sized grains, whereas A-type stars lack such small grains. They base this interpretation on the absence of fine dust features in Spitzer/IRS spectra of A and late B-type stars \citep{Morales2009}, which they attribute to a larger blowout size around A stars compared to FGK stars.
Building on these earlier studies, our blowout size results suggest that smaller dust grains are retained toward later-type stars, from A through M.
The absence of blowout sizes for old systems derived in this study further implies that, for the same K--M stellar types, debris disks in older systems would appear even bluer and exhibit more fine dust features. In essence, our estimations of $s_\text{BO, eff}$ clarify the spectral-type dependence of blowout size and additionally indicate an age dependency for lower-mass stars. 

\subsection{Semi-analytical contrast analyses} \label{sec:dis_semianal}

While detailed distribution maps are essential for accurately modeling resonant structures relevant to interferometric detection, they can be complemented by a fast, generalized framework for estimating resonant ring strengths. To this end, we now turn to semi-analytical approaches, which offer a computationally efficient means of estimating exozodi levels across a broad range of system parameters without requiring full numerical simulations.  

\begin{figure*}
    \centering
    
    \includegraphics[width=0.47\linewidth]{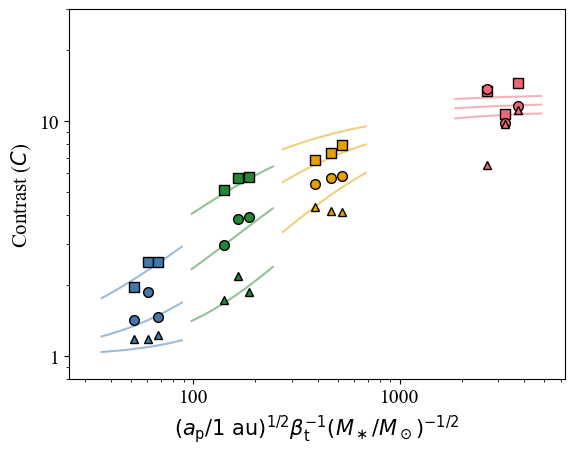}
    \includegraphics[width=0.47\linewidth]{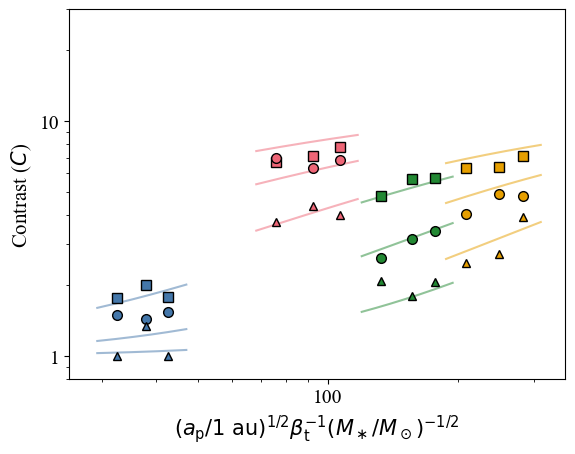}

    \caption{Same data as Fig.~\ref{fig:contrast_pilot}, but plotted against the modified term $a_\text{p}^{1/2} \left[ \beta_\text{PR} \left(1+\psi\right) \right]^{-1} M_\ast^{-1/2}$ for Model~I (\textit{left}) and Model~II (\textit{right}). The  tangential effective beta term $\beta_\text{t}=\beta_\text{PR}\left(1+\psi\right)$ (see Eq.~\ref{eq:beta_t}) is used for convenience. See the caption of Fig.~\ref{fig:contrast_pilot} for details on colors and symbols.  Solid lines show new fits (see Eq.~\ref{eq:C} for functional form), truncated to remain within the range relevant to each spectral type for clarity, as the contrast depends on both $M_\ast$ and $M_\text{p}$ unlike the original fits from \citet{SK2008}. Multiple lines of the same color correspond to different $M_\text{p}$, increasing from bottom to top. The $x$-axis scale in the right panel is reduced compared to Fig.~\ref{fig:contrast_pilot}, reflecting variations in $\psi$.}
    \label{fig:contrast_pilot_mod}
\end{figure*}

\begin{figure*}[htbp]
  \centering
  \includegraphics[width=0.47\linewidth]{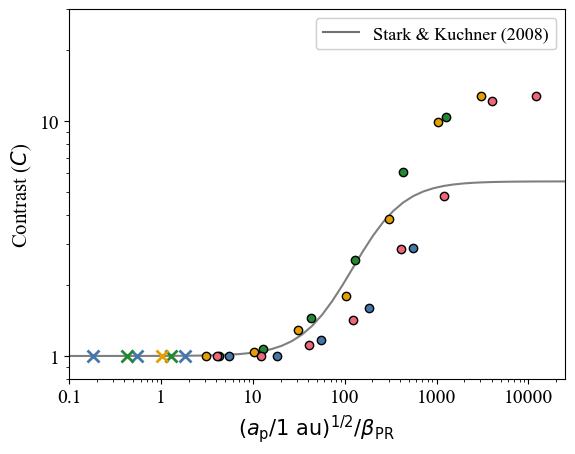}
  \includegraphics[width=0.47\linewidth]{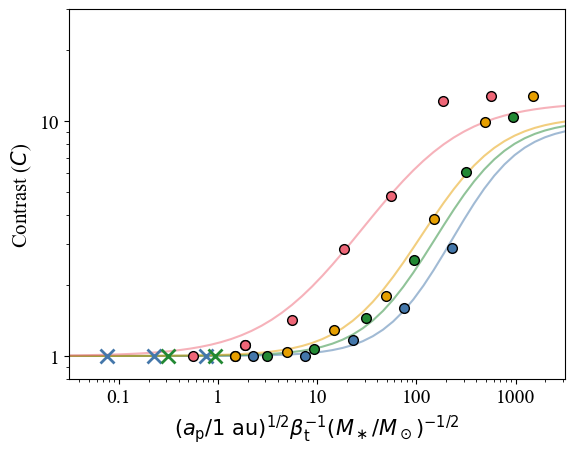}
  \caption{Same data as Fig.~\ref{fig:contrast_main_s}, but plotted against the original term $a_\text{p}^{1/2}/\beta_\text{PR}$ from \cite{SK2008} (\textit{left}) and the modified term $a_\text{p}^{1/2} \left[ \beta_\text{PR} \left(1+\psi\right) \right]^{-1} M_\ast^{-1/2}$ (\textit{right}). Since $\beta_\text{PR}$ is inversely proportional to grain size (Eq.~\ref{eq:beta_pr}) and $a_\text{p}$ is fixed, larger x-values indicate larger grains. K4 (yellow) and M4 (red) appear to have one fewer point because their $0.3~ \mathrm{\mu m}$ and $1~ \mathrm{\mu m}$ data replace the $0.1~ \mathrm{\mu m}$ points respectively, causing overlap (see Section~\ref{sec:method_main}). See the caption of Fig.~\ref{fig:contrast_main_s} for more details on colors and symbols. A single solid fit line is shown per spectral type, as only $M_\text{p} = 1~ M_\oplus$ and $a_\text{p} = \text{HZ}_\text{mid}$ was considered in the main study setup.}
  \label{fig:contrast_main_mod}
\end{figure*}

The semi-analytical expression of contrast (Eq.~\ref{eq:C}) expanded from the formulation of \citet{SK2008}, were used to fit the results presented in Sections~\ref{sec:res_models_vs} and~\ref{sec:res_main_C}.
Fig.~\ref{fig:contrast_pilot_mod} show the same datasets from the pilot study plotted against the modified expression $a_\text{p}^{1/2} \left[ \beta_\text{PR} \left(1+\psi\right) \right]^{-1} M_\ast^{-1/2}$ from Eq.~\eqref{eq:mod_x}, along with the improved contrast fits. Since this modified $x$-axis expression increases with decreasing stellar mass, the data generally follow the order F4, G4, K4, to M4. However, the high $\psi$ value for the M-type star significantly shifts their data toward smaller $x$-values, altering their relative positions in the plot. 
Cases with different planetary mass other than 1~$M_\oplus$ constitute a small portion of the total dataset due to the dominance of the main study data. This contributes to some deviations of the pilot study data from the modified fits, combined with the uncertainties caused by the small number of particles, as previously mentioned. Nonetheless, the data show overall better agreement under the modified expressions, as seen by comparing Figs.~\ref{fig:contrast_pilot} and~\ref{fig:contrast_pilot_mod}. 
Similarly, Fig.~\ref{fig:contrast_main_mod} shows a better fit to the main study data using the modified parameters, supporting the utility of the semi-analytical function for approximate estimations.

From the modified contrast function,  the semi-analytical value of the optical depth contrast can be derived using Eqs.~(12) and (13) from \cite{SK2008}:
\begin{equation} \label{eq:C_tau}
\langle C_\tau \rangle = \frac{\sum  s^2 n_{\mathrm{BG}} C }{\sum s^2 n_{\mathrm{BG}}} = \frac{\int_{s_{\min}}^{s_{\max}} s^{3 - \alpha} C(s) \, ds}{\int_{s_{\min}}^{s_{\max}} s^{3 - \alpha} \, ds},
\end{equation}
where the background number density $n_\text{BG} \propto s^{-\alpha} \cdot t_\text{mig} \propto s^{1-\alpha}$, and the function from Eq.~\eqref{eq:C} with the best-fit parameters (Eq.~\ref{eq:contrast_p}) is used for contrast $C(s)$. If the minimum dust size is smaller than the effective blowout size ($s_\text{min} < s_\text{BO,eff}$), the blowout size (see Section~\ref{sec:method_main}) is used instead.
The resulting semi-analytical values of $\langle C_\tau \rangle$ are 1.71, 4.40, 5.81, and 6.55 for F4, G4, K4, and M4, respectively. These values are comparable to, but $\sim30\%$ smaller than those directly calculated from the optical depth distributions (see Section~\ref{sec:res_tau_F}). The discrepancies reflect limitations in the contrast function (Eq.~\ref{eq:C}) and fitting uncertainties, since the function is not entirely reliable for extreme values of $\beta_\text{PR}$ and $M_\text{p}$, as noted by \cite{SK2008}.
Nevertheless, the semi-analytical $\langle C_\tau \rangle$ values can be used to approximately estimate $\tau_\text{BG} \cdot \langle C_\tau \rangle$ when the background level of dust is given. Thus, the semi-analytical functions remain valuable as a first-order method to estimate resonant ring optical depths for a variety of cases without conducting intensive simulations. 

The presence of the resonant structures can also be inferred from the semi-analytical contrast function. 
Among the eight discrete sizes used in the main study, the minimum size where the resonant ring appears ($s_\text{C,min}$) is approximately 10, 3, 1, and 1~$\mathrm{\mu m}$ for F4, G4, K4, and M4, respectively (Section~\ref{sec:res_main_C}). These values are broadly consistent with the more precise estimates of 14.8, 1.78, 0.70, and 0.18~$\mathrm{\mu m}$ obtained from the fitted functions.
A resonant ring structure is expected if the smallest size of dust present in the system is below $s_\text{C,min}$, since $s_\text{C,min}$ typically exceeds the corresponding values of $s_\text{BO,eff}$ (Section~\ref{sec:method_main}).
Therefore, the fitted functions provide a way to estimate $s_\text{C,min}$ and assess the presence of resonant structures.
When stellar and planetary parameters and the dust size distribution of a system are known, the semi-analytical approach enables a quick assessment of the likelihood and strength of resonant ring structures.

\subsection{Alternative results using a different gyrochronology model} \label{sec:dis_gyro}
\begin{figure}
    \centering
    \includegraphics[width=0.95\linewidth]{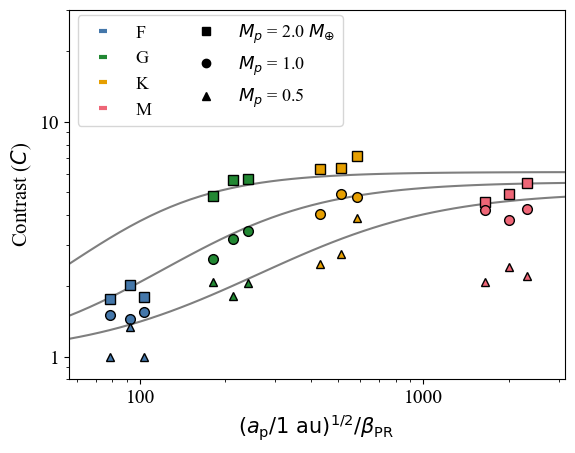}
    \caption{Right panel of Fig.~\ref{fig:contrast_pilot} re-plotted to show the estimated results when adopting a higher $\psi \sim$ 127 for M4-type stars, based on the alternative gyrochronology model of \citet{Lu2024}.}
    \label{fig:pilot_diff_gyro}
\end{figure}

While we have adopted the conservative value of $\psi \sim$ 44 for M4-type stars throughout our study, using a higher value of $\psi \sim$ 127 derived from an alternative gyrochronology model (\citealt{Lu2024}; see Sections~\ref{sec:psi} and \ref{sec:dis_psi_acc_age}) leads to changes in the trend of resonant ring contrast and optical depth across spectral types. 
With this higher $\psi$, the increase in stellar wind drag can outweigh the effect of decreasing $M_{\ast}$ (see Section~\ref{sec:dis_spt_res}), resulting in an estimated contrast reduction by a factor of $\sim$ 0.6 for dust size $ s = 50 ~  \mathrm{\mu m}$, as used in the pilot study. This rough estimate is derived from the ratio of contrasts for different $\psi$ values, using the semi-analytical function in Eq.~\eqref{eq:C}. 
Fig.~\ref{fig:pilot_diff_gyro} shows a re-plotted version of the right panel of Fig.~\ref{fig:contrast_pilot}, illustrating the expected change when applying this alternative $\psi$ value to M4-type cases in Model~II. Under this estimate, the contrasts for M4-type stars become comparable to that of G4-type, and K4-type stars exhibit the highest contrasts. These changes in trend are likewise expected to occur in the contrast values of the main study in Fig.~\ref{fig:contrast_main_s}.

Similarly, we estimate that the optical depth contrast for M4-type stars would be reduced by a factor of $\sim$ 0.7 when adopting this larger $\psi$, based on the semi-analytical function of Eq.~\eqref{eq:C_tau}. This yields an estimated value of $\langle C_\tau \rangle \sim$ 6.92 for M4-type, making it second to K4-type in contrast strength. This comparison is based on the original values of 1.95, 5.98, 8.07, and 9.85 for F4, G4, K4, and M4, respectively (see Section~\ref{sec:res_tau_F}).
The corresponding maximum optical depths approximated by $\tau_\text{BG} \cdot \langle C_\tau \rangle$ would reflect the same change in ordering, with values expected to be around $\sim$ 6, 18, 24, and 21~zodis for F4, G4, K4, and M4, respectively, after accounting for the different gyrochronology model (see Section \ref{sec:dis_spt_res}).  
However, this estimated change does not affect the trend in asymmetric or maximum fluxes in resonant structures presented in Fig.~\ref{fig:flux_max}. In fact, it would reinforce the peak appearing at K-type, since the flux from M4-type systems would further decrease due to their reduced optical depth.
In this scenario, exozodi resonant structures in M-type systems would clearly pose the least hindrance on MIR interferometric detection of exo-Earths. 

This potentially stronger influence of stellar wind drag in M-type systems would lead to even more rapid dust removal, supporting the stellar-wind-driven explanation for the lack of debris disk detections around old M-type stars, assuming a smooth disk with limited dust production (Section~\ref{sec:dis_psi}).
While our earlier analysis suggested that, in the presence of planet-induced resonant structures, warm exozodis may be more optically thick toward M-type stars than current data indicate (e.g., HOSTS; \citealt{Ertel2020}, see Section~\ref{sec:dis_spt_res}), these alternative estimates imply that resonant trapping produces a slightly larger enhancement of dust around K-type stars.
This shift indicates that the predicted optical depth for M-type systems is sensitive to the stellar wind environment and rotational evolution, and may plausibly range from values comparable to G-type stars to levels exceeding those of K-type.
Overall, our results suggest that the trend of increasing resonant contrast and optical depth with decreasing stellar mass is generally consistent, though a potential peak or inversion within the K--M spectral range remains. 

\subsection{Assumptions on stellar evolution and implications for F-type stars} \label{sec:dis_F_evol}
For all simulations and stellar wind drag estimates in this study, we used the same stellar parameters listed in Table~\ref{tab:parameters}, assuming that these properties remain largely unchanged over the main-sequence lifetime, as described in Section~\ref{sec:method_pilot}.
While this assumption is valid for low-mass stars, it is unphysical for F4-type stars at 4~Gyr, as they have evolved off the main-sequence and would no longer be classified as F4-type.
According to MIST stellar evolution models \citep{Choi2016}, the age of 4~Gyr is not applicable to stars with initial masses above $\sim$ 1.28~$M_\odot$ (earlier than F6), as these stars have evolved off the main sequence by that time. In particular, a star with an initial mass of $\sim$ 1.38~$M_\odot$ becomes a white dwarf by 4~Gyr.

Since such evolved remnants are observationally rare, and exoplanet detection surveys mostly target main-sequence stars \citep[e.g.,][]{MS2024, Menti2024}, our use of the typical main-sequence parameters remains appropriate. The $\psi$ value for a typical main-sequence F4-type star at nominal ages of 2--3 Gyr would be $\sim$1.21--0.90, which can also be approximately inferred from Figure~\ref{fig:psi_age}, slightly higher than the 4~Gyr value under the same stellar parameters. This implies a modest reduction in ring contrast, optical depth, and peak exozodi flux relative to the results of our study if the exact 2--3~Gyr $\psi$ values were used, assuming a constant background level (see Section~\ref{sec:dis_psi} and Eq.~\ref{eq:C}). 
Consequently, using $\psi$ values more representative of typical main-sequence ages for F-type stars does not alter the overall trend of increasing optical depth toward later spectral types, nor the conclusion that asymmetric exozodi flux is brightest around K-type stars. 

Therefore, while this approximation introduces some uncertainty, our adopted stellar parameters hold for the scope of this study, and our primary conclusions remain robust.

\subsection{Inner disk emission and IWA effects} \label{sec:app_IWA}
\begin{figure*}
    \centering
    \includegraphics[width=1\linewidth]{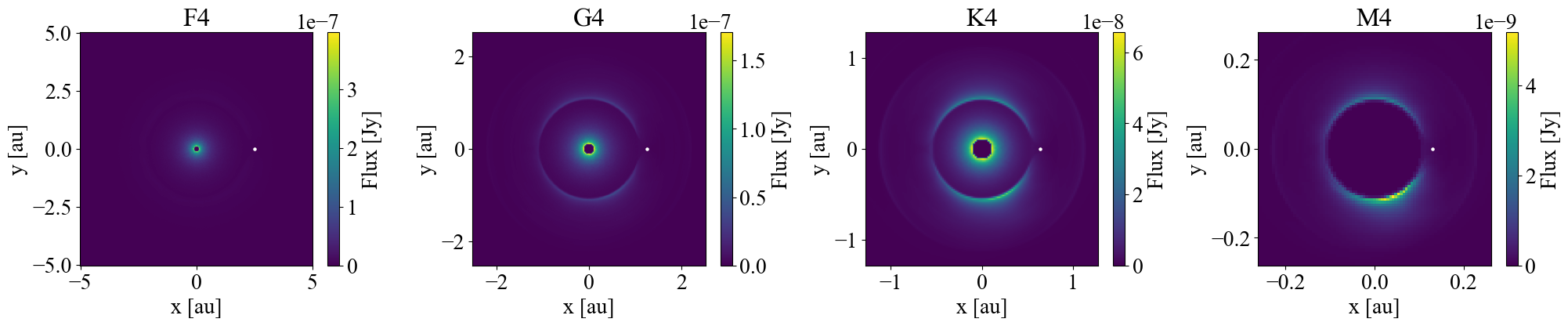}
    \caption{Flux maps (10~$\mathrm{\mu m}$) same as the lower row of Fig.~\ref{fig:maps} but reproduced with the region within the IWA of LIFE obscured. }
    \label{fig:flux_IWA}
\end{figure*}
For completeness, we also generated flux maps that exclude the regions within the instrument’s inner working angle (IWA) of LIFE, computed as $\sim 0.5 ~ \lambda/B$. As noted in Section~\ref{sec:res_tau_F}, this corresponds to $\sim$~0.01$^{\prime\prime}$ at $10~\mathrm{\mu m}$ and a baseline of 100~m, which is slightly smaller than the orbital separation of the planet in the M4-type system.
In these IWA-masked maps, the bright central emission due to hot dust near the star (see Fig.\ref{fig:maps}) is partially or fully suppressed, especially in more compact systems of K and M-type stars. Although higher-mass stars also exhibit some central flux-free zones due to the IWA cutoff, these regions are comparatively small relative to the overall system scale.

\end{appendix}
\end{document}